\documentclass[preprint2]{proto}
\usepackage{graphicx,epsf,amsmath,amssymb,amsfonts,wrapfig,comment,natbib}
\usepackage[a4paper, total={172mm,252mm}]{geometry}
\graphicspath{{Figures/}{Figures/in_paper/}} 
\usepackage{xcolor}
\usepackage{lipsum}  
\usepackage{xspace}
\usepackage[normalem]{ulem}
\usepackage[utf8]{inputenc}
\usepackage[T1]{fontenc}

\let\olditemize\itemize
\renewcommand{\itemize}{
\olditemize
\setlength{\itemsep}{0pt}
\setlength{\parskip}{0pt}
\setlength{\parsep}{0pt}
}
\let\olditemize\enumerate
\renewcommand{\enumerate}{
\olditemize
\setlength{\itemsep}{0pt}
\setlength{\parskip}{0pt}
\setlength{\parsep}{0pt}
}

\newcommand{\mearth}{\ensuremath{M_{\oplus}}\xspace} 
\renewcommand{\mp}{\ensuremath{M_{p}}\xspace} 
\newcommand{\rp}{\ensuremath{R_{p}}\xspace} 
\newcommand{\rearth}{\ensuremath{R_{\oplus}}\xspace}

\newcommand{\cm}[1]{\textcolor{magenta}{{#1}}}
\renewcommand{\textcolor}[2]{#2}
\renewcommand{\color}[2]{#2}

\usepackage{times}
\usepackage{amsmath}

\begin{document}

\title{\textbf{\LARGE Architectures of Compact Multi-planet Systems: \\ 
Diversity and Uniformity}}

\author {\textbf{\large Lauren M. Weiss}}
\affil{\small\em Department of Physics, University of Notre Dame, Notre Dame, IN 46556, USA}
\author {\textbf{\large Sarah C. Millholland}}
\affil{\small\em Department of Astrophysical Sciences, Princeton University, Princeton, NJ 08544, USA}
\author {\textbf{\large Erik A. Petigura}}
\affil{\small\em Department of Physics and Astronomy, University of California, Los Angeles, CA 90095, USA}
\author {\textbf{\large Fred C. Adams}}
\affil{\small\em Department of Physics, University of Michigan, Ann Arbor, MI 48109, USA}
\author {\textbf{\large Konstantin Batygin}}
\affil{\small\em  Division of Geological and Planetary Sciences, California Institute of Technology, Pasadena, CA 91125, USA}
\author {\textbf{\large Anthony M. Bloch}}
\affil{\small\em Department of Mathematics, University of Michigan, Ann Arbor, MI 48109, USA}
\author {\textbf{\large Christoph Mordasini}}
\affil{\small\em Physikalisches Institut, Universit{\"a}t Bern, Gesellschaftstrasse 6, 3012 Bern, Switzerland}

\begin{abstract}
\baselineskip = 11pt
\leftskip = 0.65in 
\rightskip = 0.65in
\parindent=1pc
{\small 
``I do not believe that even in a snowflake this ordered pattern exists at random.'' -- Johannes Kepler \\~\\}

One of the most important developments in exoplanet science in the past decade is the discovery of  multi-planet systems with sub-Neptune-sized planets interior to 1~AU. This chapter explores the architectures of these planetary systems, which often display a remarkable degree of uniformity: the planets have nearly equal sizes, regular orbital spacing, low eccentricities, and small mutual inclinations. This uniformity stands in sharp contrast to the diverse nature of the exoplanet sample considered as a whole (as well as our inner solar system). We begin with a critical review of the observations --- including possible biases --- and find that these peas-in-a-pod planetary systems are {\color{red}apparently }a common outcome of the planet formation process. 
Modest departures from exact uniformity suggest additional patterns, such as the planet mass slowly increasing with semi-major axis. The star formation process naturally produces circumstellar disks with the properties required to produce these planetary systems, although the solid material must move inward from its initial location. We discuss primary modes of planetary assembly, the role of orbital migration, and post-nebular atmospheric loss. Mature planetary systems are found to be near their minimum energy (tidal equilibrium) configurations; this finding provides a partial explanation for their observed properties and indicates that efficient energy dissipation must occur. Finally, we consider population synthesis models and show that peas-in-a-pod {\color{red}patterns emerge} with reasonable choices for the input parameters. Nonetheless, interesting observational and theoretical challenges remain in order to understand how these surprisingly organized planetary systems arise from the disorder of their formation processes. 

$$\,$$

$$\,$$

$$\,$$

\end{abstract}  

\section{Introduction}
\label{sec:intro} 

With thousands of extrasolar planets now detected, the inventory of the possible architectures for planetary systems is now coming into focus. One defining feature of this ensemble of data is that the planet formation process can produce an enormous diversity of outcomes. The possibilities include large planets found in tight orbits ($P\lesssim10$ days; hot Jupiters), smaller planets with ultra-short periods ($P\lesssim1$ day), planets in binary star systems, eccentric Jovian planets, planets much larger than Jupiter on distant orbits ($a\sim100$ AU), Earth-sized planets in potentially habitable orbits, and even some systems roughly analogous to our own solar system. Significantly, the collection of detected planets also includes a large number of super-Earth and sub-Neptune planets {\color{red} --- bodies with sizes ranging from 1 to 4~$R_\oplus$ and masses ranging from roughly 1 to  15~$M_\oplus$ --- sizes and masses not represented in our solar system. }

In contrast to the diversity found in the exoplanet sample taken as a whole, a large fraction of {\color{red}sub-Neptune }sized planets reside in multi-planet systems that display unexpected uniformity. These latter systems have planets with nearly equal sizes {\color{red}(and, where measurements are available, masses)}, with regularly-spaced orbits that are nearly circular and coplanar. Moreover, planetary pairs in these systems are generally not found in mean-motion resonance. These well-ordered compact planetary systems, and their degree of intra-system uniformity, are the primary focus of this review.

To date, the majority of known multi-planet systems were discovered by the NASA {\it Kepler} space telescope during its the prime mission (2009--2013). {\it Kepler} discovered thousands of planets with sizes between that of Earth and Neptune, with $\sim 40\%$ of them in systems containing more than one planet (\citealt{Batalha2013}; for a comprehensive review of results from {\it Kepler}, see the chapter by Lissauer, Batalha, \& Borucki). The {\it Kepler} population is particularly valuable due to the observing strategy of the mission; $\sim$150,000 stars were monitored nearly continuously for 4 years, regardless of whether or not they hosted transiting planets. Because the observing strategy was straightforward and well-documented, it is possible to model how it has shaped planet detections, and one can make projections to characterize the underlying planet population.  For example, \citet{Mulders2018} estimate that $\gtrsim$42\% of Sun-like stars have nearly coplanar planetary systems with seven or more exoplanets. Although the occurrence rate is subject to some uncertainty, such multi-planet systems represent one of the most common outcomes of planet formation.  For completeness, we note that subsequent transit missions (the {\it Kepler} extended mission and the NASA {\it TESS} mission) and radial velocity surveys have found additional multi-planet systems, with the latter important for making mass estimates and detecting long-period giant planets. 

\begin{figure}[h!]
\centering
\includegraphics[width=1\columnwidth]{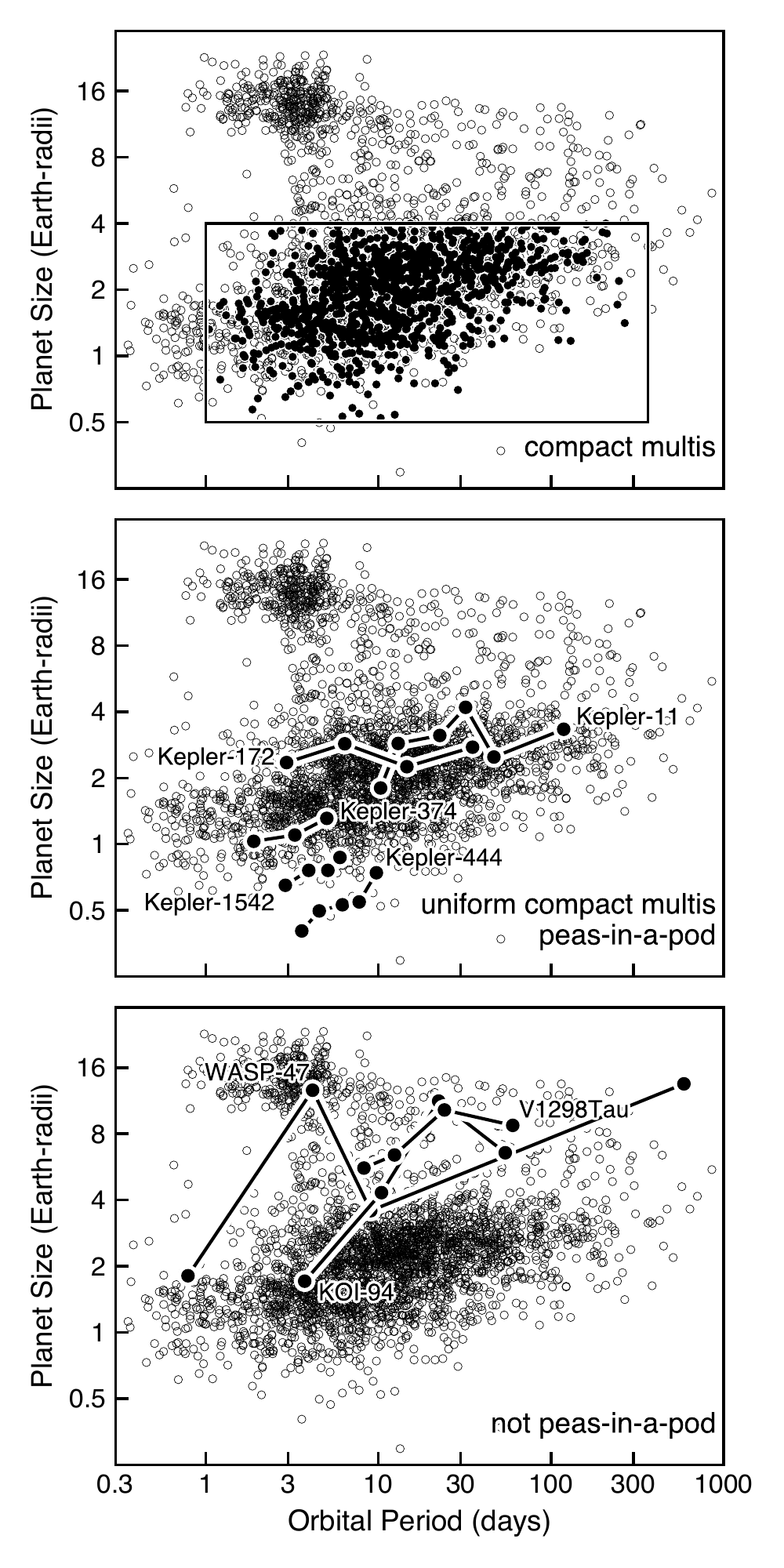}
\caption{Top: Census of confirmed extrasolar planets discovered by the transit method. Top: we define a compact multi to be a system with at least two planets with sizes and periods residing in the box. Middle: a few examples of the many compact multis that have a narrow dispersion of size and orbital spacing. Bottom: WASP-47 and KOI-94 are two systems that meet our definition of compact multi but have a large dispersion in size. V1298 Tau fails our definition, but may be evolving toward a compact multi in the distant future. (Source: NASA Exoplanet Archive, 2021-06-03).} 
\label{fig:context}
\end{figure}

Before characterizing these compact multi-planet systems in detail, we must specify the parameter space spanned by the systems of interest. For the sake of definiteness, we adopt the following working definition: {\em We define a compact multi-planet system (`compact multi') to be a planetary system containing multiple planets with radii between 0.5 and 4.0~$\rearth$ and with orbital periods between one day and one year.} In Figure~\ref{fig:context}, we show these planets in the period-radius ($P$-$R_p$) plane, alongside the broader population of known planets. Our definition is agnostic about the presence (or absence) of giant planets and/or additional stars\footnote{\color{red}The role of giant planets in shaping the architectures of compact multis is certainly interesting, but is beyond the scope of this review.}. In spite of its utility, this observationally-motivated definition has the following shortcomings: (1) many single transiting systems may indeed be compact multis viewed from glancing angles, and (2) some systems may be closely related to the compact multis from a formation standpoint, but may fail our size and period criteria. 

With the above definition in place, we note that most --- but not all --- of the compact multi-planet systems contain planets with nearly equal sizes (masses) and regularly spaced orbits. {\color{red} This so-called ``peas-in-a-pod'' pattern of intra-system uniformity is an emergent property of compact multi-planet systems.  Note that ``peas-in-a-pod'' is a pattern, not a particular group of planetary systems, although it is possible to define subsets of the compact multis that adhere to this pattern based on a variety of metrics.}  
The specification of these well-ordered systems, and the mechanisms that lead to their formation, represent a defining theme of this chapter.

For purposes of illustration, it is useful to highlight particularly significant compact multi-planet systems according to the definitions given above. Kepler-11 is a noteworthy and historically significant system \citep{Lissauer2011-kepler11}. This example, along with other early {\it Kepler} discoveries, showed that the inner regions of many planetary systems differed from that of the solar system and from the predictions of the leading theoretical models of the day. Kepler-11 hosts five transiting planets interior to Mercury's orbit with masses of $\mp \approx$ 2--8~$\mearth$. The total mass in these planets is $M_T\approx22~\mearth$ compared to only $0.05~\mearth$ in Mercury itself, an enhancement relative to the solar system by a factor of $\sim$400 \citep{Lissauer2013}. The planets themselves range from 1.8~$\rearth$ to 4.2~$\rearth$ and provide early examples of close-in Neptunes and sub-Neptunes.
%
%
The planets have a 0.11~dex dispersion in size (about 30\% fractional dispersion). Many compact multi-planet systems have an even narrower dispersion in size. Figure~\ref{fig:context} shows three of the most uniform Kepler systems, having size dispersion of 0.04~dex or less. Systems like these motivate the theoretical work presented here. 

On the other hand, not all compact multis have such a narrow size dispersion. Figure~\ref{fig:context} also highlights WASP-47 and Kepler-89 (a.k.a. KOI-94), two well-known systems that host both sub-Neptune and Jovian-size planets \citep{Becker2015,Weiss2013}. Such systems are relatively rare, but theoretical models must allow for occasional exceptions to uniformity. 

As noted earlier, some systems that meet the definition for being a compact multi could have formed differently from the majority of such systems, whereas others that fail the definition could have formed through similar mechanisms.  V1298 Tau is one such system \citep{David2019}. With four planets with sizes $\rp = 6$--$10~\rearth$, it does not meet our compact multi definition. However, given that the system has an age of only 20~Myr, it is plausible that these planets are actively undergoing Kelvin-Helmholtz contraction and/or atmospheric loss. The V1298 Tau system could thus be on its way to becoming a compact multi in the future.

Our own solar system is another useful point of reference. Here, the planetary orbits are  mostly circular and coplanar, with orbital  eccentricities $e<0.2$ and inclinations $i<7^\circ$. The terrestrial planets are close to the Sun (with $a<1.6$ AU), whereas the gas and ice giants inhabit the space beyond the snow line ($a>5$ AU).  
Our solar system is thus relatively well-ordered, and has been used in the past as the ultimate example of a deterministic system or `clockwork' universe \citep{newton1687}.\footnote{Of course, the advent of chaos complicates this description.} As outlined below, however, a large fraction of the compact multi-planet systems of interest here are {\it even more well-ordered} than the solar system. 

This chapter is organized as follows. First, we review the prevailing physical and orbital patterns found in compact multis (\S\ref{sec:observations}), including the degree of uniformity in their masses, radii, and orbital spacing. Next, we review key aspects of star/disk formation that set the boundary conditions for disk properties and planet formation (\S\ref{sec:stardisk}). We then explore the specific processes by which planets form in their disks, with a particular focus on the physics that sets the characteristic mass scale for compact multi-planet systems (\S\ref{sec:planet-formation-theory}). We also explore the planet-planet interactions that regulate mass growth, and we develop a pairwise energy optimization model that naturally produces peas-in-a-pod architectures (\S\ref{sec:pp interactions}). With these results in place, we consider population synthesis models (\S\ref{sec:popsynth}) and show that peas-in-a-pod patterns can be produced under reasonable assumptions. The chapter concludes (\S\ref{sec:conclusion}) with a summary of results and a discussion of how future work can move forward our understanding of these well-ordered planetary systems. 

\section{Properties of Compact Multi-Planet Systems}
\label{sec:observations}

\subsection{Homogeneous Catalogs}
{\color{red}The discovery of several striking attributes of the compact multi-planet systems relied on large, homogeneous catalogs of stellar and planet properties.  A particularly important advance came from the California Kepler Survey (CKS, \citealt{Petigura2017}), in which high-resolution spectra from the W. M. Keck Observatory were used to derive a homogeneous catalog of Kepler stellar and planet properties for 2025 planets orbiting 1305 stars.  This was the first homogeneous catalog of over a thousand planet-hosting stars.  More recently, thanks to the ESA Gaia mission, other large catalogs of homogeneously determined Kepler stellar and planet properties have been constructed \citep[e.g.,][]{Fulton2018,Berger2020}.  Their results are generally consistent with findings based on the CKS catalog.\footnote{Note that these catalogs focus on planets from the Kepler Mission; the K2 and TESS missions found fewer than 10 high-multiplicity (4+) planetary systems total, and thus have little to contribute regarding the planet-planet statistics of compact multis.}}

\begin{figure}[h!]
\centering
\includegraphics[width=1\columnwidth]{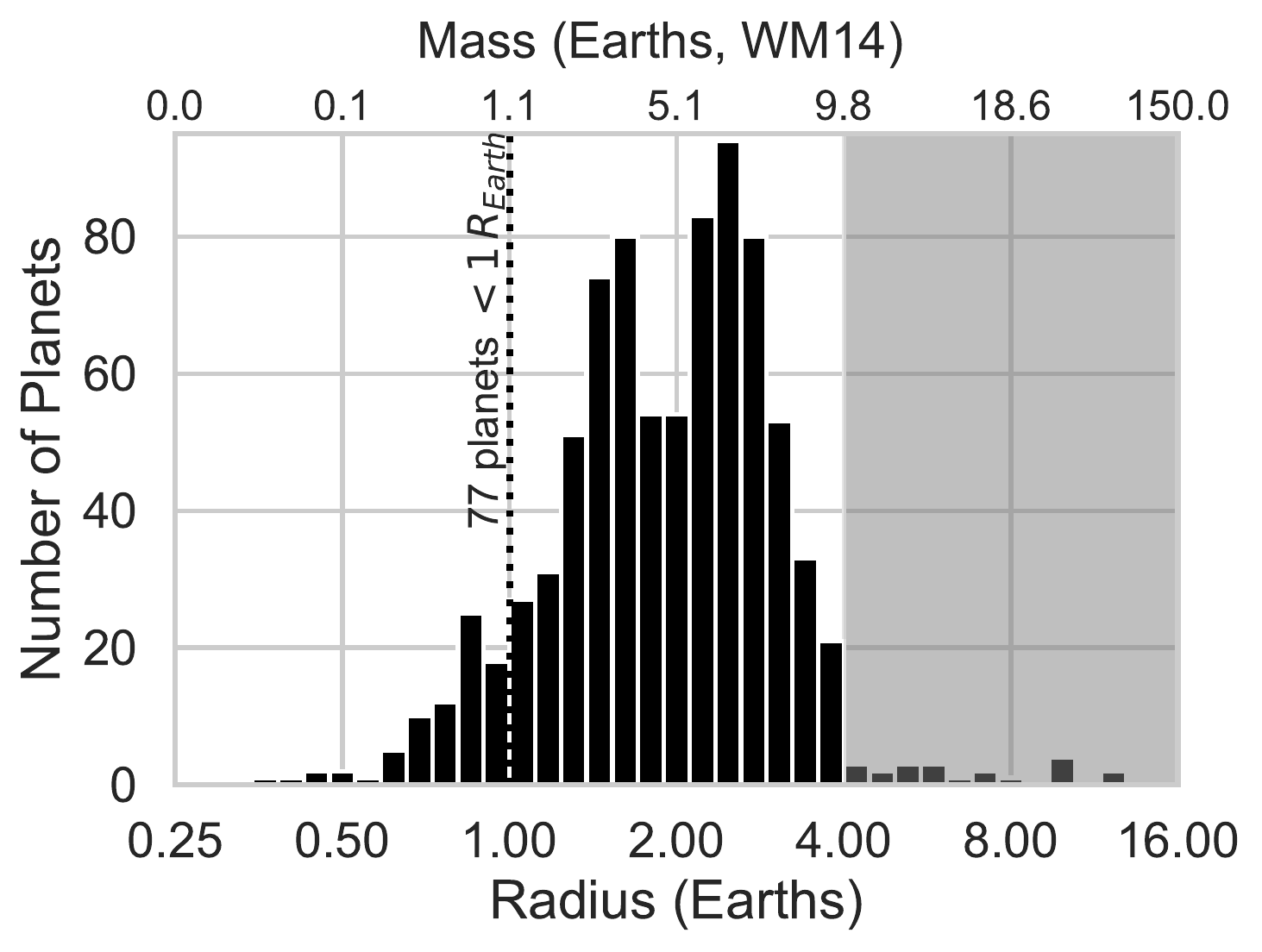}
\caption{The distribution of {\color{red}transiting} planet radii in the compact multis {\color{red}from the California Kepler Survey (CKS)}.  There is a significant gap at $\sim1.8\,\rearth$.  Planets larger than $4\,\rearth$, which sometimes occur in systems that have multiple planets smaller than $4\,\rearth$, are grayed out.  Empirically motivated estimates of the planet masses, based on \citet{Weiss2014}, are provided along the top axis.  Assuming planets smaller than $1.5\,\rearth$ are rocky, the distribution includes planets as low-mass as $0.1\,\mearth$.  The majority of planets $\sim1.5\,\rearth$ and larger, which likely have $\sim5\%$ of their mass in volatile envelopes of hydrogen and helium, are 3 to $10\,\mearth$.}
\label{fig:rp_mp_dist}
\end{figure}

\begin{figure*}
\centering
\includegraphics[width=0.82\textwidth]{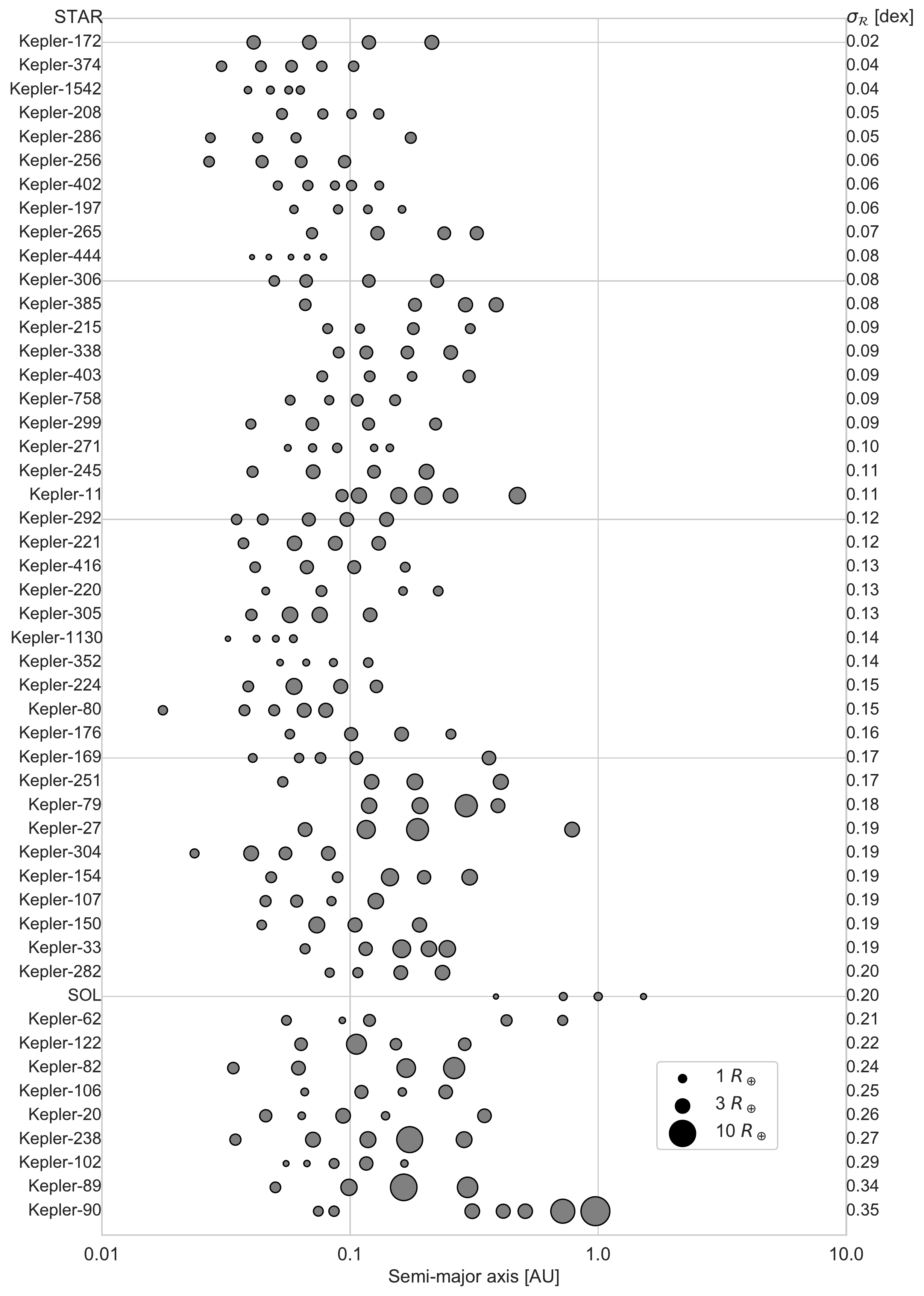}
\caption{Compact multi-planet systems with four or more transiting planets interior to 1.52 AU \citep[data from the California Kepler Survey,][]{Weiss2018a}.  The systems are ranked by their planet radius dispersions $\sigma_R$, with the most uniform sizes at the top and the greatest size diversity at the bottom.  The point sizes represent the planet radii on a logarithmic scale (see legend).  The solar system terrestrial planets are included for comparison; their $\sigma_R$ is the bottom quintile of size uniformity.  Although we do not know how many solar system analogs Kepler missed (since it was insensitive to Mars-sized planets at Mars-like orbits), the peas-in-a-pod architecture that is prevalent in the Kepler compact multis did not emerge as strongly in the final architecture of the solar system.}
\label{fig:peas-ranked-sigmaR}
\end{figure*}

Patterns emerging from the observed architectures of compact multis, including aspects of either diversity or uniformity, provide important clues as to how these ubiquitous planets formed. Such patterns relate to planetary physical properties (sizes and masses), orbital properties (spacing, eccentricities, inclinations), or a mix of the two (e.g. size-spacing relationships). Moreover, we can also examine whether planetary patterns are correlated with the properties of their host stars. This section reviews the primary patterns that have been identified in the observed sample of compact multi-planet systems. For a preview of the full range of observations that will be discussed, one may skip ahead to a schematic summary diagram in Figure \ref{fig:summary-of-observations}. 


\subsection{Uniform Sizes and Masses}
\label{sec: uniform sizes and masses}

\paragraph{Radius and Mass Distributions.}
The overall distribution of the Kepler planet radii is broad, with the sizes of detected planets ranging from the size of Mars to Jupiter. However, the size distribution is far from uniform: Earth-to-Neptune-size planets outnumber giants by an order of magnitude. As a point of reference, there are roughly $\sim$90 planets per 100 stars with radii \rp = 1--4~\rearth and periods less than 1 year, compared to only $\sim$7 planets per 100 stars with radii \rp = 8--32~\rearth in the same period range \citep{Petigura2018}. 

Thanks to the large number of transiting planets discovered by Kepler and intensive follow-up efforts, fine details have been resolved within the planet size distribution. Using the CKS sample, \cite{Fulton2017} identified a paucity of planets in the size range 1.5--2.0~\rearth compared to slightly larger and smaller planets. This gap aligned with a previously observed transition in planetary bulk density over the same size range \citep{Weiss2014}. Smaller planets (known as ``super-Earths'') have densities consistent with iron/silicate compositions while larger planets (``sub-Neptunes'') have low densities and require volatile envelopes that add significant volume to the planets \citep{Rogers2015}.  The planet radius gap is present in both the overall CKS sample and in the CKS compact multis \citep{Weiss2018b}. 

Although masses have been measured for only $\sim$5\% of compact multis, we can obtain mass estimates using a mass-radius relationship. Here we adopt the piecewise mass-radius relationship of \cite{Weiss2014}, but note that there is an astrophysical dispersion about this relationship of $\sim$4.2~\mearth.%
\footnote{We do not model mass dispersion in this chapter.}
Using this empirical mass-radius function, we find that the planets in compact multi-planet systems have masses ranging from 0.1 to 20 \mearth, with the majority of the detected planets having \mp $\approx$ 3--10~\mearth. Figure \ref{fig:rp_mp_dist} shows the full distributions for both radius and mass estimates for the sample. Although most of the planets are super-Earths and sub-Neptunes, the distribution has a tail extending below 1~\mearth (containing $\sim$8.6\% of the planets). 
 
\paragraph{Uniformity in Planet Sizes.}  Various studies have found that planets within a given multi-planet system have similar sizes to their neighbors.  The size similarity was first noted in \citep{Lissauer2011-architecture}, and was more robustly confirmed of 909 planets homogeneously characterized planets in 355 Kepler multi-planet systems \citep{Weiss2018a}. Most of these planetary systems fit the definition of compact multis provided in \S\ref{sec:intro} (833 planets in 322 systems). The peas-in-a-pod pattern is visible by eye in the high-multiplicity (4+ planet) systems shown in Figure \ref{fig:peas-ranked-sigmaR}. In the most uniform systems, the planets appear to have a characteristic size that is well-correlated with their neighbors. 

\begin{figure*}
\centering
\includegraphics[width=1.0\textwidth]{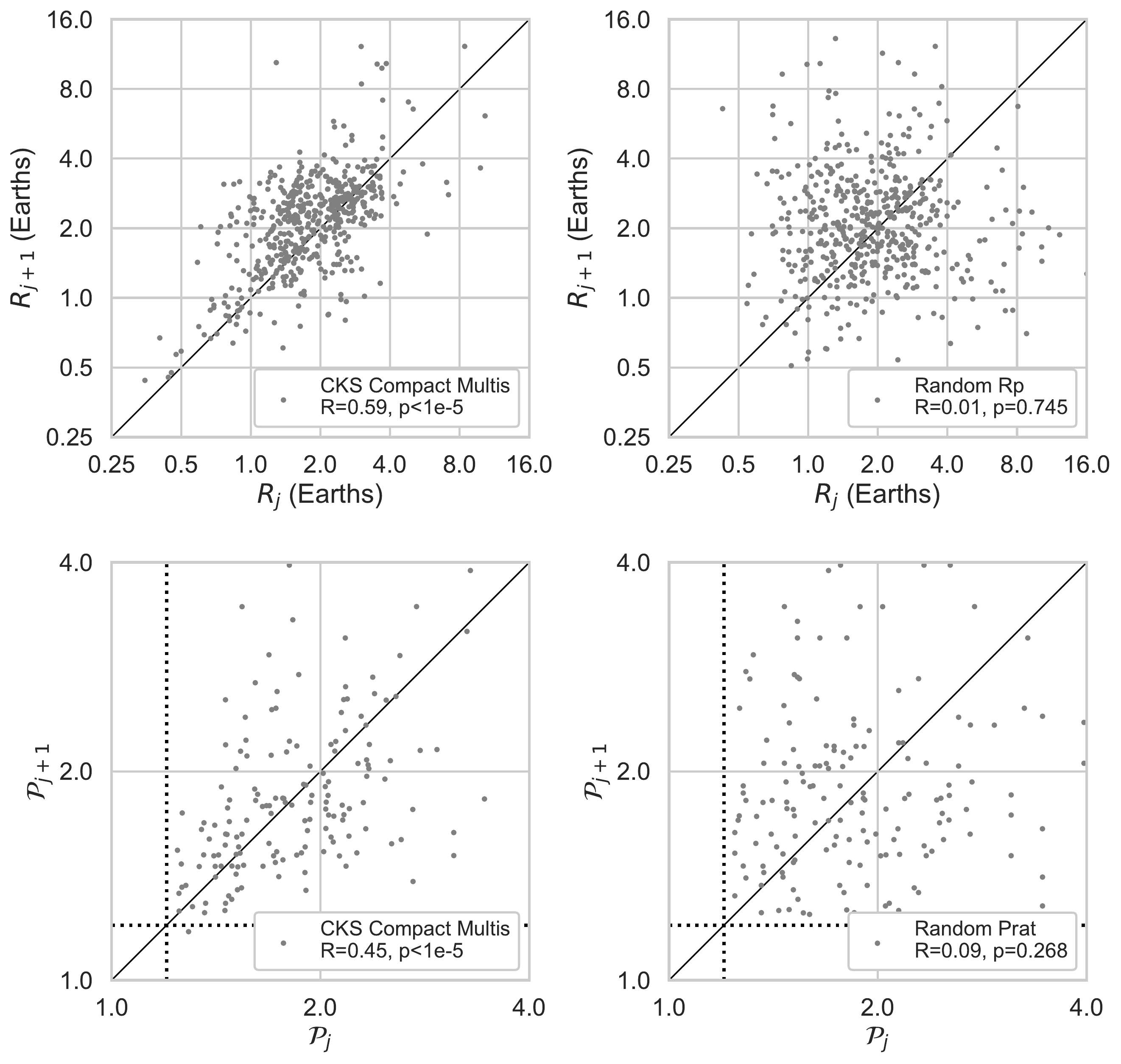}
\caption{Top left: The relationship between the radii of adjacent transiting planets, $j$ and $j+1$, from the CKS compact multi-planet systems. Top right: The same, but for planet radii drawn at random with replacement from a log-normal distribution that survive a mock Kepler detection pipeline and yield synthetic compact multis.  Bottom left: The relationship between the period ratios of adjacent pairs of transiting planets from the CKS compact multis.  Note the absence of period ratios $<1.2$ (dotted lines), which is likely due to instability.  Bottom right: The same, but for period ratios drawn at random from the observed distribution, with replacement (i.e., bootstrap). The random draws of planet radius and period ratio, when subjected to detection bias, selection criteria, and the stability requirement $\mathcal{P} > 1.2$, do not resemble the observed planetary systems.}
\label{fig:random_draws}
\end{figure*}

One might expect the size of each detected planet to be consistent with a random draw from a single underlying distribution, with a combination of detection and selection biases sculpting the pattern.  However, the size similarity found among the compact multis is inconsistent with random draws.  \citet{Weiss2018a} and \citet{Weiss2020} performed a simple controlled experiment in which they held fixed the stellar, orbital, and noise properties of the observed planets, but drew synthetic planet radii at random from both the observed and log-normal radius distributions. As shown in Figure \ref{fig:random_draws}, the observed planetary systems show more uniformity in radius than the systems produced through random sampling. Several studies have assumed more complicated underlying radius distributions (including covariance between planet radius and other properties) in an attempt to simultaneously reproduce a variety of observed distributions of the compact multis \citep{Mulders2018,He2019,Sandford2019,He2020}. These studies also found that the planet sizes are more correlated within a given planetary system than would be produced through random draws.  

Just how similar are the planet sizes in a given system?  A useful way to approach this question is to define a metric for size similarity.  The metric should be independent of the typical planet size.  For a given system of $N$ detected planets indexed by increasing orbital distance ($j=1,2,\dots N$), we consider the fractional dispersion of the planet radii,
\begin{equation}
    \sigma^2_{\mathcal R} = {\rm Variance} \left\{ \mathrm{log_{10}}(R_j/R_\oplus) \right\} \,, 
    \label{eqn:size-disp}
\end{equation}
where we take the logarithm of planet radius to ensure that the dispersion is fractional and we use $R_\oplus$ as a reference scale to make the variable dimensionless. 
The choice of log-base 10 is arbitrary.  Note that the size dispersion is only defined for systems with three or more planets; a similar metric for the dispersion in planet spacing (presented below) requires four or more planets.  In this review, we compute the size and spacing variances for systems with four or more transiting planets.  For completeness, we note that the literature contains many additional metrics for size diversity \citep[e.g.,][]{Millholland2017, Gilbert2020}.

In the sample of 4+ planet systems plotted in Figure \ref{fig:peas-ranked-sigmaR}, the systems are ranked in order of increasing $\sigma_{\mathcal R}$, with the systems with the lowest size dispersions at the top. The median value of the fractional dispersion is $\sigma_{\mathcal R}=0.14$~dex. This value is lower than the median size dispersion in systems with radii drawn at random (0.18 dex).  As another point of comparison, the inner solar system {\color{red}(Mercury, Venus, Earth, and Mars)} has $\sigma_{\mathcal R} = 0.20$~dex, placing it in the most disordered quintile of high-multiplicity systems.  Several multis with more diverse sizes than the terrestrial planets have a giant planet within 1 AU that is partially responsible for the size dispersion (e.g., KOI-834; KOI-94, \citealt{Weiss2013}; and KOI-351 \citealt{Schmitt2014, Cabrera2014}), although some systems have sub-Neptunes with diverse sizes (e.g., KOI-70, \citealt{Buchhave2016}; KOI-82, \citealt{Marcy2014}).  

Although there is some size dispersion in all systems, the variations in planet radii tend to be orderly. For example, \citet{Ciardi2013} found that the outer planet {\color{red}is larger in $68\pm6\%$ of }pairs of transiting planets (where this trend is limited to cases where the smaller planet would be detectable with the orbital period of either planet).  {\color{red}For comparison, this pattern holds for 66\% of pairs in the inner solar system (with Mars breaking the pattern).}  \citet{Kipping2018} reproduced this result and also found that size ordering in high-multiplicity systems tends to be monotonic, indicating that the planet sizes are in a low-entropy state.

\paragraph{Uniformity in Planet Masses.}
The intra-system similarity found for planetary radii extends to the planet masses as well. This finding suggests that the bulk compositions of planets, including  their envelope-to-core fractions, are also likely to be relatively uniform.  \citet{Millholland2017} first demonstrated mass uniformity using a sample of 37 systems with masses determined from Transit-Timing Variations (TTVs), as measured by \cite{Hadden2017}. They found that masses of planets were more similar to the masses of neighboring planets than would be expected from random draws. \cite{Millholland2017} also identified radius uniformity, radius ordering, and mass ordering within this sample. \cite{Wang2017} later studied a sample of multi-planet systems with mass measurements from radial velocities and showed that this sample also exhibited intra-system mass uniformity. The tendency for uniformity in both radii and masses implies that the significant scatter in the mass-radius relation (the $M_p$-$R_p$ plane) is dominated by system-to-system variance rather than intra-system variance.  

For completeness, we note that the current determinations of mass uniformity rely primarily on mass estimates made from TTVs (with the exception of \citealt{Wang2017}). TTVs are most readily detected for large planets with ultra-compact (period ratios $\lesssim2$) or near-resonant orbital spacing \citep{Mills&Mazeh2017}, which might necessitate that they have low masses to maintain stable orbits. Thus, it remains possible that the systems that display TTVs are systematically different from those systems that do not \citep{Weiss2014,Lee2016}. \cite{Millholland2019b}, for instance, showed that planets wide of resonance have systematically larger radii than non-resonant planets, suggesting some degree of physical uniqueness of the TTV planet sample. Exploring whether the similarity in planet masses and compositions robustly extends beyond the TTV systems will be a major avenue of investigation in the era of sub-meter-per-second RVs. 

Despite this caveat, there is ample reason to suggest that intra-system mass uniformity is a widespread trend. First, the intra-system uniformity of planet radii exists both for systems with larger planets (which are likely to have envelopes) and for the 72 planets with radii smaller than Earth, including a tail of Mars-sized planets.  At an orbital distance of 0.1 AU, a Mars-sized planet is unlikely to hold a volatile atmosphere. We can thus assume that the Mars-sized planets are rocky, and the size similarity in systems such as KOI-3158 (see Figures \ref{fig:context} and  \ref{fig:peas-ranked-sigmaR}) is indicative of similarity in mass (in addition to radius).  
Moreover, even among planets with envelopes, atmospheric mass loss (see \S \ref{sec: atmospheric mass loss}) tends to reduce planet radii down to $R_p \approx 2 \ R_\mathrm{core}$ \citep{Owen2017}, such that their uniform radii are indicative of uniform core sizes (and thus masses, assuming compositional similarity). Lastly, mass uniformity within systems is a fundamental outcome of the planet formation process (as discussed in \S 3--6).




\subsection{Uniform, Non-Resonant Orbital Spacing}
\label{sec:spacing}

\begin{figure*}[h!]
\centering
\includegraphics[width=0.42\textwidth]{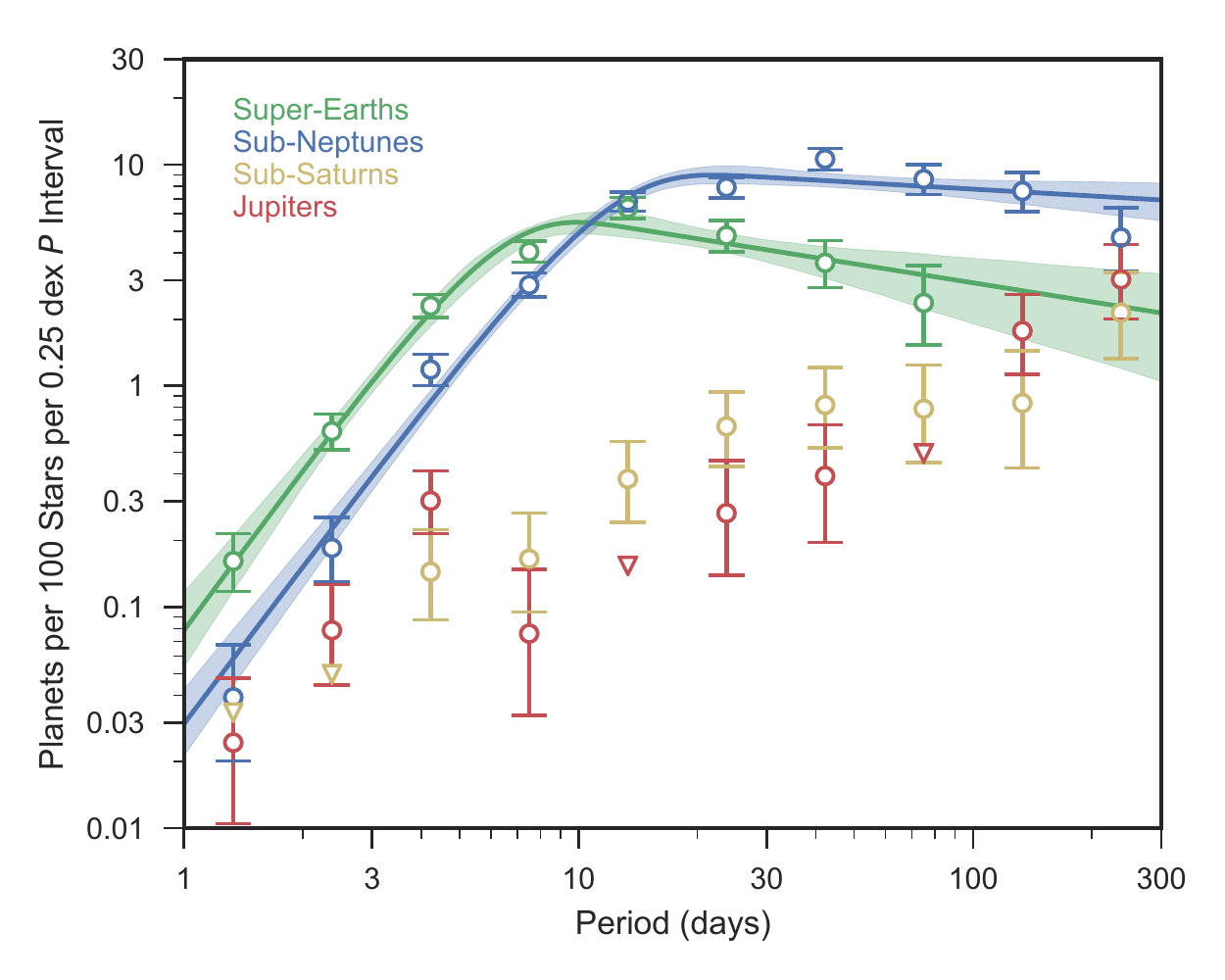}
\includegraphics[width=0.55\textwidth]{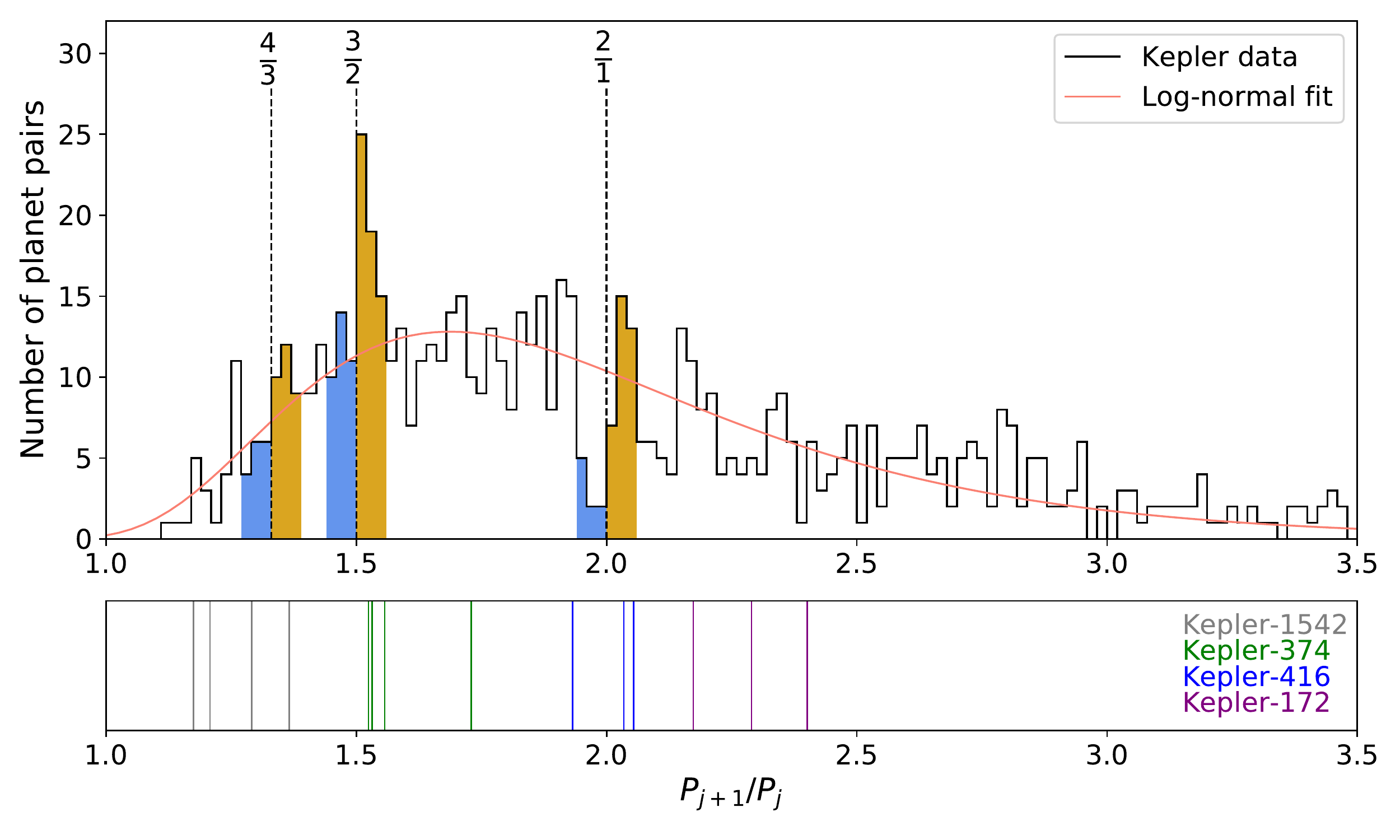}
\caption{{\color{red} Left: Planet occurrence as a function of orbital period for the following size classes: super-Earths ($R_p = 1.0{-}1.7 \ R_{\oplus}$), sub-Neptunes ($R_p = 1.7{-}4.0 \ R_{\oplus}$), sub-Saturns ($R_p = 4{-}8 \ R_{\oplus}$), and Jupiters ($R_p = 8{-}24 \ R_{\oplus}$). The Figure is from \cite{Petigura2018} who removed Kepler's observational biases when computing the distributions. Right: Period ratios of adjacent pairs of planets discovered by Kepler. Top panel: The full period ratio distribution of Kepler DR25 compact multis (containing planets with $0.5 \ R_{\oplus} < R_p < 4 \ R_{\oplus}$ and $P < 1$ year, as defined in \S\ref{sec:intro}). The smooth red line is a fit} of a shifted log-normal distribution.
{\color{red} Kepler's observational biases have not been removed}. Pairs with period ratios within 0.06 of the three strongest first-order MMRs are shaded. While there are overabundances wide of MMRs, such pairs make up a small fraction of the total distribution.  Bottom panel: Period ratios of pairs of planets within four example 4+ planet systems, demonstrating that individual systems tend to have period ratios that cluster within the period ratio distribution. }
\label{fig:period ratio distribution}
\end{figure*}

Kepler's four-year observational baseline and high-quality photometry enabled precise measurements of orbital periods, with typical fractional errors of $\mathcal{O}(10^{-5})$. Consequently, there is a rich literature not only on planet periods themselves, but also on the pairwise relationships between periods (i.e., orbital spacing) and higher-order relationships (i.e., orbital spacing dispersion). 

\paragraph{Period and Period Ratio Distributions.} Figure~\ref{fig:context} shows that compact multis span a broad range of orbital separation. However, the density of points in this figure does not visually convey the intrinsic period distribution, because of the  declining sensitivity with period due to falling transit probability and signal-to-noise. \cite{Petigura2018} accounted for these effects and found that the period distribution of super-Earths and sub-Neptunes rises over the range $P \approx 1$--10~days according to $d N / d \log P \propto P^\alpha$ with $\alpha \approx 2$. For longer periods, the occurrence function is roughly log-uniform, i.e. $\alpha \approx 0$ (see Figure~\ref{fig:period ratio distribution}).

Beyond the overall period distribution, further insight can be gained from the the distribution of period ratios, which reveals pairwise relationships. Figure \ref{fig:period ratio distribution} shows the distribution of period ratios of pairs of adjacent planets in compact multis. Here, we concentrate on the tightly-packed region with $\mathcal{P}_j \equiv P_{j+1}/P_j < 3.5$, leaving out $\sim 20\%$ of planet pairs. This region is well-described by a shifted log-normal distribution of the form 
\begin{equation}\label{eq:pratiodistfit}
f(\mathcal{P} ; \delta, \mu, \sigma) = \frac{1}{(\mathcal{P}-\delta)\sigma\sqrt{2\pi}}\exp\left[\frac{-(\ln(\mathcal{P}-\delta)-\mu)^2}{2\sigma^2}\right],
\end{equation}
with best-fit parameters given by $\delta = 0.61$, $\mu = 0.25$, and $\sigma = 0.39$. The mode of the distribution falls at $\mathcal{P} \sim 1.7$. 

Two primary features of the period ratio distribution stand out. First, the distribution is smooth and broad overall, as evidenced by its close agreement with a log-normal. This indicates that the compact multis have predominantly non-resonant orbital architectures. A second feature, however, is the existence of narrow peaks just outside (within about 5\%) of the first-order mean-motion resonances and deficits just inside of the nominal resonance locations. These near-resonance features were first noticed through studies of the Kepler multi-planet population \citep{Lissauer2011-architecture, Fabrycky2014}, and they have been shown to be statistically significant \citep[e.g.,][]{Steffen2015}. Both the paucity of resonances and the near-resonance features offer important challenges for planet formation theories. We will thus defer theoretical discussion to \S \ref{sec:migrate}, where we will review efforts to understand their origins.

\paragraph{Uniform Orbital Spacing.}
While the period ratio distribution provides a population-level summary of pairwise orbital spacing, another level of sophistication comes through the study of spacing variations from system-to-system. \cite{Weiss2018a} found that compact multis tend to have regular orbital spacing.\footnote{\citet{Weiss2018a} considered planets with $P_{j+1}/P_j < 4$ to mitigate the effects of detection bias mainly because the probability of a long-period planet transiting is low.} That is, planets in a given system are roughly evenly spaced in terms of log-orbital period in a manner that is inconsistent with random draws from the ensemble period ratio distribution (Figure \ref{fig:random_draws}). Because planets in such systems orbit a common host star, regular log-period spacing also implies regular separations in terms of log-semi-major-axis. It is important to note that the tendency for uniform orbital spacing is not in conflict with the smoothness of the period ratio distribution. Rather, the broader distribution is composed of individual systems with period ratios that are clustered within sections of the distribution. This effect is illustrated with four example systems shown in the bottom right panel of Figure \ref{fig:period ratio distribution}. 

How regular is the typical planet spacing? A slight modification of our radius dispersion metric, equation (\ref{eqn:size-disp}), yields the definition 
\begin{equation}
    \sigma^2_{\mathcal{P}} = {\rm Variance}\left\{\mathrm{log_{10}}(\mathcal{P}_j)\right\}\,,
    \label{eqn:period-disp}
\end{equation}
where we take the logarithm of each period ratio before computing the dispersion to keep the comparison fractional.  Note that there are only $N-1$ adjacent period ratios in an $N$-planet system.  For completeness, we note that other spacing metrics exist, such as the gap complexity defined in \citet{Gilbert2020}.  In the CKS compact multis with four or more transiting planets, the median spacing dispersion is 0.05~dex (0.04~dex in systems with $\sigma_R < 0.1$). In synthetic 4-planet systems where $\mathcal{P}$ is drawn at random from the ensemble of observed period ratios with $\mathcal{P} < 4$, the median value of $\sigma_\mathcal{P}$ is 0.10~dex.

\begin{figure}
    \centering
    \includegraphics[width=0.9\columnwidth]{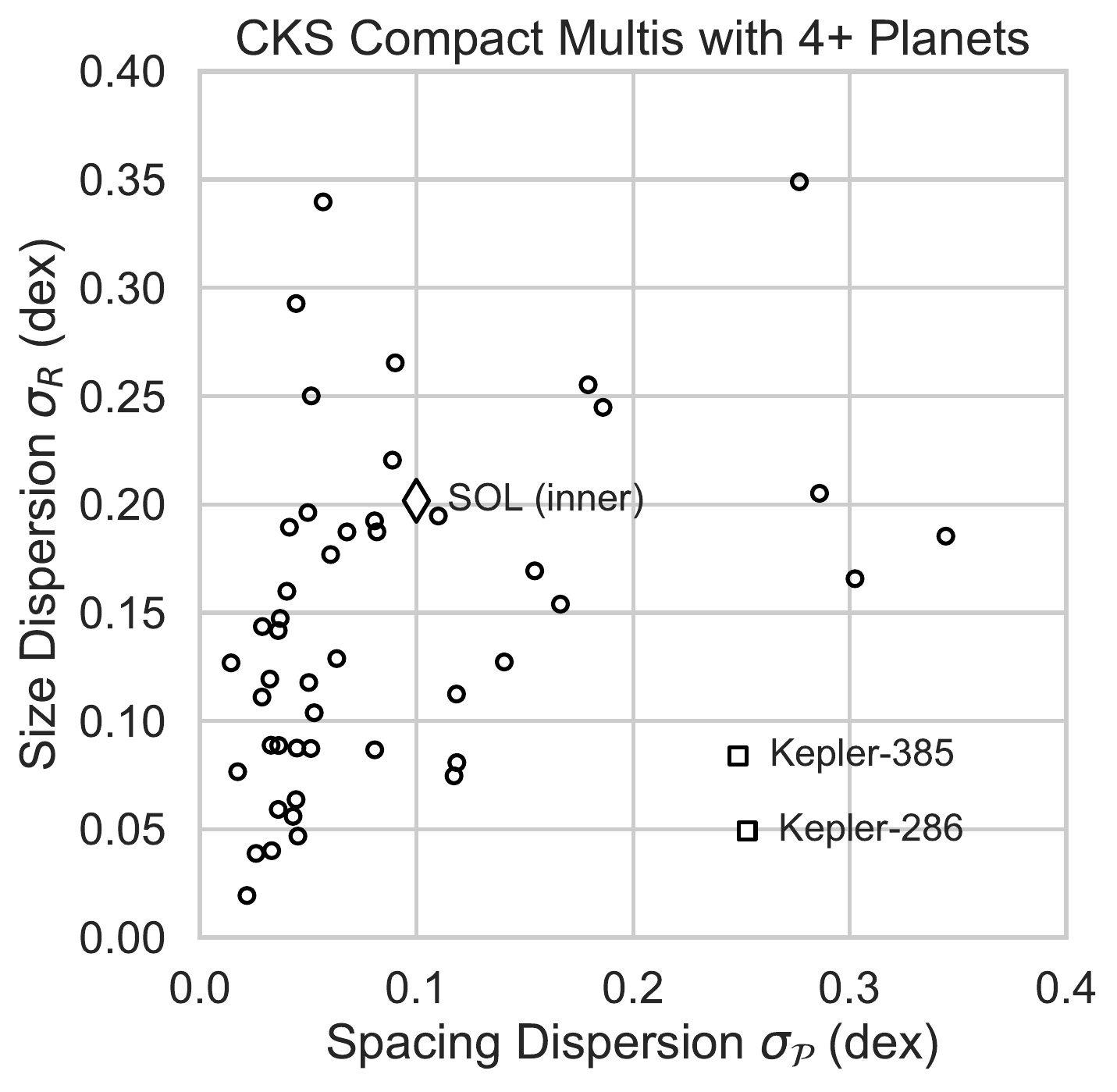}
    \caption{The typical size vs. spacing dispersions, $\sigma_R$ and $\sigma_\mathcal{P}$, of the CKS compact multis with 4 or more planets. The inner solar system (terrestrial planets only) is not atypical, but is more disorderly in both its size dispersion and spacing dispersion than the majority of compact multis.  Two systems with low size dispersions but high spacing dispersions are highlighted; these systems have clear gaps in Figure \ref{fig:peas-ranked-sigmaR} where an additional planet likely resides.} 
    \label{fig:dispersion}
\end{figure}

The metric $\sigma_\mathcal{P}$ is especially efficient at identifying high-multiplicity systems that mostly follow the peas-in-a-pod pattern, but have one or more planets missing.  For instance, systems KOI-2433 and KOI-1306 in Figure \ref{fig:peas-ranked-sigmaR} have some of the most regular planet sizes, but each has a clear gap that could be filled by a planet.  These two systems emerge as outliers in the $\sigma_R$-$\sigma_\mathcal{P}$ plane (Figure \ref{fig:dispersion}).

The solar system has long been known to have regular planet spacing \citep{Bode1768}, with ${a_{j+1} /a_j}\approx1.73$ or $\mathcal{P}\approx2.27$ \citep{Blagg1913}.  The inner solar system (Mercury-Mars) has $\sigma_{\mathcal{P}} = 0.1$~dex.  Many of the high-multiplicity CKS compact multis have lower size dispersion (80\%) and lower spacing dispersion (68\%) than the inner solar system.  Thus, the Titius-Bode rule apperas to be a special case--and yet not the strongest example--of the more general empirical rule that planets tend to have similar sizes and regular orbital spacing.

Within the solar system, the slight deviations to the regular spacing appear to have some structure that might be astrophysical in origin \citep{Blagg1913,Nieto1970}.  The majority of the compact multis have too few planets to investigate whether the deviations from regularity are orderly, but such an investigation will be valuable when higher-multiplicity exoplanet systems are discovered.

\begin{figure*}[h!]
\centering
\includegraphics[width=0.9\columnwidth]{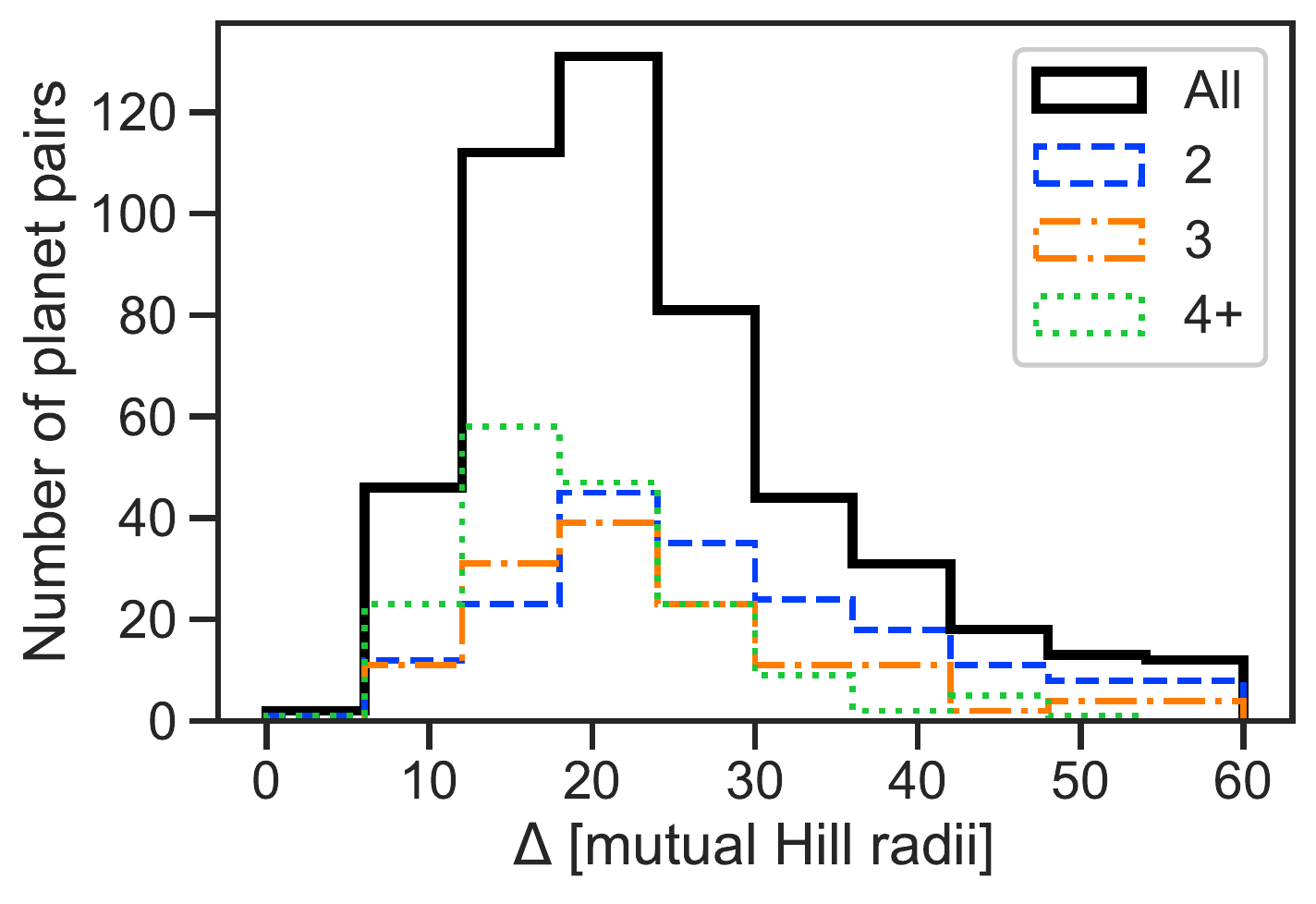}
\includegraphics[width=0.9\columnwidth]{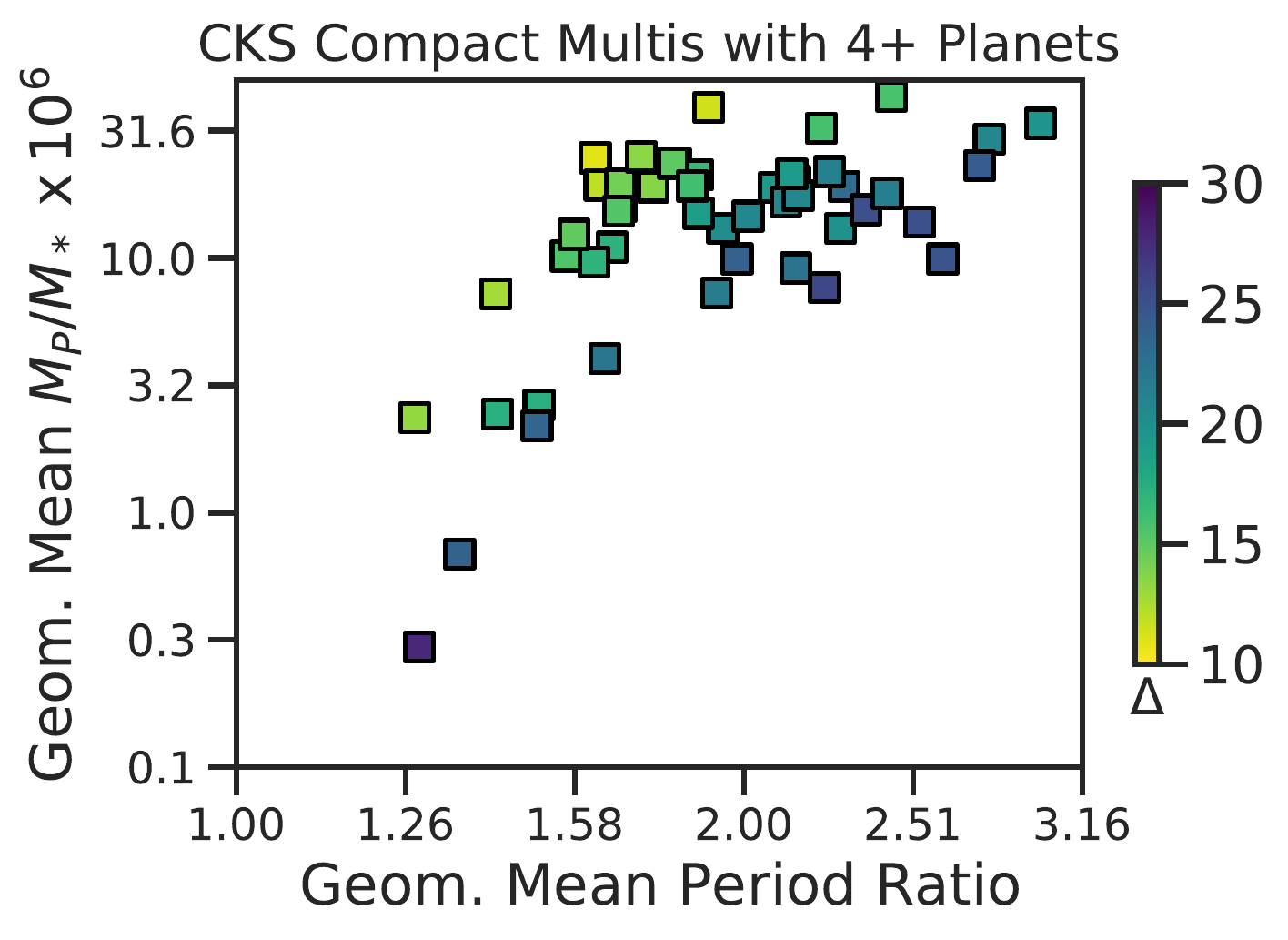}
\caption{Left: Distribution of orbital separations between 812 pairs of adjacent planets, plotted in units of mutual Hill radii. Planet masses were estimated using the empirical mass-radius relations from \cite{Weiss2014}. Planet pairs with $\Delta>60$ (21) are not shown.  The high-multiplicity systems (4+ planets) tend to have closer packing than lower-multiplicity systems.  Note that planets with apparent large separations might have an undetected planet in between.  Right: The average planet-to-star mass ratio vs. the average orbital period ratio in the CKS  peas-in-a-pod (systems with 4 or more transiting planets with $\sigma_R < 0.1$).  Both the system-wide geometric means (squares) and individual planet pairs (circles) are shown.  The grayscale represents their separations in units of mutual Hill radii.}
\label{fig:dynamical_packing}
\end{figure*}

\subsection{Ordered Distributions of Spacing and Size}

One notable attribute of the high-multiplicity (4+ planet) systems is that the typical period ratio is related to the typical planet size, especially for the planets that have the least radius dispersion ($\sigma_{\mathcal R} < 0.1$).  We convert the planet radii to masses through a nominal mass-radius relationship \citep{Weiss2014} so that we can compute the planet-to-star mass ratios, which are dynamically relevant via the mutual Hill radius,
\begin{equation}
R_H = \left({M_{p,j}+M_{p,j+1}\over3M_{\star}}\right)^{1/3}\left(\frac{a_j + a_{j+1}}{2}\right).
\label{eqn:Hill}
\end{equation}
The separation between two planets in units of mutual Hill radii is
\begin{equation}
\Delta = (a_{j+1} - a_j)/{R_H}.
\label{eqn: Delta}
\end{equation}
{\color{red}Histograms of the pairwise separations in mutual Hill radii are shown in the left panel of Figure \ref{fig:dynamical_packing}.  The typical separations are $\sim20$ mutual Hill radii.  Systems with more transiting planets tend to have smaller separations.  However, an interesting architectural feature emerges among the highest multiplicity (4+ planets) systems when} we plot the planet-to-star mass ratios versus the period ratios, representing the typical mass and spacing in each system with the geometric mean.  The {\color{red}geometric mean Hill} separation in each system, computed via equation \ref{eqn: Delta}, is indicated in {\color{red}color}.

{\color{red}The right panel of} Figure \ref{fig:dynamical_packing} displays several striking features.  First, there are no systems with typical period ratios $<1.2$.  The absence of planets closer than this minimum period ratio spacing is likely because resonance overlap becomes important at $\mathcal{P} < 1.2$ for $\sim1\,M_\oplus$ planets, leading to chaotic orbital perturbations and eventual instability \citep{Deck2013}. 
Second, the smallest planets (with $R_p\lesssim1\,\rearth$ and $M_p/M_{\star}\lesssim5\times10^{-6}$) have orbital period ratios confined to $1.2 < \mathcal{P} < 2$, although the planets would have been both stable and detectable at larger period ratios \citep{Weiss2018a}.  We will explore possible origins for the tight spacing, and also relationships between planets spacing and size, in \S3-6. 


\subsection{Low Eccentricities}

As discussed in the previous two sections, the architectural uniformity of the {\color{red}compact multis} extends beyond physical planet properties (masses and radii) to orbital properties such as semi-major axis spacing. This finding immediately raises the question of whether additional orbital properties, such as eccentricities and inclinations, also exhibit orderly structure. The answer appears to be affirmative; the orbits of the {\color{red}compact multis} are nearly circular (and nearly coplanar --- see the following section). Moreover, the orbital configurations are consistent with being the dynamically coolest extreme of a spectrum of compact multi-planet systems. In this section and the next, we review the observations that have led to this conclusion, focusing first on eccentricities and then proceeding to inclinations.

Eccentricities are difficult to measure for individual compact multi-planet systems. The multis have RV signals that are small and complex, and their host stars are generally faint, thus prohibiting strong RV-derived eccentricity constraints. The majority of individual measurements have been made using Transit Timing Variations (TTVs). \cite{Hadden2017} performed a systematic analysis of 145 Kepler planets with detectable TTVs and found predominantly small but nonzero eccentricities, with a median value of $e\sim0.02$. 

Beyond TTV analyses in individual systems, further eccentricity constraints have been made using population-level techniques. One approach is to exploit the planetary transit duration $T$, which depends on eccentricity along with period, mean stellar density $\rho_\star$, argument of periapse $\omega$, and impact parameter $b$ according to {\color{red}\citep[e.g.][]{2010exop.book...55W}:}
%
\begin{equation}
T = \left(\frac{3}{\pi^2} 
\frac{P (1-b^2)^{3/2}}{G\rho_{\star}}\right)^{1/3}\frac{\sqrt{1-e^2}}{(1+e\sin\omega)}. 
\label{eq: Tdur using rho_star}
\end{equation}
For individual planets, measurements of the duration yield weak constraints on eccentricity due to uncertainties in $b$. However, one may constrain the {\em distribution} of eccentricities in a sample of transiting planets with measured $\rho_{\star}$ \citep{Ford2008}, since the distribution of $b$ is known.

This analysis has been performed for a variety of sub-populations of Kepler systems. \cite{VanEylen2015} and \cite{VanEylen2019} analyzed transit durations of $\sim$100 planets with asteroseismically constrained $\rho_\star$ and modeled the eccentricity distribution with a Rayleigh form with a scale parameter (mode) of $\sigma_e$. Taken together, these two studies found that mulits have lower typical eccentricities than single planet systems: $\sigma_e \sim 0.05$ versus $\sigma_e \sim 0.25$. Similar trends have been observed in larger samples that used different techniques to measure $\rho_\star$. In a sample of $\sim$700 planets with spectroscopic determination of $\rho_\star$, \cite{Xie2016} found $\sigma_e \sim 0.03$ for multis and $\sigma_e \sim 0.25$ for singles. Similarly, in a sample of $\sim$1300 planets with hosts with spectroscopic/astrometric determination of $\rho_\star$, \cite{Mills2019} found $\sigma_e \sim 0.04$ for multis and $\sigma_e \sim 0.17$ for singles.


Taken together, ensemble studies of Kepler planets have found typical eccentricities of $e \approx 0.15$--0.25 among singles and $e \approx 0.02$--0.05 among multis. In some sense, the low eccentricities of the compact multis are not surprising given their tight orbital spacings. However, the eccentricities are even lower than that required by orbital stability. Among a sample of TTV-active systems with 3 or more planets, \cite{Yee2021} found that their TTV-derived eccentricities were smaller than the maximum eccentricities allowed by long-term orbital stability by factors of 2--10, suggesting that {\color{red}compact multis} attained their final orbital configurations in the presence of efficient dissipation. We will revisit the role of dissipation when we discuss the origin of intra-system uniformity from the perspective of energy optimization in \S \ref{sec:eoptimize}.

\subsection{Low Mutual Inclinations}

In addition to being nearly circular, the orbits of the compact multis are approximately coplanar, in line with the expectations of dynamical equipartition. This can be seen merely from the fact that the planets are co-transiting, which requires approximate coplanarity (unless there is a rare chance alignment of the nodes). For individual systems, quantitative measurements of mutual inclinations are even more challenging than eccentricities, and thus almost all constraints are from population-level analyses. 

In the previous section, we described how the distribution of transit durations provides population-level constraints on eccentricities. Similarly, the distribution of transit duration \textit{ratios} encodes information about mutual inclinations. The {\color{red} ratio of the transit chord lengths, $\xi \equiv (T_{\mathrm{in}}/T_{\mathrm{out}})(P_{\mathrm{in}}/P_{\mathrm{out}})^{-1/3}$,} between pairs of transiting planets in the same system is sensitive to the planets' mutual inclinations through their relative impact parameters \citep{Steffen2010}. 
{\color{red}Early analyses of these ratios showed that compact multis typically have low mutual inclinations \citep{Fang2012, Fabrycky2014}. They described inclination dispersion as Rayliegh distributions with scale parameters of $\sigma_i \sim 1^\circ-2^\circ$.}

A natural question is whether the small $\sigma_i$ of the co-transiting compact multis is fully representative of the mutual inclination distribution of the underlying (intrinsic) population of multi-planet systems (including both transiting and non-transiting planets). Studying this broader, ``parent'' population gives context to the compact multis. The observed transiting multiplicity distribution (the number of systems with $1, 2, \dots N$ transiting planets) depends on both the  underlying mutual inclination and multiplicity distributions \citep{Tremaine2012}. 

Early in the Kepler mission, \cite{Lissauer2011-architecture} noted that a single mutual inclination distribution of $\sigma_i \sim 1^\circ-2^\circ$ and a constant number of planets per system did not reproduce the transiting multiplicity distribution; there was an overabundance of singles relative to multis, a discrepancy now known as the ``Kepler dichotomy'' \citep[e.g.,][]{Johansen2012, Ballard2016}. The dichotomy is likely exaggerated (although not entirely explained) by biases due to planet detection order \citep{Zink2019}. The leading interpretation of the dichotomy is that it signals the existence of a sub-population of systems with mutual inclinations larger than a few degrees.

Several authors have constrained the joint inclination-multiplicity distributions through forward modeling \citep[e.g.,][]{Mulders2018, He2019, He2020}. Here, one generates synthetic populations of planets from parameterized size, period, inclination, and multiplicity distributions and then ``observes'' these samples with a simulated Kepler mission. The population parameters are adjusted until the observed population agrees with the actual Kepler census.  The latest models indicate that the transiting multiplicity distribution can be well-described by a mixture of high- and low-inclination sub-populations with  $\sigma_{i,\mathrm{low}} \approx 1^{\circ}-2^{\circ}$ and $\sigma_{i,\mathrm{high}} \approx 30^{\circ}-65^{\circ}$ making up $\sim60\%$ and $\sim40\%$ of systems, respectively (\citealt{He2019}; see also \citealt{Mulders2018}). However, the data are also consistent with a continuous (i.e., non-dichotomous) distribution of relatively low ($i\lesssim 5^{\circ}-10^{\circ}$) mutual inclinations with a dispersion that is inversely-correlated with the system's intrinsic multiplicity, $\sigma_i\propto n^{\alpha}$ (where $n$ is the multiplicity and $\alpha<0$;  \citealt{He2020}; see also \citealt{Zhu2018a}).


These dichotomous and continuous models reproduce the transiting multiplicity distribution roughly equally well, on account of the degeneracy between the intrinsic multiplicity and mutual inclinations. However, this degeneracy can be broken with further observational inputs, making it possible to distinguish the models. Using the statistics of Transit-Duration Variations (TDVs) of the Kepler planet population as an additional constraint, \cite{Millholland2021} showed that the continuous distribution of low ($i\lesssim 5^{\circ}-10^{\circ}$) mutual inclinations \citep{He2020} is most favored by the data. Specifically, long-term TDV signals are driven by orbital precession between inclined planets, and they increase in magnitude with mutual inclination. The Kepler planet population cannot contain a large fraction of systems with large mutual inclinations, since there would have been a larger number of TDV detections than actually observed. 

Cumulatively, the data show that systems with higher intrinsic multiplicity are dynamically cooler in both eccentricities and inclinations. The high multiplicity compact multis appear to be the ``coolest'' extreme of a dynamical continuum of close-in, multi-planet systems. In addition, if we consider the $e=0$ and $i=0$ configurations as the lowest energy state, then the excitation is roughly comparable in the two dynamical variables. Specifically, the degree of inclination excitation (expressed in radians) is given by  $\sigma_i\approx0.02-0.04$, which compares well with the degree of eccentricity excitation $\sigma_e\approx0.03-0.05$. In other words, existing data are roughly consistent with equipartition in energy excitation as measured by $(e,i)$. This finding poses an interesting avenue for future exploration: {\color{red} If one considers the degree of mass uniformity, size uniformity, or  orbital spacing uniformity as additional ``dynamical variables'' one could investigate whether or not excitation in these variables is correlated with excitation in eccentricity and inclination.}

\subsection{Weak Correlation Between Stellar Properties and Architectures of Compact Multis}
{\color{red} As we review in \S\ref{sec:stardisk}, the planet formation process begins with the collapse of a star and the formation of a circumstellar (or protoplanetary) accretion disk. For nearly all known exoplanets, the circumstellar accretion disks in which they formed are long gone. Only the host stars remain. Nonetheless, the present-day host star properties offer a window into the environment at the epoch of formation. The exoplanet literature contains many studies that have investigated correlations between star and planet properties. Such studies treated Jovian planets first because they were discovered first. 

{\color{red} As a point of comparison for the small planets, it is noteworthy that} giant planets with $a = 0.1{-}1.0$~AU  are more common around massive and metal-rich stars, and they are more eccentric than their counterparts orbiting sub-solar hosts \citep{Gonzalez97,Santos2004,Fischer2005,Johnson2007,Dawson2013}. We expect that both stellar mass and metallicity correlate with the total inventory of solids in the protoplanetry disk and conclude that solid-rich disks produce giant planets more efficiently. The prevalence of planets smaller than Neptune only slightly increases with stellar metallicity (see, e.g., \citealt{Buchhave2012,Petigura2018}) and {\em decreases} with stellar mass (see, e.g., \citealt{Howard2012,Mulders2015}). These trends suggest a different relationship between disk solid inventory and the formation of small planets.

The connection between {\color{red}stellar} metallicity and {\color{red}fundamental properties} for smaller planets is less clear due to {\color{red}the challenge of obtaining an accurate inventory of the small planets, particularly in systems that appear to have just one transiting planet}. An analysis of transit durations by \citet{Mills2019} suggested a positive correlation between eccentricity and stellar metallicity for small planets. \cite{Dong2018} found a separate line of evidence for dynamically excited small planets orbiting metal-rich stars. Sub-Neptunes with $P < 10$ days are preferentially found around stars with super-solar metallicities and are rarely members of multi-transiting systems, suggesting a correlation between stellar metallicity and mutual inclination.}  No strong correlation has been found between host star properties and the {\color{red}radius and period distributions of small planets as a function of planet multiplicity}  \citep{Xie2016,Weiss2018b}.

\begin{figure*}[h!]
    \centering
    \includegraphics[width=0.8\textwidth]{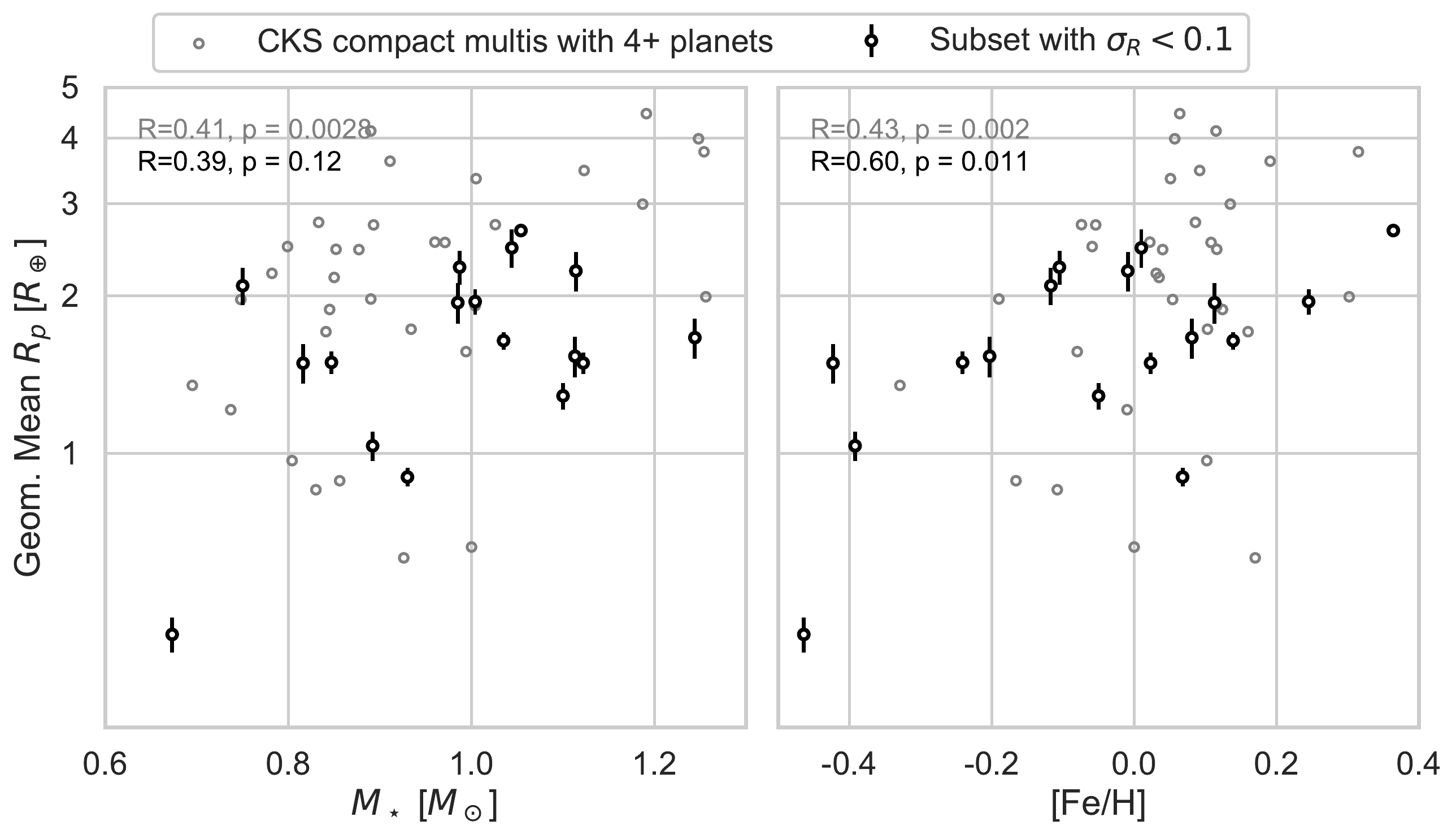}
    \caption{Left: Mean planet radius vs. stellar mass in the CKS compact multis with four or more planets, including a subset of planets with {\color{red} low size dispersion} ($\sigma_R < 0.1$ dex).  The error bars represent $\sigma_R$ in units of Earth radii.  Right: Same as left, but as a function of host star iron abundance.  Although the typical planet size is marginally correlated with stellar mass and [Fe/H] (Pearson-R and p-values indicated), the scatter in typical planet size greatly exceeds the intra-system radius dispersion, indicating that stellar mass and metallicity are not the dominant factors in setting the {\color{red}planet} sizes in the peas-in-a-pod {\color{red}pattern}.}
    \label{fig:stars-vs-peas}
\end{figure*}

Here, we investigate the relationship between host star properties and the {\color{red}fundamental properties of the compact multis, including a subset of systems with very low size dispersion} ($\sigma_R < 0.1$, Figure \ref{fig:stars-vs-peas}).  The correlations between host star properties and the typical planet size {\color{red}for the most uniform systems are weaker than for the population as a whole.  This is perhaps because we have selected a smaller sample of planets, or because the compact multis with the most uniform sizes are slightly smaller on average than the typical sub-Neptune.  Nonetheless, }the lack of strong relationships between the host star properties and the emergent peas-in-a-pod architecture suggests that the host star itself has little influence on the planetary patterns that form.

{\color{red}Note that the host stars occupy a narrow range of mass and metallicity ($0.6<M_\star/M_{\odot}<1.3$ and $-0.5<\mathrm{[Fe/H]}<0.5$).  However, if we were to include, e.g., TRAPPIST-1, which has a typical planet size of $\sim$1\,\rearth for a $0.1\,M_\odot$ star, this would only contribute to the flatness of the planet radius vs. stellar mass panel, underscoring that stellar mass is not the primary driver of the planet radii.  This flatness of planet size as a function of stellar mass was also noted in \citet{Dai2020}, who interpreted it as evidence that planet formation is more efficient in disks around low-mass stars.}

\subsection{Potential Biases}

Before transitioning to theoretical discussions of the peas-in-a-pod patterns, let us pause to consider the potential influence of observational biases and selection effects. To our knowledge, there are two recent papers that have interpreted the observed intra-system uniformity in planet sizes as the result of observational bias. 

In one study, \cite{Zhu2020} considered an alternative randomization procedure aimed at assessing the significance of intra-system uniformity. They generated random multi-planet systems by starting with the \cite{Weiss2018a} sample of planet sizes and rescaling them according to $R_p^\prime = R_{p,\mathrm{obs}} \, (\mathrm{SNR}_{\mathrm{rand}}/\mathrm{SNR}_{\mathrm{obs}})^{1/2}$ where $R_{p,\mathrm{obs}}$ is the observed planet radius,  $\mathrm{SNR}_{\mathrm{obs}}$ is the observed signal-to-noise ratio of the transit and $\mathrm{SNR}_{\mathrm{rand}}$ is drawn at random from the ensemble distribution of $\mathrm{SNR}_{\mathrm{obs}}$. \cite{Zhu2020} observed an intra-system uniformity in radius, albeit with a weaker correlation than reported in \cite{Weiss2018a}, and interpreted this uniformity as evidence that the entire peas-in-a-pod pattern was due to SNR effects. However, \cite{Weiss2020} showed that this procedure does not prove that uniform radii are due to selection effects alone. If the radii are indeed correlated, as shown in Figure~\ref{fig:random_draws}, then adding random multiplicative scalings will blur --- but not entirely erase --- the correlation. As a result, the correlation observed by \cite{Zhu2020} does not falsify an astrophysical explanation for the peas-in-a-pod pattern.

In another study, \cite{Murchikova2020} found that if the underlying radius distribution is  sufficiently steep, $d N /d R_p \propto R_p^{-4}$, then random draws can reproduce the observed size similarity, since planets just above the detection limit are much more common than slightly larger planets. Although this approach provides, in principle, a plausible way for selection effects to produce size uniformity, the assumed power-law is inconsistent with the true distribution of planet sizes. Measuring the size distribution of Kepler planets while accounting for selection effects has been the subject of many independent analyses (see, e.g., \citealt{Howard2012}, \citealt{Fressin2013}, \citealt{Petigura2013}, \citealt{Fulton2017}, or the review by \citealt{Winn2018}). Smoothing over the fine structure observed by \cite{Fulton2017}, the radius distribution is roughly log-uniform from $R_p$ = 1--3~\rearth, i.e., $d N /d R_p \propto R_p^{-1}$ instead of $d N /d R_p \propto R_p^{-4}$. In addition, the \cite{Murchikova2020} model predicts that most planets will be detected near the minimum detectable SNR, which is $\sim$ 10 \citep{Christiansen2015}. However, $\sim$70\% of the \cite{Weiss2018a} multis have SNR $>$ 20. In summary, the \cite{Murchikova2020} model can only produce the peas-in-a-pod pattern through selection effects by invoking a planet size distribution and a planet signal-to-noise distribution that are inconsistent with the Kepler data. 

\citet{Murchikova2020} also considered a second model, with a less steep radius dependence, but with 9 free parameters, including a prescription that the planets in a given system tend to be either $1.5$ or $2.4\,R_\oplus$ (based on the peaks of the planet radius distribution from \citealt{Fulton2017}), with a tuneable size dispersion for planets of either radius type.  This model somewhat reproduces the size similarity observed in CKS, but it does not reproduce the other patterns described above, including the spacing similarity and the size-spacing relation.  Also, the prescription that planets in the same system tend to be centered around $1.5$ or $2.4\,R_\oplus$ is not a null hypothesis because it assumes that planets within the same system have similar sizes.

The role of detection biases in shaping our understanding of patterns in planetary systems is indeed important, but the case for the peas-in-a-pod pattern having an astrophysical origin remains strong. Many different groups --- using independent methods and tools --- find evidence for patterns of uniformity {\color{red} \citep[e.g.][]{Weiss2018a, He2019, Gilbert2020, He2020, Mishra2021, MillhollandWinn2021, Otegi2021}}. Nonetheless, the question of the veracity of the peas-in-a-pod phenomenon will ultimately be decided by the continued accumulation of observational data.

\begin{figure*}[h!]
\centering
\includegraphics[trim={0cm 0cm 0cm 0cm}, width=0.92\textwidth]{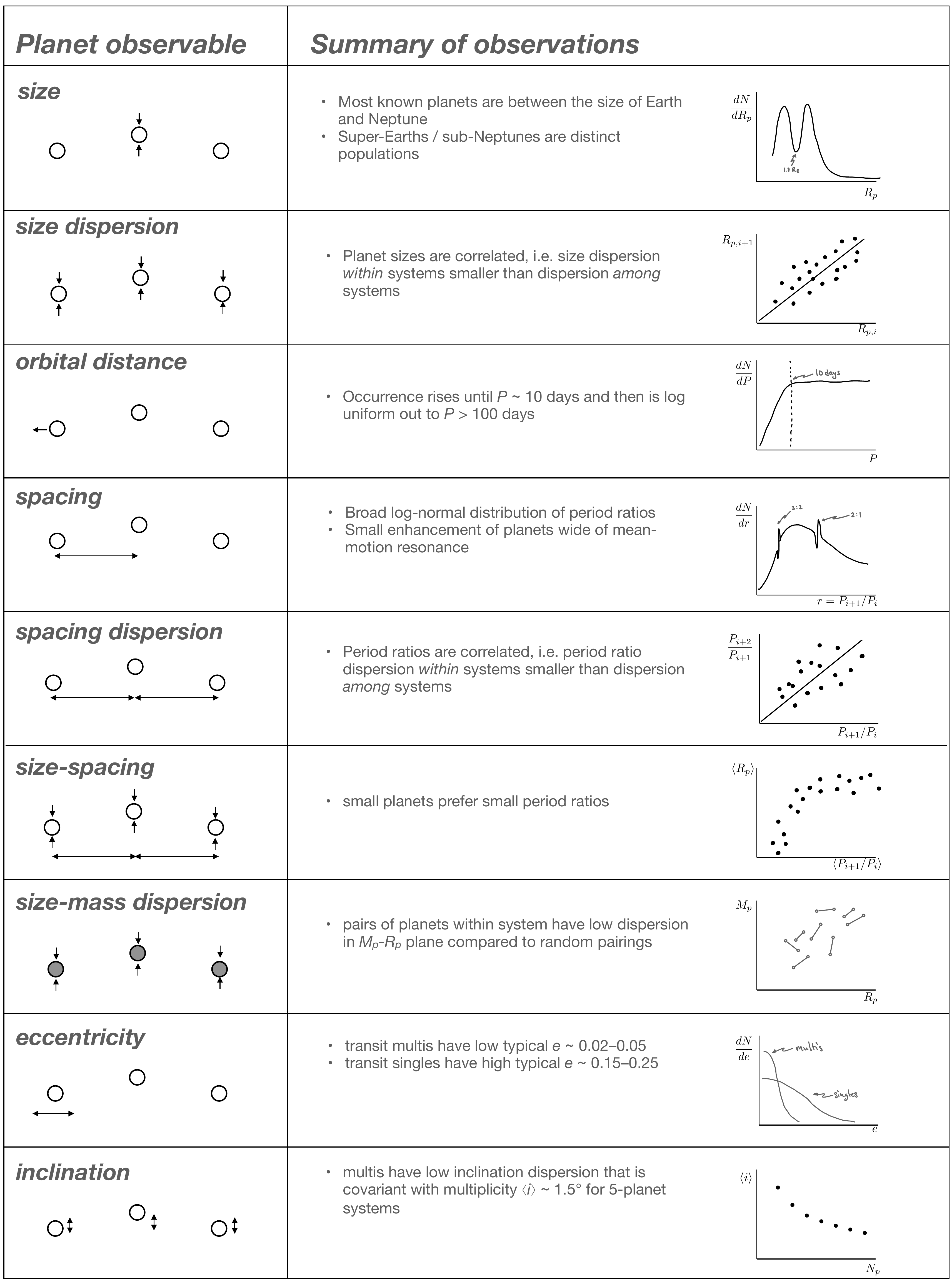}
\caption{Summary of physical and orbital properties of compact multis reviewed in \S\ref{sec:observations}. For each planet observable we summarize the observations and provide a sketch of a key figure from the following references: {\em size}---\cite{Fulton2017} Figure~7, 
{\em size dispersion}---\cite{Weiss2018a} Figure~2, {\em orbital distance}---\cite{Petigura2018} Figure~7, {\em spacing}---\cite{Fabrycky2014} Figure~4, {\em Spacing dispersion}---\cite{Weiss2018a} Figure~7, {\em size-spacing}---this work Figure~\ref{fig:dynamical_packing}, {\em size-mass dispersion}---\cite{Millholland2017} Figure~2, {\em eccentricity}---\cite{VanEylen2019} Figure~5, {\em inclination}---\cite{He2020} Figure~7.}
\label{fig:summary-of-observations}
\end{figure*}

\subsection{Summary of Observations: Peas-in-a-Pod}
Let us take stock of the physical and orbital properties of compact multi-planet systems described in this section. Figure~\ref{fig:summary-of-observations} lists a number of planet observables and summarizes the current state of the observations. For each item in the list, we provide a reference to a key figure in the literature. To summarize: The prevalence of planets increases with decreasing size, with a gap in the distribution at $\rp=1.5-2.0$~\rearth. Compact multis have correlated sizes and masses. The occurrence of small planets is roughly log-uniform down to $P \approx 10$~days, and below that value it falls precipitously. Planetary pairs have a relatively broad distribution of orbital spacing with slight over-abundances wide of first-order mean-motion resonances. Despite the overall range in observed spacing parameters, however, the spacings within a given system are more highly correlated. The orbital spacing distribution within a planetary system depends on the size of the constituent planets,  the smallest planets preferring tighter spacing. Compact multis have low eccentricities and inclinations.  Finally, the host star masses and metallicities do not strongly correlate with the peas-in-a-pod architectural features.

Taken together, these findings indicate that planets in compact multi-planet systems are dynamically cool. The planets have somehow coordinated their sizes, masses, and orbital spacings in a manner that depends only weakly on the properties of the host star. In the following sections, we consider the physical mechanisms that can account for these features.  We explore how the formation of the star-disk system (\S\ref{sec:stardisk}) and the emergence of planets in the disk (\S\ref{sec:planet-formation-theory}) sets the characteristic mass of the planets.  In the final stages of planet formation and evolution, planet-planet interactions dominate (\S\ref{sec:pp interactions}), and we consider how the dynamics in mature planetary systems sculpt both the regularity of the peas-in-a-pod pattern and the conditions for departures from that pattern. Figure \ref{fig:schematic} provides a schematic roadmap for our review of these physical processes.

\section{Connection to Star Formation and Circumstellar Disk Properties} 
\label{sec:stardisk} 


Planets form within circumstellar disks, which are naturally produced alongside their host stars. As a result, the star formation process places important constraints on the subsequent process of planet formation, including the compact multi-planet systems of interest here. More specifically, the properties of disks, which provide the initial conditions of planet formation, are sculpted by the earlier action of star formation. Here we briefly review these constraints --- including the mass,  radius, and surface density distributions of the disks --- with a focus on properties relevant to {\color{red}the emergence of the peas-in-a-pod pattern and other observed properties of the compact multis} (for a more comprehensive treatment, see the Chapters by Miotello et al. and by Manara et al.). We consider these properties at the epoch corresponding to the end of star formation (\S \ref{sec:starform}) and over the subsequent epochs when planets form (\S \ref{sec:diskevolve}, \S\ref{sec:angconstraint}). This span of time thus covers the transition from protostars to planets. 


\subsection{Constraints from Star Formation}
\label{sec:starform} 

Star formation itself occurs within molecular clouds that can include millions of solar masses of material \citep{shu1987}. However, individual star-forming events take place within molecular clouds cores, which are much smaller sub-units containing several times the mass of the stars that they produce. At intermediate scales, stars form within embedded clusters that contain 100s to 1000s of stars \citep{lada2003}. While these cluster systems can sculpt planetary systems after their formation \citep{adams2010}, this background environment is subdominant during the collapse phase that actually produces a star/disk system. 

\begin{figure*}
\centering
\includegraphics[width=0.9\textwidth]{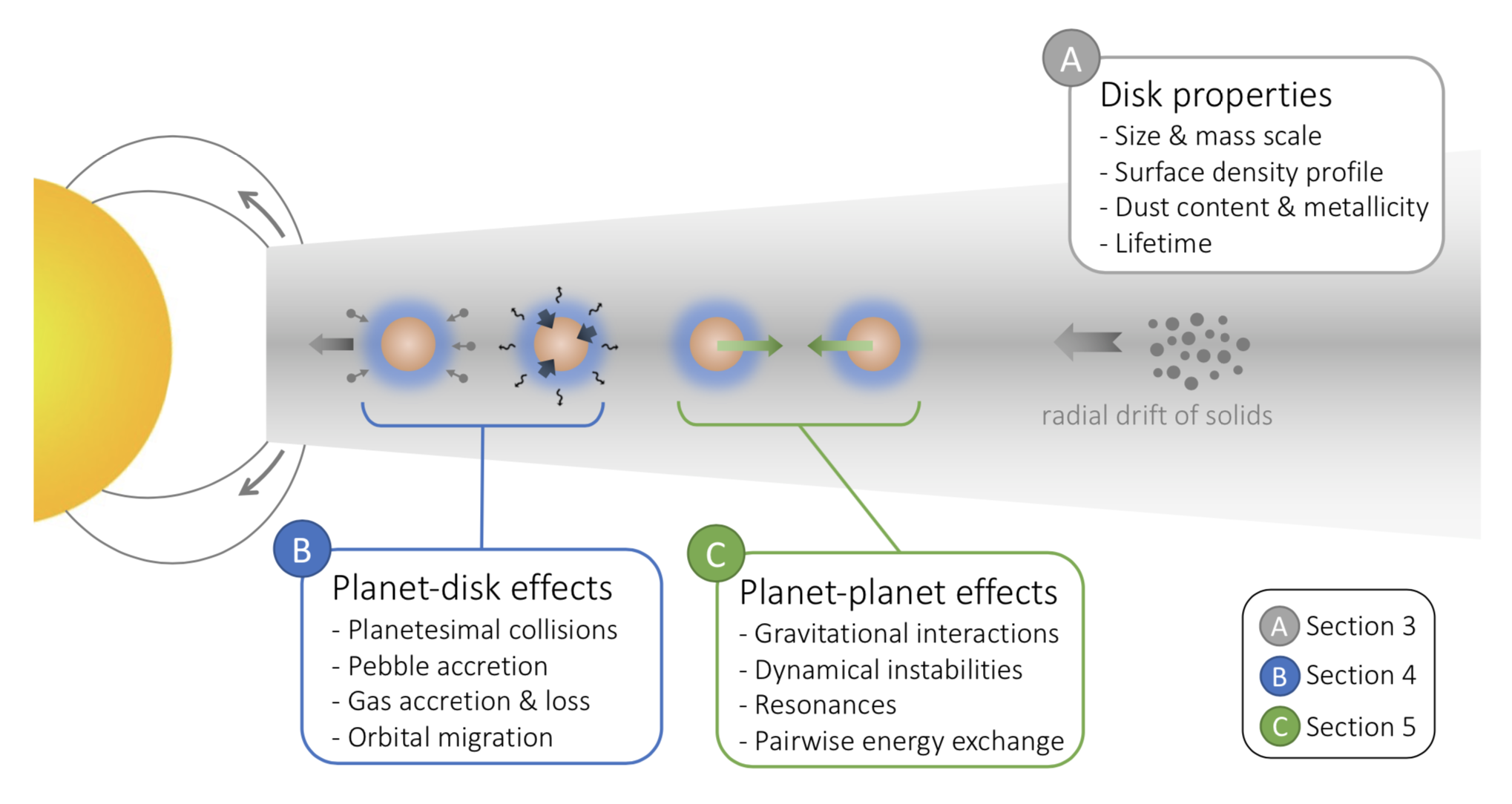}
\caption{Schematic diagram showing the various physical processes involved in the formation of {\color{red} compact multi-planet systems and the possible emergence of  peas-in-a-pod architectures}. These processes can be broadly divided into three categories: disk properties, planet-disk effects, and planet-planet effects. Disk properties are set by the formation of the star-disk system, and they include the global characteristics and evolution of the accretion disk. Planet-disk effects include the conglomeration of dust and rocky material in larger bodies, accretion of nebular gas onto the planet (and subsequent gas loss), and migration driven by planet-disk interactions. Lastly, planet-planet effects include any pairwise interactions, such as gravitational effects leading to enhanced long-term stability. These three stages cover the full planet formation process roughly temporally, forming the roadmap for \S\ref{sec:stardisk}, \S\ref{sec:planet-formation-theory}, and \S\ref{sec:pp interactions}, respectively. Finally, all of these stages will be explored together in the population synthesis models of \S\ref{sec:popsynth}. }
\label{fig:schematic}
\end{figure*}

We consider a simple model for the collapse of the molecular cloud core, from which a few key physical scales of the planet formation environment emerge.  We can thus describe the molecular cloud core with two variables: the sound speed $v_s$ sets the degree of thermal support and the rotation rate $\Omega$ determines the amount of angular momentum (e.g., see the review of \citealt{shu1987}). From these quantities, we can specify the maximum mass available 
\begin{equation}
M_{\rm max} = {v_s^3 \over G\Omega}\,,
\end{equation}
the size of the region for an individual star formation event 
\begin{equation}
R_{\rm core} = v_s/\Omega \,,
\end{equation}
and size of the disk $R_{\rm disk}$ as determined by the centrifugal barrier $R_C$ in the collapse flow
\begin{equation}
R_{\rm disk}\sim 
R_C = {G^3 M^3 \Omega^2 \over 16 v_s^8}\,.
\label{rcent} 
\end{equation}
This radius should be considered as the starting radius for a disk when a total mass $M$ has fallen to form the star/disk system. {\color{red} Under the action of viscosity,} subsequent disk evolution will spread out the disk, so that the outer disk radius is expected to grow.

The collapse conditions for star formation define a mass infall rate ${\dot M}\sim v_s^3/G$. As a result, the formation of a star with mass $M_{\star}$ must take place over a time scale $t_{\star} \sim M_{\star}/{\dot M}\sim GM_{\star}/v_s^3 \sim 0.1-0.3$ Myr. 

In the collapse flow that produces stars, nearly all of the mass falls to radii much larger than a stellar radius, i.e., material fall first onto the disk. Gravitational instabilities in the disk (along with other mechanisms) act to transfer angular momentum and facilitate accretion of disk material onto the star. Gravitational instabilities grow on a dynamical time scale when the parameter $Q=v_s\Omega/\pi G \Sigma\sim1$ \citep{toomre1964ApJ}, which results when the disk mass is comparable to the stellar mass. In order for the disk instability time scale to become longer than the collapse time of a cloud core (about 0.1 Myr), the disk needs $Q\sim10$ which corresponds to a disk mass
\begin{equation}
    M_{\rm disk} \sim {1\over10} M_{\star} \,. 
    \label{diskscale} 
\end{equation}
Note that this mass scale for the disk is the starting disk mass and/or the maximum disk mass at birth. Observed disk masses have an upper envelope bounded by this value (e.g., \citealt{hartmann2008}). In other words, the data show a range of disk masses for a given stellar mass, but the maximum observed disk mass scales according to equation (\ref{diskscale}). More specifically, the range of disk masses for a given stellar mass spans an order of magnitude \citep{andrews2013}.

For completeness, we note that the disk must have an approximate minimum mass for a given disk radius. The initial molecular cloud cores that form stars contain more angular momentum than can be carried by a star (the star would have to rotate much faster than its breakup angular velocity to contain that angular momentum -- this finding represents the well-known angular momentum problem of star formation). As a result, the disk must {\color{red} initially} contain most of the angular momentum of the initial cloud core {\color{red} (note that subsequent outflows can remove angular momentum)}. In approximate terms, the disk angular momentum is given by $J\sim M_{\rm disk} \sqrt{GM_{\star}R_{\rm disk}}$, where $R_{\rm disk}$ is the outer disk radius and where we ignore dimensionless factors of order unity. For typical stellar masses $M_{\star}\sim1M_\odot$ and observed disk radii $R_{\rm disk}\sim100$ AU, the disk must contain roughly $\sim0.01M_{\star}$ in order to carry the angular momentum. Again, these values represent the disk properties at the end of star formation, and the beginning of planet formation (coincident with the T Tauri phase of stellar evolution). As the system evolves, the disk mass decreases due to accretion, evaporation, and planet formation. Accretion causes the disk radius to grow, evaporation carries away some of the angular momentum with the flow, and the remaining angular momentum is locked up in the planets (primarily in their orbits, with some accounting for planetary rotation). 

The maximum and minimum disk masses discussed above indicate that the {\it starting} mass available for planet formation falls in the range $0.01-0.1M_{\star}$. If we consider typical stars with mass $M_{\star}=M_\odot$ and typical metallicity $Z=0.01$ (near solar), the available mass in solids thus falls in the range $M_{\rm solid}=30-300$ $M_\oplus$ {\color{red} (e.g., see \citealt{Tychoniec2018} for observational constraints).} Given that {\color{red}compact multis} typically contain at least 4 planets with $\sim10M_\oplus$ of rocky material, the heavy metal inventory must be greater than about 0.1 of the solar value to produce such {\color{red} typical} systems. In other words, systems with sufficiently low $Z$ are predicted to not have {\color{red} multiple $\sim10M_\oplus$ planets. Systems with moderately lower $Z$ can in principle produce systems with lower mass planets, but the planet masses must vanish in the limit $Z\to0$.}


In addition to the total mass of solids in a disk, the radial distribution of solids is relevant for planet formation.  When disks are built from the infall-collapse flow that produces their stellar hosts, the surface density distribution has a nearly power-law form 
\begin{equation}
    \Sigma(r) = \Sigma_0 \left({r \over r_0}\right)^{-p}\,,
\end{equation}
where $r_0$ is a reference orbital radius and the coefficient $\Sigma_0$ is determined by the total mass of the disk. The power-law index depends on the manner in which the disk forms. If the starting angular momentum profile of the core corresponds to an isothermal sphere with uniform rotation, and if the material does not change radius as it become incorporated into the disk, then the power-law index $p=5/3$. However, the incoming material does not, in general, have the correct azimuthal velocity to become part of a Keplerian disk. If one assumes that parcels of gas lose energy, at constant angular momentum, as they join the disk, then the power-law index $p\approx7/4$ {\color{red} (\citealt{cassen1981,terebey1984}; see also \citealt{adamsshu1986})}. For comparison, the well known Minimum Mass Solar Nebula (MMSN) also has a power-law form (starting with \citealt{hayashi1981}). This profile, obtained by augmenting the observed masses of solar system planets and by using their current orbits, has power-law index $p=3/2$. 

Observations of circumstellar disks tell a similar story, but suggest some differences. A full review of the data is beyond the scope of this contribution (again, see the Chapters by Miotello et al. and Manara et al.), but the bottom line is that observed surface density estimates also show a power-law form, in agreement with theoretical considerations and estimates made from the exoplanet sample. On the other hand, observed disks tend to have shallower density profiles -- smaller indices -- where current {\color{red} SMA and ALMA maps suggest that $p\approx1$ (e.g., \citealt{andrews2009})}.

Finally, we note that various attempts have been made to use the sample of observed exoplanets to construct a minimum-mass extrasolar nebula. These analyses also generally give power-law forms with varying indices and normalizations. Using all of the planets available at the time, \cite{chiang2013} find a power-law index $p=1.6$ for both the gaseous disk component and the solids. Although this form is widely used, it is inconsistent with the peas-in-a-pod configuration (note that a purely peas-in-a-pod system, which has equal mass in equal units of $\mathrm{dln}a$,  has $p=2$, whereas the energy optimized configuration of \citealt{Adams2019} has $p$ = 11/6; see \S\ref{sec:eoptimize}). Earlier work \citep{kuchner2004} find similar results with $p\sim3/2$, consistent with the minimum mass solar nebula, but with large uncertainties (due to the smaller amount of data available at the time).

The above collection of power-law indices for circumstellar disks can be summarized as follows: 
\begin{table}[h!]
\centering
\begin{tabular}{l r}
Source &  $p$ \\
\hline
ALMA observations & $\sim 1$ \\
Minimum Mass Solar Nebula & $3/2$ \\ 
Minimum Mass Extrasolar Nebula & $\sim3/2$ \\ 
Star formation theory & $5/3 - 7/4$ \\
Pair-wise energy optimization & $11/6$ \\
Pure peas-in-a-pod configuration & $2$
\end{tabular}
\label{tab: index summary}
\end{table}

\noindent Both circumstellar disks and planetary systems are observed to have nearly power-law surface density distributions, with indices falling in the range $p=1-2$. Disk configurations of this general form are produced naturally from the collapse flow that forms stars. Moreover, these results are consistent with known (expected) properties of the angular momentum profiles of the molecular cloud cores that provide the initial conditions. In spite of the apparent consistency, one important discrepancy remains. The peas-in-a-pod {\color{red}pattern has a} steeper surface density distribution than the disks that gave rise to it (as well as other planetary systems). {\color{red} This difference suggests that either the raw materials for planet formation, or the planets themselves, must move relative to the positions indicated by the initial disk profiles.} 

With the disk masses, outer radii, and surface density profiles specified, we now consider the inner boundary condition. The inner disk edge is controlled by the presence of strong magnetic fields generated within the star. The circumstellar disks are truncated at the radius $R_X$ where the inward pressure due to the accretion flow is balanced by the outward pressure from stellar magnetic fields. For dipole field structures, the radius $R_X$ can be written in the form  
\begin{equation}
R_X = \zeta_X \left( {B_{\star}^4 R_{\star}^{12} \over 
G M_{\star} {\dot M}^2} \right)^{1/7} \,,
\label{eqn:trunk} 
\end{equation}
where ${\dot M}$ is the mass accretion rate through the disk and $B_{\star}$ is the magnetic field strength on the stellar surface. The dimensionless parameter {\color{red}$\zeta_X$} is expected to be of order unity and depends on the model assumptions \citep{ghosh1978,blandford1982}. The magnetic field strengths have been measured for a collection of T Tauri stars \citep{Johnskrull2007} and fall in the range $B_{\star}\approx1-2$ kilogauss. As a result, the truncation radii that define the inner disk edge have typical values $R_X\sim0.1$ AU.  Note that this size scale is comparable to the semimajor axes of the observed {\color{red}compact multis} (where the latter are observed at much later ages). {\color{red} Notice also that as the mass accretion rate decreases, the truncation radius $R_X$ increases, so that this coincidence only holds at early epochs of disk evolution (ages of few Myr).}

One thus obtains the ordering of radial scales
\begin{equation}
    R_{\rm core} \gg R_{\rm disk} \sim R_C \gg R_X > R_{\star} 
\end{equation}
and the corresponding ordering of mass scales
\begin{equation}
    M_{\rm core} > M_{\star} \gg M_{\rm disk} \gg M_p \,,
\end{equation}
where $M_p$ is the mass scale of planets ($\sim10$ $M_\oplus$ for the {\color{red}typical planet in a compact multi}, of interest here). Another defining characteristic of the disks is that they are geometrically thin, so that the scale height of the disk much smaller than the radius, i.e., $h/r\ll1$. Since the scale height ratio is equivalent to the ratio of speeds, $h/r=v_s/v_{\rm orb}$, the disks are {\it cold}. These scales and their ordering sets the stage for the subsequent process of planet formation.\footnote{In spite of the disks being geometrically thin, their vertical extent is far greater than the radii of planets, $h\gg R_p$.}  

\subsection{Evolution of Disks, Stars, and Clusters} 
\label{sec:diskevolve}

The disk properties outlined above result from the process of star formation and correspond to the time when the star is first formed, i.e., the epoch when the star/disk system has gathered the majority of its final mass. The time scale for this process, and hence the system age when the disk has the aforementioned properties, is of order 0.1 Myr. Significantly, this collapse time scale is much shorter than several other scales of interest. Estimates for the expected time required for planet formation vary widely, but extend up to 100 Myr for the formation of Earth in our solar system. The observed lifetimes of circumstellar disks typically fall in the range 1 -- 3 Myr. The pre-main-sequence contraction time for stars of solar mass and smaller are also of order 10 Myr. Finally, the lifetimes of embedded stellar clusters are $\sim10$ Myr, with the more robust open clusters lasting $\sim100$ Myr. All of these astrophysical processes taking place in the background can influence planet formation, as outlined below. 

Perhaps most importantly, lifetimes for circumstellar disks are relatively short. Several observational surveys have studied disk signatures, such as infrared excess emission and veiled emission lines, in star forming regions with measured ages \citep{hernandez2007,mamajek2009,fedele2010}. The fraction of stars with disk signatures is found to be a decreasing function of time with an approximately exponential form. Roughly half of the disks (or at least their observational signatures) are gone by an age of 3 Myr, with only $\sim10\%$ of the systems retaining disks at 10 Myr \citep{meyer2007,williams2011,hartmann2016,manzomartinez2020}. These observations correspond to the presence of gas and accretion processes in the disks, and thus indicate that gas is only available for planet formation over 3 -- 10 Myr. On the other hand, for the production of smaller rocky planets, with little or no gas, this time constraint is less problematic. The {\color{red}compact multis} fall in an intermediate regime. Although they are thought to have a primarily rocky composition, some of the planets retain a hydrogen/helium atmosphere {\color{red}\citep{Weiss2014,Rogers2015}}. Although the gaseous mass is small, only a few percent of the total, its presence indicates that {\color{red} the planets are forming before the disk fully dissipates. On the other hand, if large rocky planets form while the disk retains a large gas supply, the planets can readily accrete thicker atmospheres than indicated by observations (for further discussion and possible solutions, see \citealt{Lee2014,Ormel2015,Lambrechts2017}). } 

Disks evolve through a variety of processes, including viscous accretion, evaporation, {\color{red} and disk winds.} The latter can be driven by radiation fields from the central star itself \citep{owen2012} and/or by external sources \citep{adams2004}. In addition, some mass becomes locked up within forming planets. All of these effects combine to make the disk mass decrease with time. In addition, viscosity acts as a diffusive process \citep{shu1992} so that the disk surface density is expected to spread out with time. In other words, the natural evolution of the disk is to become less centrally condensed with time, whereas {\color{red}the peas-in-a-pod pattern corresponds} to a state of greater concentration. This mismatch indicates that the rocky material that makes up planets must move (migrate) relative to the gas that comprises most the disk mass. How this redistribution takes place represents an important unresolved issue. Note that this movement can take place at different stage of planetary development, from early times when the rocky material is still small (rocks) to late times after the planets have already finished fully forming. Nonetheless, by an age of $\sim5$ Myr, the outskirts of disks are depleted in dust mass, suggesting that the mass has already been consolidated into pebbles or planetary-sized bodies or migrated inward \citep{Ansdell2015}.


As the disks spread out and lose their mass, and planets form within them, the host star evolves on a similar time scale. During the time in which planetary systems are forged, the central star contracts in radius and decreases its luminosity. Over the same time, it slows down its rotation rate and decreases its quadrupole moment $J_2$, thereby changing the manner in which the star couples dynamically to the disk and/or planets. 

\subsection{Global Angular Momentum Constraints} 
\label{sec:angconstraint} 

Circumstellar disk properties are determined to a large extent by conservation of angular momentum, as discussed in \S\ref{sec:starform}. The resulting configurations have roughly solar system sizes ($R_{\rm disk}\sim100$ AU) and moderate masses (starting at $M_{\rm disk}\sim0.1M_{\star}$). To constrain how such star/disk systems subsequently evolve, consider the following question: What is the lowest energy state accessible to a star/disk system subject to conservation of angular momentum? The optimum energy is achieved by placing essentially all of the mass in the central object, while leaving one small particle with a large orbit to carry the angular momentum. Note that this optimum energy state is nearly realized by our solar system: Essentially all of the mass is contained in the Sun, with the majority of the angular momentum carried by Jupiter in its orbit.  Since any mechanism for energy dissipation causes star/disk systems to evolve toward a similar state, exoplanetary systems are expected to have similar properties. 

Disks can lose angular momentum through the action of photoevaporative winds, which carry away gaseous material. If the solids remain in small entities, they can be swept away with the gas. If, instead, the solids grow to size scales large enough to decouple them from the gas, they remain behind as the gas evaporates. As a result, stars {\it must} have `solar systems' consisting of some rocky material. Disks typically contain a mass $M_{\rm solid}=30-300M_\oplus$ of solid material (\S\ref{sec:starform}). Suppose that a fraction ${\cal F}$ of the original solid content of the disk is left behind. Even if all of the gas can be successfully evaporated, the disk (planetary system) will contain a mass ${\cal F}M_{\rm solid}$. {\color{red} If the evaporating gas does not carry away the specific angular momentum of the rocky material, the remaining system will have} characteristic size $R\sim R_C/{\cal F}^2$ (where $R_C$ is given by equation [\ref{rcent}]).

{\color{red} Compact multis} typically consist of 3 or 4 {\color{red}detected} planets, each with mass $\sim10M_\oplus$ and orbital periods less than $\sim100$ days. The total mass in rocky material is thus a significant fraction of the expected mass $M_{\rm solid}$ from the initial state. Since the semimajor axes are small, however, the angular momentum of the system is only a fraction $\sim(a/R_C)\sim1/10$ of the total expected value. This discrepancy implies that the {\color{red}compact multis} must either have additional material in orbit at larger radii, or have been formed with highly efficient removal of angular momentum.

\section{Formation of Sub-Neptune-Sized Planets}
\label{sec:planet-formation-theory}

The formation of the star-disk system sets the global conditions in which planet formation occurs, as detailed in the previous section. Within this environment, dust and rocky material coalesces into progressively larger bodies, eventually generating proto-planets (recall the schematic outline in Figure \ref{fig:schematic}). The specific processes by which this growth occurs is a topic of vigorous research. Nevertheless, numerous new ideas pertaining to the formation of sub-Neptune-sized planets offer insights into how planets might emerge at a variety of physical scales. Here we briefly review the detailed formation processes with an eye towards quantifying the characteristic planet mass and size scale (for a more general treatment, see the Chapter by Drazkowska et al.). We proceed temporally, focusing first on the growth of plantesimals (\S\ref{sec:dust to plantesimals}) and then planets (\S\ref{sec:planetesimals to planets}), before considering the role of orbital migration (\S\ref{sec:migrate}) and atmospheric loss (\S\ref{sec: atmospheric mass loss}).

\subsection{From Dust to Planetesimals}
\label{sec:dust to plantesimals}

The first step towards quantifying how planetary objects coalesce lies in understanding the genesis of $\sim10-100\,$km planetary building blocks, i.e., the planetesimals within the protoplanetary disk. Rocky and icy grains within the nebulae originate as $\sim\mu$m dust grains that are delivered to the circumstellar disks along with the infalling gas (\S 3). Like snowflakes in the Earth's atmosphere, these grains gradually grow to become $\sim$mm-cm ``pebbles". Their growth, however, does not continue beyond this characteristic size scale due to the so-called \textit{fragmentation barrier}. Both modeling and experimental studies have shown that beyond a characteristic particle size on the order of a centimeter (millimeter), even low-velocity collisions among icy (rocky) particles lead to breakup or mass-transfer among impactors \citep{Blum1993, 2012Windmark}. Accordingly, the planetesimal formation process must boil down to the recurrent and rapid conversion of innumerable number of pebbles into massive asteroid-like bodies.

A number of empirical lines of evidence pertaining to the surviving small-body populations of the solar system -- including the size-distribution of the asteroid belt \citep{morby2009} as well as the inclination distribution of binaries within the Kuiper belt \citep{nesvorny2019} -- indicate that the physical machinery that drives planetesimal formation is driven by direct \textit{gravitational collapse}. In other words, pebbles within the protoplanetary disk concentrate into vast clouds that grow massive enough to become gravitationally unstable. In turn, concentration of pebbles into these clouds is facilitated by a distinct hydrodynamical process -- the \textit{streaming instability} \citep{youdin2005,johansen2007}. 

Although the detailed picture of the streaming instability (and resonant drag instabilities in general) is intricate \citep{squire2018}, its basic physics can be understood in a straightforward manner. As a starting point, recall that due to internal pressure support, the orbital velocity of gas within the circumstellar disk lags the Keplerian speed by a small margin $\Delta v = \eta_w \,v_{\rm{kep}}\sim (h/r)^2\,v_{\rm{kep}}$. Due to interactions with a sub-Keplerian nebular flow, dust grains within protoplanetary disks gradually drift inwards. Among other factors, their drift rate is controlled by the ratio of the particle cross-sectional area to the  mass. If we consider the evolution of two nearby dust grains in a protoplanetary disk, their cumulative drift rate can be diminished if one particle ``hides" in the hydrodynamic wake of the other, such that the mass of the two-particle aggregate is doubled while leaving its effective cross-sectional area unchanged. The differential drift among solids created by this process allows the two-particle aggregate to encounter and capture a third particle, and so on. Extending this thought experiment to an entire cloud of pebbles, one can understand how a sufficiently massive peloton of icy and rocky grains can locally accelerate the gas towards the Keplerian value, thereby creating a broadly favorable environment for continued capture of drifting particles and facilitating its own continued growth. 
 
 The purely hydrodynamic accumulation of solids into massive filaments facilitated by the streaming instability cannot continue indefinitely, and the nonlinear evolution of pebble clouds within protoplanetary nebulae culminates in their gravitational collapse. Drawing upon the pronounced analogy of this process with the formation of stars, {\color{red}\citet{Klahr2021}} derived an effective Jeans mass for planetesimals by considering the competition between the timescales for pebble turbulent diffusion and gravitational collapse. This relation sets the characteristic mass scale for bodies generated from self-gravitational collapse:
\begin{align}
m\approx100\,\bigg(\frac{\alpha}{\mathcal{Z}_{\rm{Hill}}} \bigg)^{3/2}\,\bigg(\frac{h}{r} \bigg)^3\,M_{\star},
\end{align}
where $\alpha\sim10^{-4}-10^{-2}$ denotes the turbulent diffusivity parameter, $\mathcal{Z}_{\rm{Hill}}\sim10-100$ is the mid-plane dust-to-gas ratio upon reaching Hill density, and $h/r\gtrsim10^{-2}$ is the disk's geometric aspect ratio. 

{\color{red}\citet{Liu2020} took a different approach to computing $m$, and obtained a characteristic planetesimal mass by extrapolating the results of numerical streaming instabiity simulations. For their preferred fit parameters, the relevant expression takes the form:
\begin{align}
m\approx200\,\big(h\,r\,\Sigma \big)^{3/2}\,\sqrt{\frac{Z}{M_{\star}}}\,\frac{M_{\oplus}}{M_{\odot}},
\end{align}}
Given nominal {\color{red}disk properties}, the above expressions yield characteristic planetesimal radii on the order of {\color{red}one to} a few hundred kilometers, {\color{red}with the analytical estimate systematically overestimating the numerical results. Nonetheless}, the emerging paradigm suggests that the formation of planetary building blocks stems from a near-universal gravity-hydrodynamic mechanism and instills a pronounced degree of mass uniformity onto newly born planetesimals.

\subsection{From Planetesimals to Planets}
\label{sec:planetesimals to planets}

Planetesimals generated through gravitational instabilities of pebble clouds can perpetuate their growth through two distinct pathways: pairwise collisions with other planetesimals, and/or continued accumulation of small rocky bodies. At a basic level, both of these processes can be understood from analytic considerations, and the characteristic rate of mass-accretion can be derived from simple $n$-$\sigma$-$v$ type relations. We begin by reviewing the former process.

\paragraph{Pairwise Collisions.} To leading order, the collisional mass accretion rate can be approximated as the product of the rate of collisions $\Gamma$ and the impactor mass $m$ such that $\dot{M}\sim m\,\Gamma$.  Replacing the number density with the planetesimal surface density $n\,m\sim\Sigma_{\rm{pl}}/h$, and approximating the velocity dispersion as $\langle v \rangle\sim (h/r)\,v_{\rm{kep}}$, the mass accretion rate takes the form \citep{lissauer1993,kokubo1996}:
\begin{align}
\dot{M}=\Sigma_{\rm{pl}}\,\pi\,R^2\,(1+\Theta)
\,\Omega.
\label{eqn:Mdot}
\end{align}
Note that in the above expression, the physical cross-section for collisions is augmented by the gravitational focusing factor $(1+\Theta)$, where $\Theta=(v_{\rm{esc}}/\langle v \rangle)^2$ is the Safronov number. This factor accounts for the fact that in the limit of very low velocity dispersion, the Safronov number approaches $\Theta\rightarrow\infty$, and accretion ensues at an accelerated rate due to enhanced gravitational focusing. Alternatively, if the velocity dispersion of the planetesimal swarm becomes comparable to the escape velocity of the constituent planetesimals, the collision cross-section becomes comparable to the physical radius.

For a given value of $\Theta$ (with the caveat that $\Theta$ does not exceed unity by a large margin), it is convenient to express $\dot{M}$ as $4\,\pi\,\rho_{\star}\,R^2\,dR/dt$, and note that the dependence on the physical radius cancels, yielding a constant $dR/dt$ (equivalently,  $M\propto t^3$). Taking advantage of this simplification, we can rewrite equation (\ref{eqn:Mdot}) as an accretion timescale. The time required for the radius to evolve from $R_0$ to $R$ is then:
\begin{align}
\mathcal{T}_{\rm{accr}}=\frac{4\,\rho_{\star}\,(R-R_0)}{\Sigma_{\rm{pl}}\,\Omega\,(1+\Theta)}.
\end{align}

\begin{figure*}[h]
\centering
\includegraphics[width=0.95\textwidth]{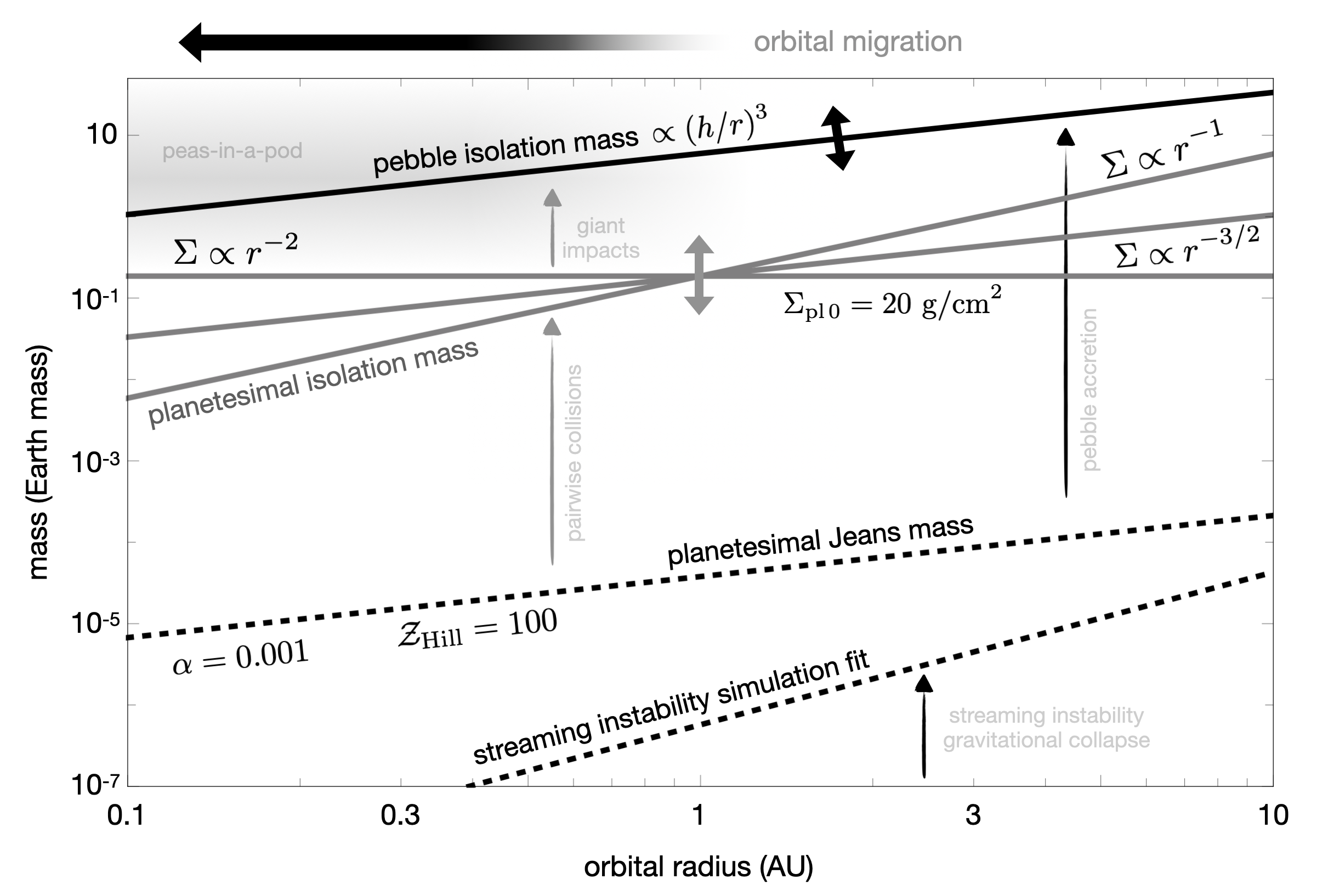}
\caption{Characteristic mass scales of planet formation. Planetesimals form within the protoplanetary disk through direct gravitational collapse. The planetesimal {\color{red}mass estimates are} shown in the figure with a dotted lines. Growth beyond planetesimals occurs through collisions with other planetesimals as well as gravito-hydrodynamic capture of pebbles. Planetesimal accretion subsides when protoplanetary embryos reach the planetesimal isolation mass -- shown on the figure with gray lines for a variety of surface density profiles. {\color{red}Beyond this scale, planetary growth can proceed via giant impacts.} Pebble accretion terminates at the considerably greater pebble isolation mass, shown on the figure as a black line. Once planets become sufficiently massive to raise significant wakes in their natal nebulae, orbital migration sets in.}
\label{fig:pebble}
\end{figure*}

Before evaluating $\mathcal{T}_{\rm{accr}}$, we remark that collisional growth cannot persist indefinitely, and stalls when the planetary mass reaches the \textit{isolation} threshold \citep{lissauer1993,kokubo1996}. This argument hinges on the fact that the planetesimal feeding zone available to the planet only extends over a few Hill spheres in either radial direction, i.e., $\Delta a \sim 3\,a\,(M/(3M_{\star}))^{1/3}$. Consequently, isolation is reached when the cumulative planetesimal mass contained within an an annulus of radial extent $2\,\Delta a$ matches the planetary mass itself: $M=4\,\pi\,a\,\Sigma_{\rm{pl}}\,\Delta a$. Rearranging this expression for the isolation mass, we have:
\begin{align}
M_{\rm{iso}}=\xi\,\pi^{3/2}\,
\frac{\Sigma_{\rm{pl}}^{3/2}}{\sqrt{M_{\star}}}\,a^3,
\label{eqn:planetesimal iso mass}
\end{align}
where $\xi\approx30$ is a dimensionless constant. 
Figure \ref{fig:pebble} shows the isolation mass $M_{\rm{iso}}$ as a function of radial distance  for a series of surface density profiles, normalized to an MMSN-like value of the solid surface density $\Sigma_0=20\,$g/cm$^2$ at 1 AU. An immediately evident result is that the isolation mass itself can be rather uniform if the solid surface density power law index approaches $p=2$. As importantly, if we assume that planetesimal formation can efficiently operate significantly interior to $1\,$AU, then the timescale needed to reach isolation in the inner-most regions of the nebula is on the order of $\mathcal{T}_{\rm{accr}}\sim10^5\,$years, although this value increases to $\sim10\,$Myr at $1\,$AU. More strikingly, however, the value of $M_{\rm{iso}}$ is on the order of the mass of Mars, and is thus significantly smaller than the typical mass-scale of short-period extrasolar planets.

The above argument indicates that pairwise-collisional mode of accretion {\color{red}within a MMSN-like profile of solids does not naturally yield multi-Earth-mass objects. Instead, such a scenario necessitates a post-isolation phase of giant impacts to consolidate the emergent embryos into bonafide planets. Of course, a Mars-like value of $M_{\rm{iso}}$ is not universal: one may invoke a substantially more solid-rich (e.g., MMEN-type; \citealt{ChiangLaughlin2013}) disk, and rely on the $M_{\rm{iso}}\propto \Sigma_{\rm{pl}}^{3/2}$ proportionality to boost the isolation scale (thereby bringing the planet formation paradigm closer to the standard model of giant-planet satellite formation; \citealt{CanupWard2002,BM20}). Yet another alternative is that of a disk with a markedly non-uniform distribution of solids, where $\Sigma_{\rm{pl}}$ can achieve high values at discrete locations while maintaining a low overall solid mass-budget within the disk. To this end, recent work has shown how radial concentration of solids within protoplanetary nebulae can lead to the formation of planetesimals within dense narrow annuli \citep{Draz2016,Morby2022}. Although such a scenario has gained traction within the solar system formation literature (see e.g., \citealt{Izidoro2021} and the references therein), the viability of collisional accretion of mass-uniform sub-Jovian planets within rings of solid debris remains to be demonstrated quantitatively.}


\paragraph{Pebble Accretion.} A {\color{red}different} mode of planetary growth proceeds through the capture of mm- to cm-size particles by a growing body --- a regime of conglomeration known as \textit{pebble accretion} \citep{Ormel2010,Lambrechts2012}. Counter-intuitively, {\color{red}under certain conditions this process can lead} to a mass accretion rate that is considerably faster than that given by pairwise collisions, even though the mass of each captured particle is very small in comparison with the mass of a planetesimal. 

The most important distinction between planetesimal accretion and pebble accretion arises from the accretion cross-section. Particularly, in the case of pebble accretion, the physical radius (augmented by gravitational focusing) present in equation (\ref{eqn:Mdot}) is replaced by a characteristic length-scale over which a drifting particle experiences a large-angle deflection by the planetary gravity. For low planetary masses, this length scale is given by the modified Bondi radius (that accounts for the fact that smaller particles are more tightly coupled to the gas and are therefore difficult to capture; see \citealt{Lambrechts2012}). For larger planets, the critical impact parameter for accretion becomes comparable to the Hill radius. 

{\color{red}If vertical settling of solids is efficient (such that pebbles form a thin sub-disk), accretion can proceed in 2D,} and in analogy with equation (\ref{eqn:Mdot}), the rate of mass accretion in the Hill regime can be written as
\begin{align}
\dot{M}_{\mathrm{2D}}\sim2 \,\Sigma_{\rm{pebble}}\,R_{\rm{Hill}}^2\,\tau^{2/3}\,\Omega,
\label{eqn:Mdotpeb}
\end{align}
where $\tau=\Omega\,(\rho_{\rm{pebble}}/\rho_{\rm{gas}})(s/v_{\rm{th}})$ is the Stokes number (dimensionless stopping time) of the particles. Given that the Hill radius exceeds the planetary physical radius by a factor of order $\sim a/R_{\star}$, it is easy to see how {\color{red}under the favorable assumption of $\tau\sim1$,} the process of pebble accretion can facilitate the formation of multi-Earth-mass bodies on a timescale that is much shorter than the age of the protoplanetary disk (e.g., $\sim10^4$ years; \citealt{Lambrechts2012, Morbidelli2012}). Moreover, unlike the case of planetesimal accretion where the local mass budget plays a limiting role, {\color{red}pebble accretion is facilitated by the radial drift of solids, and can therefore ``access" the mass reservoirs that initially reside in the outer disk region and continuously drift inward to cross the planet's orbit.}

{\color{red}Although the 2D regime of gravito-hydrodynamic dust capture can in principle be staggeringly efficient, the advantages of pebble accretion largely fade if the solid component of the disk is vertically well-mixed with the gas. In this case, accretion proceeds in the 3D regime, at an exponential, but slow rate \citep{ormel2017}:}
\begin{align}
\dot{M}_{\mathrm{3D}}\sim6\,\pi\,\rho_{\rm{pebble}}\,R_{\rm{Hill}}^3\,\tau\,\Omega,
\label{eqn:Mdotpeb3D}
\end{align}
{\color{red}where $\rho_{\rm{pebble}}\approx\Sigma_{\rm{pebble}}/(\sqrt{2\,\pi}\,h_{\rm{pebble}})$ and $h_{\rm{pebble}}$ is the dust sub-disk's scale-height. If we adopt a dust particle radius on the order of a millimeter (in accordance with the fragmentation barrier; \citealt{2012Windmark}), the Stokes number evaluates to $\tau\lesssim0.001$ in the inner disk. Given that $h_{\rm{pebble}}\sim h\,\sqrt{\alpha/\tau}$ \citep{ChiangYoudin2010}, a nebular turbulence parameter that is substantially lower than $\alpha\ll 0.001$ is required to break the system out of the comparatively inefficient 3D regime. Thus, adopting an oft-quoted radial pebble flux of $F_{\rm{pebble}}\sim10^{-4}\,M_{\oplus}/$year (which translates to a surface density of $\Sigma_{\rm{pebble}}\sim F_{\rm{pebble}}/(4\,\pi\,r\,v_{\rm{kep}}\,\eta_w\,\tau)$; \citealt{Lambrechts2019}) and setting $h_{\rm{pebble}}\sim h$, we obtain a mass $e$-folding timescale of $(M/\dot{M}_{\rm{3D}})\sim\sqrt{8\,\pi}\,(M_{\star}/F_{\rm{pebble}})\,(h/r)^3\sim6\times10^8\,$years. This timescale is much longer than the typical lifetime of a protoplanetary disk, and unlike 2D accretion -- which is expected to ensue beyond the ice-line where particles can attain much larger Stokes numbers \citep{BM22} -- 3D accretion in the inner disk is unlikely to drive significant planetary growth.}


Independent of the accretion mode, an important feature of the pebble accretion paradigm is that the protoplanetary agglomeration process itself is self-limiting \citep{Lambrechts2014}. This aspect of the theoretical picture can be understood by considering the gravitational back-reaction of the growing proto-planet upon the gaseous disk. In the vicinity of a sufficiently massive planet, planetary gravity can {\color{red}locally} accelerate the gas {\color{red}above} the Keplerian {\color{red}velocity, reversing the action of gas-drag. When} this happens, pebbles circumvent the planet, and the very effect that facilitates pebble accretion subsides.

To estimate the characteristic planetary mass at which accretion subsides we equate the Hill radius to the disk scale-height to obtain \citep{ormel2017}:
\begin{align}
\frac{M_{\mathrm{iso}}}{M_{\star}}= \zeta \bigg(\frac{h}{r}\bigg)^3,
\label{eqn:pebbleisomass}
\end{align}
where $\zeta$ is a numerical factor of order unity. Importantly, numerical experiments of \cite{Lambrechts2014} and \cite{Bitsch2015} recover the cubic scaling of the mass ratio upon the disk’s geometric aspect ratio and sharpen the above estimate. In particular, for a flared disk with an aspect ratio that scales as $h/r \propto r^{1/4}$, \cite{Lambrechts2014} argue for an isolation mass of $M_{\mathrm{iso}}\sim20\,M_{\oplus}\,(a/5\,\mathrm{AU})^{3/4}$. This estimate is shown on Figure \ref{fig:pebble} as a black line. 

A steep dependence on $h/r$ -- a quantity that can vary by a factor of $\sim$2 from system to system -- implies that pebble accretion can generate a broad range of planetary properties, while preserving a uniform mass-scale within a single system. Within the same framework, the stellar mass plays a sub-dominant role. Nevertheless, equation (\ref{eqn:pebbleisomass}) clearly predicts that despite inherent variability, planets that orbit less massive stars should also be less massive on average. 

{\color{red}Taken at face-value}, this result is {\color{red}attractive as an explanation for the architecture of} close-in extrasolar planets: beyond reproducing the appropriate mass-scale for the known super-Earth/sub-Neptune population, it further illuminates that the pebble accretion model of planet formation predicts a distinct pattern of uniformity among the generated planets. {\color{red}On the other hand, inefficient settling of solids within the inner region of protoplanetary disks insinuates that pebble accretion only operates vigorously beyond the ice-line, thereby preferentially generating water-rich planets -- an expectation that is confirmed by detailed numerical simulations \citep{Izidoro2017}. Therefore, recent determination of the broad prevalence of silicate-rich short-period planets \citep{Rogers2021,Zeng2019} fosters substantial compositional tension between the pebble accretion model and observations, casting doubt on pebble accretion as the primary process responsible for the} genesis of sub-Neptune-sized extrasolar planets.

\subsection{Orbital Migration}
\label{sec:migrate} 

An inescapable consequence of planetary growth within a massive gaseous disk is the onset of \textit{planet-disk interactions}. Distinct from aerodynamic drag (which is negligible for planetary-mass objects) the primary mode of angular momentum exchange between planets and their surrounding gas is gravitational \citep{Goldreich1979,Lin1979,Ward1997}. In particular, planetary perturbation of the gas flow in its local neighborhood manifests as a spiral density wave -- or ``wake" -- that trails the planet's orbital phase exterior to its orbit and leads the planet at interior orbital radii \citep{Ogilvie2002}. It is the gravitational back-reaction of this wake upon the planet that drives migration.

Somewhat counter-intuitively, gravitational interactions between the planet and its wake are effectively repulsive: the exterior, trailing part of the wave continuously pulls back on the planet, sapping it of orbital energy and causing it to fall deeper into the stellar potential well. The reverse effect manifests as a consequence of interaction between the planet and the leading {\color{red}wake}. Consequently, orbital changes themselves arise from a subtle imbalance of the torques exerted by the inner and outer arms of the spiral density wave.

Because the wake itself is a response to planetary gravity and planetary migration is a response to the wake, the characteristic rate of orbital transport scales as the product of the effective disk mass, $\sim\Sigma\,r^2$, and the mass of the planet, $M_p$. The relevant expression further exhibits a non-trivial inverse-square dependence on the speed of sound within the disk (via the disk aspect ratio $h/r=c_{\rm{s}}/v_{\rm{kep}}$), which stems in part from the fact that waves are only excited where the gaseous flow past the planet becomes supersonic, as well as the fact that the torque difference in leading and trailing arms itself depends on the disk scale height \citep{Tanaka2002}. Cumulatively, a general expression for the rate of migration has the form:
\begin{align}
\frac{\dot{r}}{r}=\Upsilon\,\bigg(\frac{M_p}{M_{\star}}\bigg) \bigg(\frac{\Sigma\,r^2}{M_{\star}}\bigg)\,\bigg( \frac{r}{h} \bigg)^2\,\Omega,
\label{eqn:mig}
\end{align}
where $\Upsilon$ is a dimensionless coefficient that depends on specific disk properties. 

Substituting the characteristic values for the pebble isolation mass (\ref{eqn:Mdotpeb}) into this expression, for nominal MMSN-like disk parameters, we obtain characteristic orbital migration rates that are undeniably fast. For example, the characteristic migration timescale of a $20\,M_{\oplus}$ planet at $5\,$AU is {\color{red}$\Upsilon\,r/\dot{r}\lesssim10^5\,$}years. Migration rates only accelerate for shorter-period orbits. In fact, even a Mars-mass embryo in the inner regions of the disk will be characterized by a nominal migration rate on the order of $\sim10^4\,$years. The same flavor of planet-disk interactions additionally leads to decay of planetary eccentricities and inclinations on a timescale that is even more rapid (by a factor of $\sim(r/h)^2$; \citealt{Tanaka2004, Kley2012}). Consequently, orbital migration is almost certainly an important and active process within protoplanetary disks that plays a crucial role in regulating the terminal architectures of planetary systems.

Despite the pronounced importance of planetary migration, its direct consequences are difficult to calculate in a robustly predictive manner. Being dictated by a difference in two large numbers (i.e., inner and outer spiral arm torques), both the direction and magnitude of disk-forced orbital evolution exhibit sensitive dependence on poorly constrained disk parameters -- a complexity that is captured in the highly non-trivial form of the coefficient $\Upsilon$. To this end, \cite{Paardekooper2010} have obtained the following form for this coefficient:
\begin{align}
\Upsilon = \gamma\,(p/10-17\beta/10-5/2),
\label{eqn:Upsilon coefficient}
\end{align}
where $p$ is the surface density power-law index as before, $\beta$ is the power-law index of the temperature profile, and $\gamma$ is the ratio of specific heats {\color{red}(see also \citealt{Paardekooper2011} for a discussion on the corotation torque and its saturation)}. More generally, hydrodynamic simulations reveal complex mass-period ``migration maps" that are flecked with intricate regions of both inward and outward migration, with variable rates \citep{Bitsch2015,Coleman2014,Coleman2016}. Indeed, the picture is rendered even more complex by consideration of disk turbulence -- that introduce a random component into the migration process \citep{Nelson2005, Adams2008, Rein2009} -- as well as fluctuations in the nebular entropy gradient, which are themselves controlled in part by the chemical structure of the disk. All of these intriguing aspects of migration theory remain an active area of research.

Returning to the specific consideration of close-in sub-Neptunes, let us highlight two key results pertinent to orbital migration. First and foremost, the disk's inner edge, which is presumably set by the magnetic truncation radius (equation \ref{eqn:trunk}), appears to act as a trap to inward-migrating planets {\color{red}\citep{Masset2006,Liu2017,Romanova2019}}. This is not because the planets are pushed into the cavity where the gas density is tenuous, but instead because the sharp gradient in the surface density profile dramatically enhances the corotation torque, which causes the inner edge to repel the planets outward \citep{Paardekooper2018}. On the other hand, irregularities in the migration portraits of typical disks tend to diminish over the disk's lifetime, leading to an expectation of an overall tendency towards inward migration in mature disks \citep{Lyra2010}. Thus, the architectures of close-in exoplanetary systems {\color{red}should be} broadly consistent with a formation scenario that is shaped by sustained inward decay of planetary orbits that terminates at the disk's inner edge.
However, some challenges still remain for the migration paradigm, particularly the question of whether it can explain the largely non-resonant orbital architectures of the compact multis (\S\ref{sec:spacing}) {\color{red}as well as the radial mass-ordering of the planets (see e.g., \citealt{Ogihara2015,Coleman2019})}. We now revisit this observation in the context of its implications for understanding sub-Neptune formation and migration. 

\paragraph{Paucity of Resonances and Near-Resonance Features.}


Convergent orbital migration naturally shepherds planets into mean-motion resonances (e.g., \citealt{Mustill2011,Batygin2015}). Thus, the question of why resonances are intrinsically rare among the compact multis (recall \S\ref{sec:spacing}, Figure \ref{fig:period ratio distribution}) remains an active area of research. Multiple ideas have been put forth to explain this puzzle over the last decade and a half. One suggestion is that, in a sufficiently turbulent protoplanetary disk, resonant capture can be impeded through stochastic fluctuations of the disk’s gravitational potential \citep{Adams2008,Rein2009}. Alternatively, \cite{Goldreich2014} argued that a specific choice of planet-disk interaction parameters can render resonances metastable within the nebulae.  {\color{red}Finally, \citet{McNally2019} have argued that in sufficiently laminar disks, emergence of vortices can prevent resonant trapping of planets.}  While these models can indeed provide a resonance disruption mechanism under certain conditions \citep{Deck2015,Batygin2017}, a scenario which has had the most success in matching the period ratio distribution on a \textit{quantitative} level is one where primordial resonances are disrupted through a phase of post-nebular instabilities {\color{red}\citep{Izidoro2017,Izidoro2019, MatsumotoOgihara2020}. }

{\color{red}Beyond reproducing the typical orbital architectures themselves, the resonant-chain instability scenario is broadly consistent with the observed pattern of uniformity. To this end, the numerical experiments reported in \citet{GB22} show that if the majority of planetary systems originate as resonant chains that relax through a transient period of planet-planet scattering, the process of collisional consolidation that ensues during the dynamical instabilities does not degrade the mass uniformity of the individual systems beyond that exhibited by the data. }Numerous ideas have been put forward to explain how sub-Neptunes can enter a phase of planet-planet scattering (e.g., \citealt{Johansen2012,Spalding2016,Spalding2018,Pichierri2020,Petit2020}). However, a specific understanding of how dynamical instabilities are triggered in the proportion needed to explain the period ratio distribution remains elusive. 

While resonances are uncommon overall, they are not completely absent from the population. In \S\ref{sec:spacing}, we showed that the period ratio distribution contains small but significant overabundances of planet pairs wide of resonance (see Figure \ref{fig:period ratio distribution}). There have been a wide range of efforts to understand the origin of these near-resonance features. Toward this end, many authors \citep{Papaloizou2010,Lithwick2012,Batygin2013,Bart2013,Delisle2014,ChatFord2015} have demonstrated that these features can be accounted for by slow divergence of initially resonant orbits, facilitated by sustained eccentricity damping. Pertinently, \cite{Millholland2019} have shown that obliquity tides provide a natural source of this dissipation (see also \citealt{Choksi2020,GB21AJ}). Other proposed ideas involve various (higher order) planet-disk interactions effects \citep[e.g.][]{Baruteau2013, Migaszewski2015, Ramos2017}. Nevertheless, even though the dominant sculptors of these near-resonance features are still uncertain, any comprehensive theory of sub-Neptune formation must somehow account for them. We will return to this point when reviewing population synthesis models in \S \ref{sec:popsynth}.

\subsection{Atmospheric Escape and Sculpting}
\label{sec: atmospheric mass loss}


Regardless of their orbital architectures, planets that form within H/He nebulae capture gaseous atmospheres. The planets in the {\color{red}compact multis}, however, do not always retain the atmospheres they accrete during the nebular phase. Instead, their highly-irradiated orbits render these planets susceptible to thermally-driven atmospheric mass loss. Planets with sufficiently low masses and close orbits can lose their initial envelopes entirely, and thus end up as stripped cores. Atmospheric escape sculpts the radii of the {\color{red}compact multis} and weakens their size uniformity from its primordial state. 

At the population-level, evidence for atmospheric escape can be seen in the distribution of radii of $1-4 \ R_{\oplus}$ planets. The distribution is bimodal, containing a dearth of planets with radii between $1.5 - 2 R_{\oplus}$ that is known as the ``radius valley''. The existence of the radius valley was predicted by models of atmospheric escape \citep{Owen2013, Lopez2013,Jin2014}, which tend to produce planets that either retain their $\sim 0.5\%-1\%$ H/He-dominated envelopes (resulting in sub-Neptunes) or lose them entirely (resulting in super-Earths). The radius valley was later confirmed through observations from CKS (\citealt{Fulton2017}; see \S \ref{sec: uniform sizes and masses}). 

Two types of thermally-driven atmospheric mass loss are influential for planets in {\color{red}compact multis}: photoevaporation and core-powered mass loss. They differ primarily in their heat source. With photoevaporation, the energy source is the EUV and X-ray flux from the host star, which photoionizes hydrogen and heats the upper atmosphere to high temperatures ($T\sim10^4$ K), producing a hydrodynamic outflow that is strongest in the first $\sim100$ Myr \citep{Lammer2003, Baraffe2004, MurrayClay2009, Owen2013, Lopez2013}. With core-powered mass loss, the energy source is a combination of the stellar bolometric flux and remnant thermal energy from formation, which is slowly released from the core to the atmosphere, producing a cooler hydrodynamic outflow that is sustained over billions of years \citep{Ginzburg2016, Ginzburg2018, Gupta2019}.

Both the photoevaporation and core-powered mass loss models reproduce demographic features of the close-in super-Earths/sub-Neptunes as a whole and the {\color{red}compact multis} in particular. They predict similar slopes of the radius valley as a function of period, incident flux, and stellar mass, and these agree with the data \citep{Owen2017,Jin2018,Gupta2019, Gupta2020}. They also predict a breakdown of perfectly-uniform peas-in-a-pod {\color{red}patterns}, given that the least massive and most irradiated planets in these systems will lose their atmospheres. {\color{red} The data show a degree of size diversity that is consistent with atmospheric escape; {\color{red}systems exhibit a tendency towards size ordering  \citep{Ciardi2013, Millholland2017, Kipping2018, Weiss2018a, MillhollandWinn2021}, with systems much more frequently hosting super-Earths interior to sub-Neptunes than vice versa.}
}
For completeness, we note that mass loss can be enhanced by large radioactive abundances in close-in planets. The magnetic truncation radius given by equation (\ref{eqn:trunk}) is coincident with the semimajor axes of planets in {\color{red}compact multis}. In the region where the disk is magnetically truncated, the fields continually wrap up, short out, and reconnect. This activity leads to the acceleration of particle radiation (cosmic rays) that can drive spallation processes in the reconnection region \citep{lee1998,shu1997}. A number of short-lived radionuclides (SLRs) can be generated via spallation, and the predicted {\color{red}cross sections} are large enough that the {\color{red}compact multis} could be enhanced {\color{red}in SLR abundances} \citep{Adams2021final}. {\color{red}Energy released from SLR decay can}, in turn, amplify mass loss mechanisms from young and forming planets, as well as remove volatile components (e.g., water). 


Finally, it is important to note that thermally-driven mass loss is not the only driver of intra-system size diversity via atmospheric loss. Late-stage giant impacts, particularly those occurring after disk dispersal, can significantly reduce the H/He envelopes of sub-Neptunes and sometimes strip the cores entirely {\color{red} \citep[e.g.][]{Inamdar2015, Inamdar2016, Liu2015, Biersteker2019, Kegerreis2020}}. Despite their efficiency, the frequency of atmosphere-stripping giant impacts is still poorly-constrained. The observed degree of intra-system uniformity of planet sizes and masses could feasibly provide a constraint on their prevalence; quantitative efforts towards this goal remain to be seen.

\section{Planet-Planet Interactions} 
\label{sec:pp interactions} 


The planet formation process, as detailed in the previous sections, is exceedingly efficient at producing close-in planets with a characteristic mass set by a series of planet-disk interactions. So far, however, we have not comprehensively addressed the interactions between planets that occur during their growth and after they are fully formed (recall the schematic outline in Figure \ref{fig:schematic}). These planet-planet interactions dominate at the later stages of formation, and they can also play an important role in establishing intra-system uniformity of planet masses and orbital spacings. This section discusses the influences of dynamical interactions, in the form of instabilities (\S\ref{sec:stability}) and pairwise energy exchange via dissipation (\S\ref{sec:eoptimize}). We also review how these dynamics may sculpt the planetary mass function (\S\ref{sec:pmf}).

\subsection{Dynamical Stability Constraints} 
\label{sec:stability} 

Once planets have fully formed, with given values for their masses and well-defined orbital elements, the resulting planetary systems must be dynamically stable. In general, planetary orbits cannot be spaced too closely without rendering the system unstable, where the minimum spacing depends on the planet masses (and orbital  eccentricities). This constraint can be written in terms of the ratio of semi-major axes of adjacent planets, 
\begin{equation}
{a_2 \over a_1} \equiv \Lambda > 
1 + \Delta {R_H \over a}\,,
\end{equation}
where $R_H/a$ is determined by the mutual Hill radius (equation \ref{eqn:Hill}).   For two planet systems, the minimum value of $\Delta$ can be calculated $\Delta=2\sqrt{3}$ \citep{gladman1993}. For systems of three or more planets, the minimum value of $\Delta$ required for stability is larger, where numerical simulations typically find $\Delta=10$ (e.g., 
\citealt{Pu2015}; see also \citealt{Petit2020,Pichierri2020,Tamayo2021} for detailed theoretical inquiries). Note that one way to characterize planetary systems is by their distributions of spacing parameters $\Lambda$. The key features of {\color{red}compact multis} are that [1] the spacing parameters are large enough to imply dynamical stability, [2] the distribution of spacing parameters for the entire sample shows a broad peak at values just larger than those required for stability, and [3] the variation of spacing parameters for pairs of planets within a given system is smaller (tighter) than for the distribution as a whole (see \S \ref{sec:spacing}). 

\subsection{Pairwise Energy Optimization}
\label{sec:eoptimize} 

Many of the observed properties of the peas-in-a-pod {\color{red}architectures} can be understood by considering the tidal equilibrium states for a pair of planets \citep{Adams2019,Adams2020}. Suppose that two planets are forming from an annulus in a circumstellar disk. Under a range of circumstances, we expect that the total mass $M_T$ available to make planets will be fixed, and that the total angular momentum $L$ will be conserved. Let us further constrain the spacing parameter $\Lambda=a_2/a_1$ of the planetary orbits to also be determined (but different values of $\Lambda$ can be considered after the fact). The properties of the system can be specified by the planet masses $M_{p,1}$ and $M_{p,2}$, the semi-major axes of the orbits $a_1$ and $a_2$, the orbital eccentricities $e_1$ and $e_2$, and the mutual inclination $i$ between the orbits. The tidal equilibrium state corresponds to the lowest energy state available to the system subject to the constraints. In other words, we can find the values of the system parameters $(M_{p,1},M_{p,2},a_1,a_2,e_1,e_2,i)$ that provide the lowest energy state for given fixed values of total mass, angular momentum, and spacing $(M_T,L,\Lambda)$. 

The solution to this optimization problem shows that the minimum energy state corresponds to co-planar and circular orbits ($i=e_1=e_2=0)$ with nearly equal masses. The optimized value of the mass fraction is given by 
\begin{equation}
f = {M_{p,1} \over M_T} = 
{\Lambda + \sqrt{\Lambda} - 2 \over 
3(\Lambda-1)} = {\sqrt{\Lambda} + 2 \over 3(\sqrt{\Lambda}+1)} \,.
\end{equation}
The corresponding ratio of masses $\eta=M_{p,2}/M_{p,1}$ can be written in the form 
\begin{equation}
\eta = {2 \sqrt{\Lambda} + 1 \over \sqrt{\Lambda} + 2}    \,.
\label{etadef} 
\end{equation}
In the limit of close separations $\Lambda\to1$, the mass fraction $f\to1/2$ and the mass ratio $\eta\to1$. In other words, for close orbital spacing, the minimum energy state accessible for a pair of forming planets has exactly equal masses. The mass fraction $f$ decreases slowly with increasing $\Lambda$, but only falls to $f=1/3$ ($\eta\to2$) in the limit $\Lambda\to\infty$. For any reasonable finite value of the spacing parameter, energy optimization predicts {\it nearly} equal mass planets, as suggested by observations (which typically measure nearly equal radius planets). 

The above considerations apply directly to pairs of forming planets. For the case of three or more planets, energy optimization can be carried out in two conceptually different ways. Consider a three planet system: In the first case, one can find the global energy minimum for all three planets. In the second case, one can find the minimum energy state for both pairs of planets and then match boundary conditions (such that the middle planet is the outer planet of the inner pair, as well as the inner planet of the outer pair). The second approach leads to systems like those observed. Specifically, in their optimal state, the planets all have zero eccentricity and inclination, and have nearly equal masses. The planet mass grows slowly with increasing semimajor axis (by a factor of $\eta$ given by equation (\ref{etadef})). Note that this slowly increasing progression of planetary masses corresponds to an underlying surface density distribution of $\Sigma\sim r^{-11/6}$ (see \S\ref{sec:starform}). 

The energy optimization approach has another feature that also conforms to properties of observed planetary systems. The discussion thus far has not included the self-gravity of the planets in the energy budget. If we add this complication to the analysis \citep{Adams2020}, the tendency for planetary pairs to have nearly equal masses continues to hold when the total mass (of the pair) is less than a threshold value given by 
\begin{equation}
M_C = M_{\star} \left({R_p\over a_2}\right) 
{ (\sqrt{\Lambda}+2)(\Lambda-1)^2 \over 4\alpha_g \sqrt{\Lambda}} \,.
\label{bifurmass} 
\end{equation}
Here, $\alpha_g$ is a dimensionless constant of order unity that depends on the internal structure of the planet. The mass scale $M_C$ is a bifurcation parameter and has the value $M_C\sim40M_\oplus$ for the {\color{red}compact multis} of interest. For low mass planetary pairs, with total mass $M_T<M_C$, energy optimization leads to nearly equal mass bodies. For the high mass case, $M_T>M_C$, it becomes energetically favorable for one planet to experience runaway growth and consume the majority of the mass. This trend is also seen in the observed planetary sample: Systems that contain large planets (e.g, those of Jovian mass) do not display the same peas-in-a-pod behavior as systems with lower mass planets (e.g., \citealt{Wang2017}). 

Finally, we note that the bifurcation mass scale (\ref{bifurmass}) depends on the location of the planets, and varies inversely with the semimajor axis. As a result, planetary pairs forming in the outer regions of their systems will have much smaller values of $M_C$, which implies that most planetary pairs will exceed the critical mass scale. This trend implies a clean prediction: Planets with larger orbits ($a\gg1$ AU) are much less likely to exhibit peas-in-a-pod properties. 

We note that there are forms of dissipation that can decrease energy while ensuring angular momentum conservation. This type of mechanism provides a path to the optimal energy state while preserving angular momentum. One can also show that if the system is not at a minimum of the energy-momentum, then such dissipation destabilizes any other type of critical point. This property is important for understanding stability in this context for the following reason: finding a Lyapunov function for an equilibrium implies stability. However, the failure of a Lyapunov function to be definite does not preclude stability -- another such function may exist, for example, or one may be able to prove stability in some other fashion. The dissipation result implies that for practical purposes failure of the energy momentum analysis to prove stability implies instability. Angular preserving dissipation often has a classic form related to the Hamiltonian dynamics of the system. A canonical and instructive case is that for rigid body dynamics where the dissipative angular momentum preserving dynamics take the form: 
\begin{equation}
I\dot{\omega}=I\omega\times\omega +\alpha
I\omega\times(I\omega\times\omega)\,,
\end{equation}
where $I$ is the moment of inertia matrix and $\omega$ is the body angular velocity vector and $\alpha$ is a positive constant which determines the dissipation rate. Similar forms can be written down for multi-body systems and fluids; it is also possible to write down effective angular momentum preserving integrators that generalize symplectic integrators. For a history of the energy momentum method and such forms of dissipation in various contexts, including  astrophysical settings, see \cite{Bloch1996}.

\subsection{The Planetary Mass Function}
\label{sec:pmf} 

The distribution of masses of the fundamental objects of interest represents an important issue for any astronomical study. For the multi-planet systems considered here, their mass distribution is markedly different from the mass distribution of the entire exoplanet sample. It is thus useful to consider how the distributions of planetary masses for the {\color{red}compact multis} are related to those of the entire sample of exoplanets.\footnote{Note that this discussion is by necessity preliminary: The planetary mass function for the entire sample is still being determined, and many observational selection effects and biases remain. In addition, for the {\color{red}compact multis}, we generally measure the planetary radius and infer the planetary mass, and this limitation leads to additional uncertainties.}

Mass distributions often display power-law forms (e.g., the stellar initial mass function; \citealt{salpeter1955}). Current data (e.g., \citealt{cumming2008,rosenthal2021}) indicate that the planetary mass function (PMF) has the approximate form 
\begin{equation}
    {df\over dM_p} \propto M_p^{-1.3} 
    \label{pmf} 
\end{equation}
over the mass range $0.1M_J\le{M_p}\le10M_J$, or, equivalently $30M_\oplus\le{M_p}\le3000M_\oplus$. In contrast, the finding emerging from observations of {\color{red}compact multis} is that, to leading order, the planets have nearly the same mass $M_p\sim10M_\oplus$ {\color{red}\citep{Millholland2017}}. More specifically, the inferred masses of members of these multi-planet systems show a peaked distribution near this mass scale, which lies in the regime of `super-Earths'. Note that the lower end of applicability for the power-law mass distribution (30 -- 40 $M_\oplus$) is roughly coincident with the bifurcation mass scale of equation (\ref{bifurmass}). 

The typical masses of the multi-planet members thus fall just below the masses found in the global sample of larger planets that display a power-law PMF. According to the current theory of planet formation (see the following section and references therein), most planets in this mass range consist of a rocky core surrounded by an envelope of gas. The mass scale of the rocky core is of order 10 $M_\oplus$, comparable to masses inferred for {\color{red}compact multis} and comparable to the cores of Jupiter and Saturn.\footnote{For completeness, we note that Jupiter might not have a solid core {\it per se}. Nonetheless, the chemical composition of Jupiter is consistent with it having a rocky core of order $\sim10M_\oplus$ surrounded by a gaseous envelope (which can also have an enhanced metallicity).} 

We can thus organize the current observational picture as follows: Circumstellar disks readily form rocky bodies with super-Earth masses, of order 10 $M_\oplus$. In many systems, including those that ultimately produce peas-in-a-pod planets, these rocky bodies acquire relatively little additional mass. In particular, they accrete little gas. In other systems, the rocky bodies can accrete gas, slowly at first (corresponding to phase 2 in the current theory of giant planet formation; {\color{red} \citealt{Pollack1996}}) and then more rapidly (in phase 3). For these larger accreting planets, the variables of the problem conspire to produce the power-law mass distribution of equation (\ref{pmf}). {\color{red} Note that some exoplanets are inferred to have cores that are larger than the nominal $\sim10M_\oplus$ mass scale \citep{Thorngren2016}, indicating that additional rocky material is also accreted}. If the currently available data provide an accurate representation, then the rocky bodies of the {\color{red}compact multis} comprise a larger population than the gaseous (and partially gaseous) planets obeying the power-law PMF. Moreover, since the latter PMF is a steeply decreasing function of mass, large gaseous planets are rare, which suggests that planets find it difficult to sustain mass accretion from their parental disks over large spans of time. 

Part of the explanation for observed PMF is that gas accretion is relatively slow and disk lifetimes are short (only $1-10$ Myr). As a result, disks produce many more `failed' giant planets (with masses less than Saturn) than large giant planets (with masses larger than Jupiter). In addition, the ubiquity of super-Earths indicates that their formation is efficient. The question that remains concerns the timing of super-Earth production: Do all of the $\sim10M_\oplus$ objects form quickly (on time scales less than $\sim1$ Myr), but most never get a chance to accrete substantial amounts of gas? {\color{red} In other words, the complicated interface between the planetary envelopes and the background circumstellar disk could delay cooling and limit gas accretion \citep{Ormel2015,Lambrechts2017}.} Alternately, do these objects form slowly, so that only the objects on the rapid end of the timing distribution are able to accrete gas? {\color{red} In that case, gas damping is weak and super-Earth cores can grow through mergers of protoplanets \citep{Lee2014}.} Note that a third possibility also exists -- that the super-Earth planets found in {\color{red}compact multis} represent a separate population from the cores of giant planets, and the similarity in mass and rocky composition is largely a coincidence.


\section{Population Synthesis}
\label{sec:popsynth}

The previous sections showed that the formation of {\color{red}compact multis} is consistent with the initial conditions provided by star-disk formation (\S \ref{sec:stardisk}), attainable through known planet formation processes (\S \ref{sec:planet-formation-theory}), and expected from planet-planet interactions that regulate mass growth (\S \ref{sec:pp interactions}). The next step is to understand how all of these sub-processes work together --- and in opposition --- to produce not only examples of the planetary systems of interest, but also the entire population of possible systems. Toward that end, this section reviews results from population synthesis models.

\subsection{Population Synthesis Method}

Simulating the formation of entire planetary systems end-to-end --- i.e., from tiny dust grains to fully-fledged Gyr-old planetary systems --- for varying initial conditions was first attempted for our own solar system \citep{dole1970}. The more recent discovery of the exoplanet population, which offers a rich data set for statistical comparison, led to renewed interest in formation and evolution models, particularly models offering the possibility of statistically comparing theory and observation. The first modern population synthesis calculation was conducted by \cite{Ida2004}, and a number of groups have developed related models since that time. Most are based on variants of the core accretion paradigm like the Bern Model \citep{alibert2005,mordasini2009a,emsenhuber2020a}, the Lund Model \citep{ndugu2018,ndugu2019}, or the McMaster Model \citep{alessi2018,alessi2020}. 

These different approaches share a common goal --- to distill the results of specialized theoretical models of the many physical processes involved in planet formation (see the previous sections), to combine them into one global computational scheme, and to put this model to the observational test. The initial conditions for these models are varied in a Monte Carlo fashion (typically the disk dust-to-gas ratio, i.e., metallicity; disk initial gas mass; disk lifetime; see for example \citealt{Emsenhuber2020b}). This produces synthetic populations that can be statistically compared with observations, often after applying a synthetic detection bias (e.g., \citealt{Mordasini2009b,Mulders2019}). Discrepancies between synthetic and observed populations reveal shortcomings in our current theoretical understanding and highlight subjects requiring further work. The hypothesis underlying the method is that the observed diversity of planetary systems is the consequence of the  diversity of initial conditions, as provided by the observed properties of protoplanetary disks (\citealt{tobin2020}; see \S \ref{sec:stardisk}). Population synthesis models can make predictions regarding a wide range of planets, including close-in compact systems found with the transit method \citep{Mulders2019,Mishra2021}, planets detected by radial velocities \citep{schlecker2020,burn2021}, cold low-mass planets around M-dwarfs found by microlensing \citep{Suzuki2018}, and distant self-luminous massive planets detected with direct imaging \citep{Vigan2021}. This discussion concentrates on the first class of compact systems (for more general treatments of the population synthesis method, see the reviews of \citealt{Benz2014} and \citealt{Mordasini2018}).

\subsection{Comparison to Observations}
A powerful attribute of population synthesis is the ability to compare the synthetic planet populations to the observed exoplanets. This exercise provides a measure of whether the physical mechanisms included in the population synthesis model (e.g., those discussed in \S3-\S5) were valid choices.  Here we review some early discrepancies in population synthesis, subsequent improvements to the models, and finally compare current results to the observations presented in \S\ref{sec:observations}.

\paragraph{\bf Lessons from Early Discrepancies.}

Several early synthesis models \citep{Ida2008,Mordasini2009b} predicted an absence of close-in, low-mass planets, a prediction that was not borne out by observations from the Kepler mission \citep{Borucki2010}. For example, the models of \citet{Mordasini2009b} predicted a paucity of planets with masses between 2 and 10 $\mearth$ inside of about 0.4 AU. This deficit arose because of two model shortcomings: (1) early models did not include the action of non-isothermal migration (e.g., \citealt{Paardekooper2010}), and (2) these models  assumed only one planetary embryo per disk. The early synthesis models relied instead on the state-of-the-art migration models at that time (isothermal models, \citealt{Tanaka2002}) that predicted fast and always inward-directed migration. This migration scheme had the consequence that most embryos migrated into the star during the planet formation epoch. To counter this effect, it was necessary to include artificial reduction factors in the migration rates in order to reproduce the observed population of cold and temperate giant planets known at that time. One consequence of this ad-hoc solution was that the models produced very few close-in, low-mass planets.

In contrast, modern models include non-isothermal migration schemes  \citep{baruteau2010,masset2010} to (1) better match recent observational data, and (2) provide a physical (rather than ad-hoc) method for handling the complexities of planet migration. This interplay provides an example of how population synthesis and specialized models strengthen one another. In this way, the lessons learned in the comparisons of synthetic and observed exoplanet populations lead to progressively better theoretical understanding of planet formation. Modern models also simulate the concurrent growth of multiple embryos per disk \citep{alibert2013}, which reduces the extent of migration \citep{Emsenhuber2020b}. These current models predict a large population of close-in low-mass planets, similar to observations (e.g., \citealt{Lambrechts2019,Emsenhuber2020b}).

\cm{However, it is clear that modern global models, despite being more complex than early approaches like \citet{Ida2004}, still only represent a simplified picture of the actual planet formation process. In particular, they rely on the existence of specialized models for all relevant physical processes. In the model used here, it is assumed that at the beginning, planetesimals exist everywhere in the disk with a radial distribution inspired by the MMSN. This is obviously a strong assumption. To improve this, a consistent model of the evolution of the solids is needed, going from dust to pebbles to planetesimals to embryos. Such models were only recently constructed \citep[e.g.,][]{Voelkel2020,Coleman2021}. Some of these models predict an early emergence of planetesimals with a power-law-like profile throughout the disk at early times \citep{Lenz2019}, much like the assumptions made here. Other models, in contrast, predict bursts at specific disk locations \citep{Draz2017,Schoonenberg2017}. Such  different distributions would lead to formation pathways differing significantly from the ones found here, and affect the predictions regarding the compact multiplanet systems.}

\paragraph{\bf General Properties of Close-in, Small Planets.}

\begin{figure}[tb]
\centering
\includegraphics[width=0.5\textwidth]{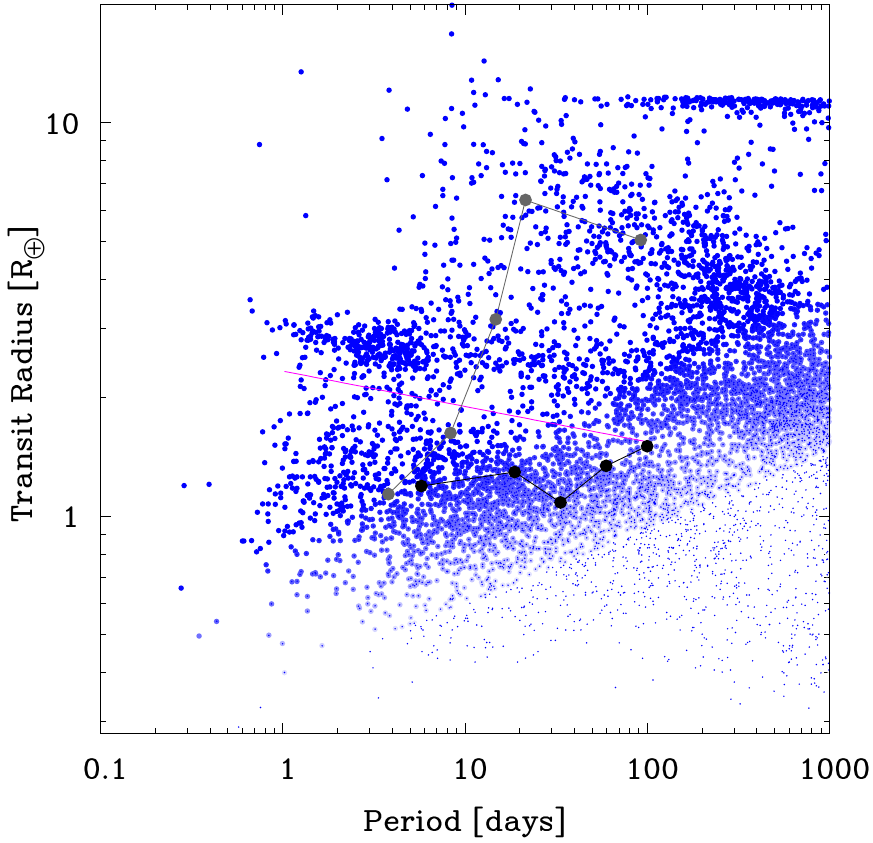}
\caption{Transit radius --- orbital period plane for synthetic planetary systems around 1 $M_\sun$ stars at 5 Gyr (updated from \citealt{Emsenhuber2020b}). Large and tiny blue dots indicate the detectable and the unbiased underlying population, respectively. The magenta line is the observed location of the radius valley \citep{VanEylen2018}. Example systems with similar (black) and dissimilar (grey) radii are shown, including planets inside of 100 days. Comparison with the observed population (Figure \ref{fig:context}) reveals a number of similarities (e.g. rich population of small close-in planets, radius valley, evaporation desert) but also differences (e.g. paucity of hot Jupiters).}
\label{fig:SyntPeriodRadiusandHisto}
\end{figure}

To illustrate the efficacy of state-of-the-art population synthesis models, we consider a simulation of an ensemble of 1000 systems around 1 $M_\odot$ stars at an age of 5 Gyr. Here we use the Generation III Bern Model (see \citealt{emsenhuber2020a} for a complete model description), but utilize an updated \cm{atmospheric evaporation \citep{Kubyshkina2018} and} internal structure  model.\footnote{\cm{The latter now uses the temperature-dependent equation of state of  \citet{Haldemann2020} for water}.} Each planetary system initially contains 100 lunar-mass embryos. Their starting positions are drawn at random from a log-uniform distribution of semimajor axes. \cm{The distributions of the five disk initial conditions (dust-to-gas ratio, disk gas mass, inner and outer radius, and lifetime) are derived from several observational constraints in order to reflect the known characteristics and diversity of actual protoplanetary disks \citep{Santos2005,Tychoniec2018,Venuti2017,Andrews2010, mamajek2009}.  Deriving these distributions of initial conditions from observations is far from trivial (e.g., \citealt{Franceschi2022}), but our disk models and initial conditions reproduce one of the most relevant observational constraints, namely the relation between disk mass and stellar accretion rate \citep{Manara2019}. The most important parameters in the simulation shown here are a turbulent viscosity $\alpha_\nu=2\times 10^{-3}$, a reduction factor of the grain opacity in the protoplanetary atmosphere relative to the ISM of 0.003 \citep{Mordasini2014,Ormel2014}, and a 0.6 km size of the planetesimals. The latter is discussed further below.  We refer the reader to \citet{Emsenhuber2020b} for details.} 

During the initial formation phase, the planets grow by accreting planetesimals and gas, as well as via giant impacts. Gas accretion, radii, and luminosities are calculated by solving the classical 1D spherically symmetric planet interior structure equations \citep{bodenheimer1986,mordasini2012}. \cm{This direct solution of the governing differential equations differs from the approach taken in most other global models which use empirical fits or semi-analytical approximations for the gas accretion rate. Such approximations cannot capture the complex behavior of the gas envelope mass, which depends not only on ambient disk conditions, but also on the formation history of the planet itself including its current and past solid accretion rate. Such approximations can lead to a severe overestimate of gas accretion relative to the approach used here \citep{Alibert2019}.} Orbits evolve via gas-driven orbital migration \citep{Paardekooper2010,dittkrist2014} and N-body interactions, which are modeled explicitly with an N-body integrator \citep{chambers1999}. During the subsequent evolution phase (100 Myr to 5 Gyr), the masses and orbits are fixed except for mass loss due to XUV-driven photoevaporation \citep{Jin2014} and orbital decay due to stellar tides \citep{benitez2011}. However, the planets still evolve thermodynamically through cooling and contraction \citep{mordasini2012}. 


Figure \ref{fig:SyntPeriodRadiusandHisto} provides a view of the synthetic detectable \cm{as well as the} underlying unbiased population. We assigned a detection probability for each planet based on an interpolation of Kepler's sensitivity map (\citealt{Petigura2018}), although we neglected the effect of mutual orbital inclinations. One can identify a number of similarities, but also differences, between the observed and synthetic planets. Most fundamentally, by varying the disk initial conditions over a range likely occurring in nature, the model predicts a large diversity in the synthetic population, similar to the one observed. The synthetic population in particular contains numerous small close-in planets. 

This population of close-in planets is dominated by planets with radii less than about 4 $\rearth$. These synthetic populations are broadly consistent with the observed occurrence distribution of small, close-in planets (\S\ref{sec:observations}). Some finer structure in the period-radius distribution, including the radius valley at about 1.8\,\rearth, is also apparent. It is noteworthy that this structure was predicted in analytical theory and numerical models \citep{Lopez2013,Owen2013} and \cm{independently} in population synthesis \citep{Jin2014}, several years before its discovery \citep{Fulton2017}.

\begin{figure}[tb]
\centering
\includegraphics[width=0.45\textwidth]{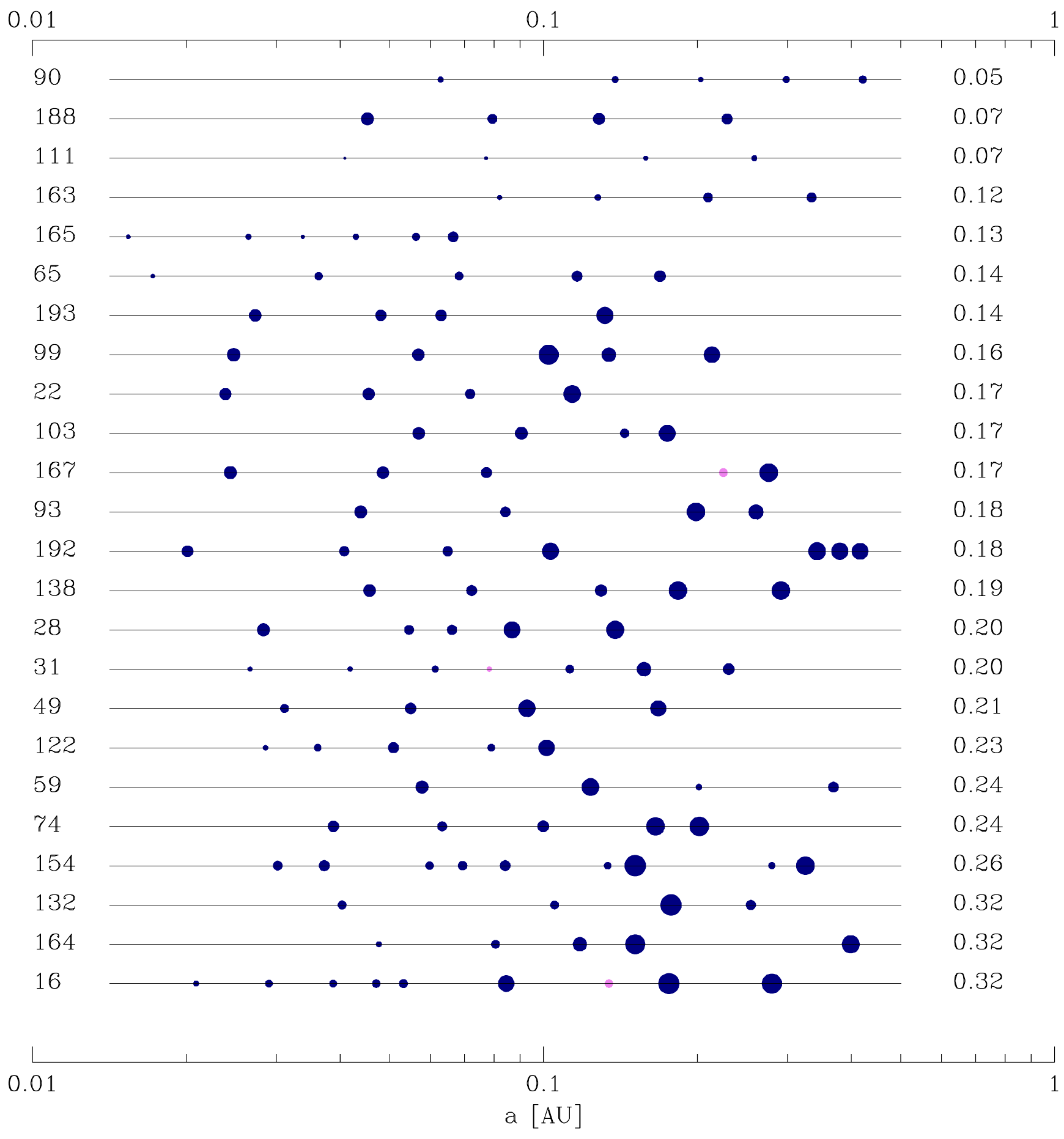}
\caption{The architectures of 24 synthetic systems with 5 or more surviving, detectable planets, analogous to the observed systems in Figure \ref{fig:peas-ranked-sigmaR}.  The planets are assumed to be coplanar. The point size corresponds to the planet radius (scaled logarithmically).  The systems are ranked by their planet radius dispersion $\sigma_{\rm R}$ (displayed on the right). The  label on the left is the system ID. The three violet points are planets inside 100 days that escaped detection because of their small sizes.}
\label{fig:SyntARpeasord}
\end{figure}

\paragraph{\bf Planet-Planet Patterns.}
Figure \ref{fig:SyntARpeasord} provides a synthetic plot analogous to the actual observed architectures of close-in compact multi-planet systems displayed in Figure \ref{fig:peas-ranked-sigmaR} by showing the planetary radii and orbital spacings. Synthetic systems at 5 Gyr with 5 or more planets and with orbital periods less than 100 days are shown. We adopted the same parameterized model of the Kepler detection efficiency as \citet{Petigura2018}, applying the bias to each synthetic planet individually to eliminate small planets. The 100 day cut roughly reflects the period range of most actual observed systems in Figure \ref{fig:peas-ranked-sigmaR}. For simplicity, we have assumed the planets are all transiting; a more realistic treatment that includes orbital inclination dispersion \citep{Mulders2019,Mishra2021} might reduce the apparent planet multiplicities in some of these synthetic systems. 

Many synthetic systems show intra-system regularities in their radii and orbits, similar to the actual systems in Figure \ref{fig:peas-ranked-sigmaR}. There is also a trend of increasing planet size  with orbital distance.  The fractional dispersion of the radii $\sigma_{R}$ given by equation (\ref{eqn:size-disp}) has been used to order the systems. The 24 examples in the plot  were taken out of the first 200 synthetic systems, covering a range in $\sigma_{R}$ from 0.05 to 0.32 dex (similar to the observed range). Systems 90 and 164 have the smallest and largest $\sigma_{R}$ and are the individual systems shown in Figs.
\ref{fig:SyntSim090} and \ref{fig:SyntSim164} below.

\begin{figure}[tb]
\centering
\includegraphics[width=0.45\textwidth]{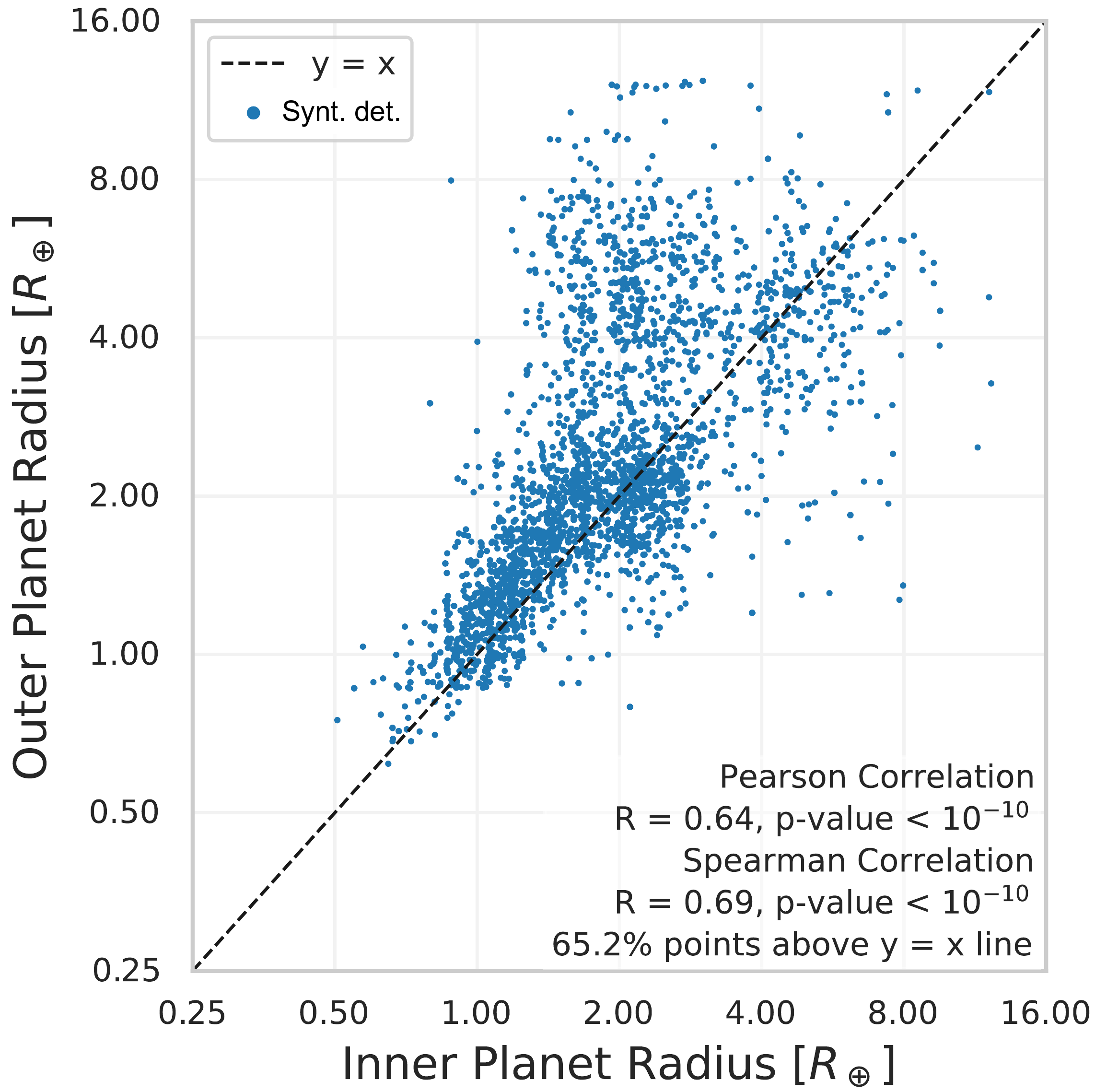}
\caption{Reproduced with permission from \citet{Mishra2021}: Peas-in-a-pod effect in the synthetic population. The plot shows the sizes of adjacent planets in the detectable synthetic sub-population, exhibiting correlations between the sizes of neighbouring planets. The degree of correlation, as quantified by the Pearson $R$ and the fraction of outer planets that are larger than the inner, are in excellent agreement with the observed values of \citet{Weiss2018a}.} 
\label{fig:SyntRadiusPeas}
\end{figure}

The intra-system uniformity, and its statistical significance, is more apparent when we consider just one pairwise property at a time. Figure \ref{fig:SyntRadiusPeas}, reproduced from \citet{Mishra2021}, shows the radii of adjacent, detectable planets in the 1000 synthetic planetary systems around 1 $M_\odot$ stars. The plot only includes detectable synthetic planets that were found by simulating the geometrical transit constraints and the detection biases of the  Kepler transit survey (see \citealt{Mishra2021} for a detailed discussion; \cm{note that the observational data in Fig. \ref{fig:random_draws} includes additional cuts on the planets, explaining the difference with Fig. \ref{fig:SyntRadiusPeas}.}). The results show a strong statistical correlation between the sizes of adjacent (detected) planets in the synthetic population (Pearson $R$ = 0.64); moreover, in 65\% of the synthetic adjacent pairs, the outer planet is larger than the inner. These findings are in excellent agreement with observations (\S\ref{sec: uniform sizes and masses}) as found by \citet{Weiss2018a} and similar to the results of \citet{Ciardi2013}. \cm{Two effects contribute to the trend of increasing radius: for systems where the final masses are mainly given by a giant impact phase (like in Fig. \ref{fig:SyntSim090}), the increase of the available solid mass ($\propto\Sigma_{\rm pl}(a) \times a^2$) with distance for the assumed MMSN-like $\Sigma_{\rm pl}(a)$ clearly plays a role. On the other hand, for systems where orbital migration is important (like in Fig. \ref{fig:SyntSim164}), atmospheric evaporation seems to be produce the trend. As discussed in \citet{Mishra2021}, detection biases only play a small role --- the trend also exists in a similar way in the underlying unbiased synthetic population.} 

\begin{figure*}
\centering
\includegraphics[width=\textwidth]{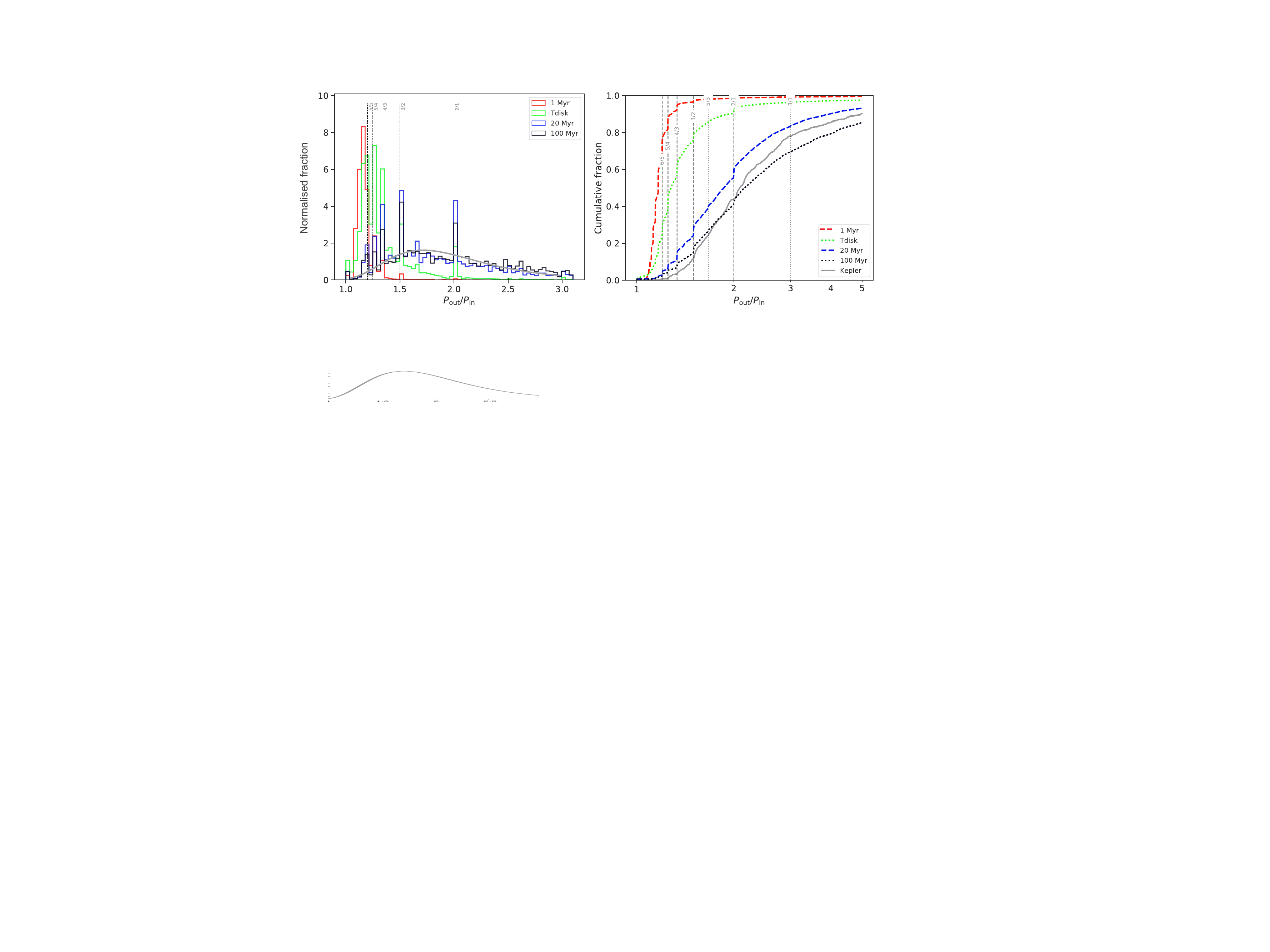}
\caption{Synthetic period ratio $\mathcal{P}$ distribution for comparison with the observed distribution shown in Figure \ref{fig:period ratio distribution}. The left and right panels show the histogram and cumulative distribution of the period ratios of synthetic planets with $P<400$ days. Different moments in time are shown: at 1 Myr, at the time of dispersal of the gas disk (on average 3 Myr), at 20 Myr, and at 100 Myr. When the planets are still embedded in the gas disk and thus experiencing eccentricity damping, they are in extremely compact configurations. Between the moment of gas disk dispersal and 20 Myr, instabilities and collisions increase  $\mathcal{P}$. Between 20 and 100 Myr, the distribution shifts to even larger values and the number of planets in MMR decreases. At 100 Myr, the distribution has an overall width similar to the observed one (the grey line on the left shows the log-normal fit to the observations from equation \ref{eq:pratiodistfit}), but a higher fraction of planets are still in MMRs. }
\label{fig:SyntPeriodRatios}
\end{figure*} 

The regularity in orbital spacing (\S\ref{sec:spacing}) is also recovered in the synthetic population \citep{Mishra2021}. The correlation in the synthetic population (with Pearson $R$ = 0.25) is somewhat weaker than in the observed population ($R$ = 0.46). Finally, the synthetic population reproduces the positive correlation between average size of planetary pairs with their spacing (\S 2.3). The degree of the correlation is in good agreement with observations. This correlation also arises, with an even slightly stronger correlation coefficient, when using the average mass instead of the radius. 



\paragraph{\bf Period Ratio Distribution.}
Figure \ref{fig:SyntPeriodRatios} shows the synthetic period ratio distribution, both in the form of a histogram and a cumulative distribution. The synthetic population offers the interesting opportunity to study the temporal evolution of the $\mathcal{P}$ distribution. 

The figure shows the period ratio distribution at a variety of times: at 1 Myr (when  all synthetic disks still contain gas), at the time of the gas disk dispersal (which differs from system to system, but is on average about 3 Myr; see \citealt{emsenhuber2020a}), at 20 Myr, and at 100 Myr, which is the maximum time during  which the orbits were integrated. One sees a consistent evolution from (extremely) compact systems during the presence of the gas disk to a distribution with an overall width comparable to the observed distribution at 100 Myr (equation \ref{eq:pratiodistfit}). At 1 Myr, the majority of period radios have $\mathcal{P}<1.2$. At this early time, the gas disk damps the eccentricities and stabilizes the orbits. Such tight packing is expected from convergent migration (discussed below). It is also expected from the oligarchic planetesimal growth phase, leading to embryos with relative spacing of about 10 mutual Hill radii (\citealt{Kokubo1998}, equation \ref{eqn:Hill}). In the simulation shown here, 100 lunar-mass embryos were initially placed in the disk. This configuration corresponds to a spacing of about 28 mutual Hill radii, with a period ratio of 1.09. At the time of disk dispersal, when damping vanishes, the frequency of pairs with $\mathcal{P} < 1.2$ strongly decreases. Instead, MMRs like 6/4, 5/4, and 4/3 are now strongly populated. Pairs with $\mathcal{P} > 2$ are still largely absent. A large change occurs  between disk dispersal and 20 Myr when many resonances break \citep{Ida2010, Izidoro2019}. Between 20 and 100 Myr, the orbits continue to evolve in a gas free environment, and the fraction of planets in and near the resonances, especially in the tighter ones, decreases even more. The fraction of resonant systems, however, still remains larger than observed. For further discussion of how systems can move in and out of resonance, see Section 4.3 (and references therein).  

One might wonder if the peas-in-a-pod pattern has been ``hard-coded'' into the initial conditions of the population synthesis model by the identical initial masses. However, the final planet masses are 2 to 3 orders of magnitude larger than the initial masses (corresponding to 7-10 mass doublings), so that the planets have ``forgotten'' their initial conditions. To test whether the similar masses (and sizes) of the planets were indeed dependent on the initial conditions, we performed an analogous set of simulations, but started the embryos with a wide range of even lower starting masses. The final correlations in the planet masses (and sizes) did not depend sensitively on the {\color{red}initial mass} distribution of embryos, {\color{red}demonstrating that the peas-in-a-pod pattern does not require fine tuning}. Similar considerations hold for the other correlations \citep{Mishra2021}.

Thus, the various aspects of the peas-in-a-pod pattern are not simply a hard-coded result in the population syntheses. Rather, they are a natural consequence of the existence of characteristic mass scales arising from the different governing physical processes and the N-body interactions. Under the assumption that the underlying theoretical formation and evolution model captures processes occurring in nature, these results -- when taken together -- give further support to an astrophysical origin of the peas-in-a-pod pattern, rather than an observational bias. 

\cm{\textbf{Other Statistical Properties.} Besides the various aspects of the peas-in-a-pod pattern and the period ratios, it is interesting to compare other statistical properties of the biased synthetic population with the observed characteristics of compact multis summarized in Figure \ref{fig:summary-of-observations}. Including synthetic planets with $R\leq 4 R_\oplus$ (the upper limit used in the compact multis definition) one finds the following results: a mean eccentricity of 0.14 and 0.06 for single and multi transiting systems, respectively (observed: about 0.15-0.25 and 0.02-0.05, respectively). The inclination distribution is also in good agreement with observations, as already found in \citet{Mulders2019}. The mean mutual inclinations decrease with multiplicity $n$ and are approximately given by $5.7^{\circ} n^{-0.9}$ in the biased synthetic population. No dichotomy in planet multiplicity is seen. The properties of the radius valley and the radius histogram are also recovered. Considering that the simulations start from 100 lunar-mass embryos per disk that are then followed in their formation and evolution over 5 Gyr with a model that includes (but simplifies) many physical processes, this is a good level of agreement. It is important to note that these results can only be obtained if there is a sufficiently high number ($\sim 100$) of initial embryos present in each disk \citep{Mulders2019}. This means that a growth mode via the interactions (collision, scattering) of many protoplanetary bodies is necessary to reproduce the observations. } 

\cm{There are also a number of differences: while the general shape of the period distribution of both small and large synthetic planets is in good agreement with observations (Fig. \ref{fig:period ratio distribution}), for the small planets, the synthetic distribution becomes flat (log-uniform) at about a period of 5 instead of 10 days as observed. The decrease at this larger distance could be caused by an ionization transition with a corresponding jump in the disk viscosity and surface density, thus halting migrating planets at 10 day periods \citep{Flock2019}. This effect is currently neglected in the Bern model. The synthetic multiplicity distribution (number of systems with $N$ transiting planets) scales in the synthesis approximately $\propto\exp(-N/0.65)$. As already found in \citet{Mulders2019, Mishra2021}, there are more high-multiplicity synthetic system than observed, which is partially linked to the previous point. Together with the higher number of planets close to or in MMR in the synthetic population relative to observations (Fig. \ref{fig:SyntPeriodRatios}), this indicates that more gravitational interactions between planets would be needed to bring theory and observation into closer agreement. A longer N-body integration time (here 100 Myr) or external perturbers \citep[e.g.,][]{Malmberg2009} could also be relevant. }

\subsection{Planet Properties over Time}

\begin{figure*}
\centering
\includegraphics[width=1\textwidth]{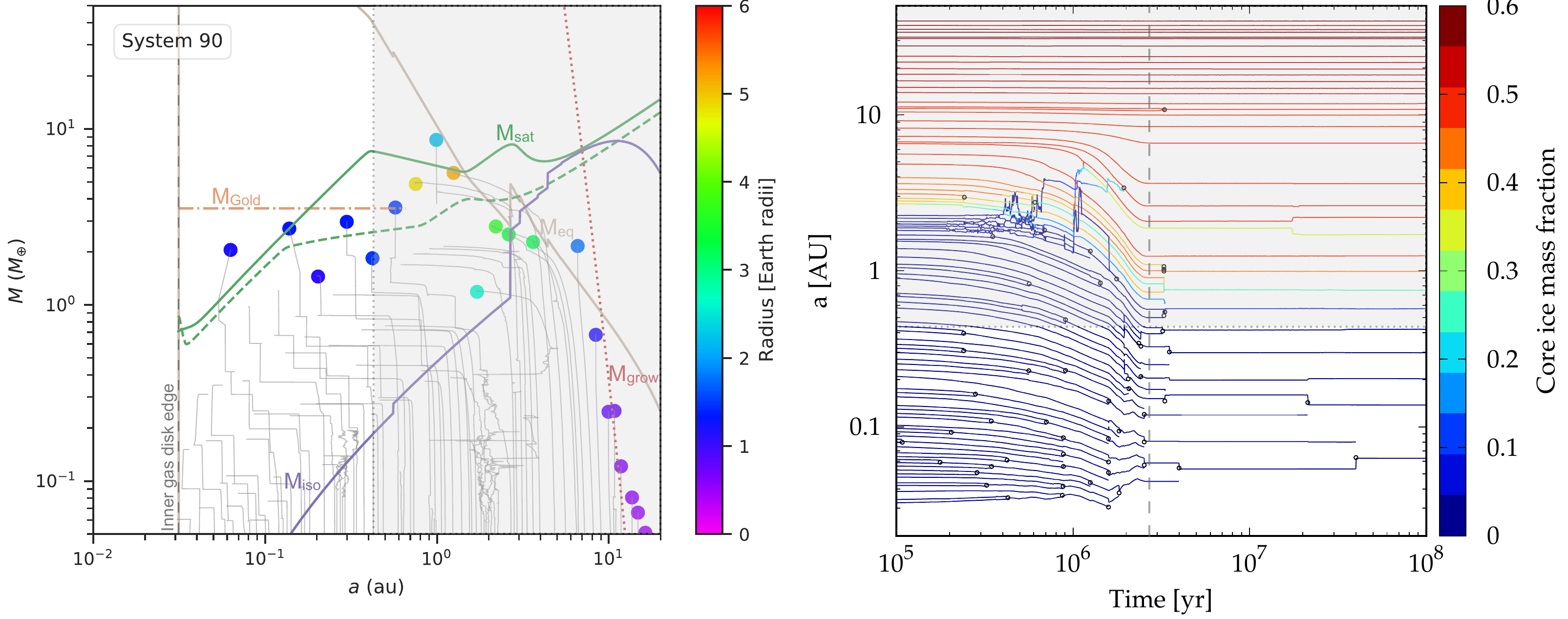}
\caption{Synthetic formation tracks of a system with similar-sized planets (System 90 with smallest $\sigma_{\rm R}=0.05$ dex in Figure \ref{fig:SyntARpeasord}). The left panel shows in the mass-distance plane the tracks (gray lines) and final positions (colored points, coding the radius at 5 Gyr). Lines ending without a point are planets that were accreted by another more massive body. Colored lines show analytical estimates of the final masses:  Mean Goldreich mass $M_{\rm gold}$ resulting from giant impacts (brown dashed-dotted, horizontal); planetesimal isolation mass $M_{\rm iso}$ (violet solid, increasing with distance); mass limited by the growth timescale $M_{\rm growth}$ (red dotted, decreasing with distance);  equality mass $M_{\rm eq}$ where planetesimal accretion and migration timescale are equal (gray solid, decreasing with distance); saturation mass $M_{\rm sat}$ where the corotation torque saturates evaluated at $t=0$ and 1 Myr (green solid and dashed). The right panel shows the planets' semimajor axes as a function of time, with color indicating the core ice mass fraction. Small black circles indicate giant impacts. The dashed vertical line indicates the moment of gas disk dispersal. The zone with orbital periods greater than 100 days is grayed out in both panels.}
\label{fig:SyntSim090}
\end{figure*}

How do the systems shown in Figure \ref{fig:SyntRadiusPeas} come into existence?  As hinted in Figure \ref{fig:SyntPeriodRatios}, the ability of population synthesis to track simulated embryos over time can reveal the conditions under which patterns emerge. Figures \ref{fig:SyntSim090} and \ref{fig:SyntSim164} illustrate the formation pathways of close-in multi-planet systems as seen in the aforementioned population synthesis simulations (from \citealt{Emsenhuber2020b}). To understand these outcomes, we compare the simulation results to different analytical mass scales, some of which are from \S\ref{sec:planet-formation-theory} and some of which --- for the later stages --- are introduced here.

Figure \ref{fig:SyntSim090} shows the emergence of the planetary system with the smallest dispersion in radii among the synthetic systems shown in Figure \ref{fig:SyntARpeasord}. The left panel shows the formation tracks of the (initially 100) planetary embryos in the mass-distance plane, whereas the right panel shows the evolution of the corresponding semi-major axes over time. \cm{The initial conditions are a  gas disk mass of 0.02 $M_\odot$ and a total planetesimal mass of 101 $M_\oplus$. This value is slightly below the mean (108 $M_\oplus$) of the distribution which covers a range from about 10 to 1000 $M_\oplus$. The disk lifetime is 2.8 Myr which is a bit less than the average lifetime of the synthetic disks (about 3.4 Myr).}   
These panels illustrate how planets migrate, are captured in resonances, and undergo collisions leading to larger orbital separations. The majority of the embryo collisions occur at around the time of gas disk dissipation. 

In the left panel, the tracks generally go upward and inward, corresponding to growth and inward orbital migration, respectively. Horizontal sections show phases were planets migrated inward without planetesimal accretion (gas accretion is inefficient for low core masses). Vertical sections correspond to giant impacts. Gravitational interactions lead some tracks to fluctuate stochastically. An important quantity when interpreting the tracks in the $a-M$ plane is the track slope, $d \ln{M}/d \ln{a}$, which is the same as the ratio $\tau_{\rm mig}/\tau_{\rm grow}$ of the migration to the mass growth timescale. An important feature is that the Type I orbital migration timescale decreases with increasing planet mass (e.g., \citealt{Ward1997}), whereas the  oligarchic planetesimal accretion timescale increases with mass \citep{thommes2003}. This difference means that (in general) accretion will dominate first, followed by migration. However, additional effects like giant impacts, resonant capture, and/or migration traps complicate this picture. 

Figure \ref{fig:SyntSim090} compares the numerical results to several analytical mass scales, as shown by the curves. They are computed numerically using the data provided by the simulation's planetesimal and gas disk model. Below we summarize the origin and importance of these various mass scales, noting that several of them build on theoretical principles discussed in \S\ref{sec:planet-formation-theory}. Specifically, the planetesimal and pebble isolation masses introduced in \S\ref{sec:planetesimals to planets} consider the growth of isolated and stationary embryos from smaller bodies, whereas embryo interactions (e.g. giant impacts) and orbital migration add further complexity to the picture.

\textbf{Planetesimal Isolation Mass.} The violet solid line shows the planetesimal isolation mass $M_{\rm iso}$ (\S\ref{sec:planetesimals to planets}, equation \ref{eqn:planetesimal iso mass}). For the assumed surface density profile,
$\Sigma_{\rm pl} \propto a^{-3/2}$ (akin to the MMSN), the isolation mass increases with distance. Jumps occur at the different condensation lines, most importantly the water ice line slightly inside of 3 AU. As discussed in \S\ref{sec:planetesimals to planets}, $M_{\rm iso}$ is small in the inner system. The final masses in the simulation are more than one order of magnitude larger than $M_{\rm iso}$ because of late-time giant impacts and inward transport of matter caused by Type I migration. The isolation mass is, however, still important, as shown by the gray tracks: in the inner system, $M_{\rm iso}$ marks the transition from a regime dominated by solid growth via oligarchic planetesimal accretion to one of inward migration and growth by giant impacts. At small orbital distances, the planetesimal accretion timescale is short and planet masses are small. As a result, the first planetesimal growth phase occurs nearly in situ, leading to virtually vertical tracks.

\textbf{Equality Mass.} The gray solid line shows the mass where the planetesimal accretion and the orbital migration timescales become equal, the equality mass $M_{\rm eq}$. Once a planet crosses this line, its track will generally start to bend inward, and migration instead of planetesimal accretion will start to dominate. This occurs for planets outside of the ice line starting between $\sim3-7$ AU, for which $M_{\rm eq}<M_{\rm iso}$ and $M_{\rm eq}\approx 5 \mearth$. Equating the planetesimal accretion timescale in the oligarchic regime \citep{thommes2003} with the migration timescale from equation (\ref{eqn:mig}), one finds 
\begin{equation}
    M_{\rm eq}=k \frac{(\Delta \, C_{\rm D} \Sigma_{\rm pl})^{3/10} f_{\rm dg}^{3/4} M_\star^{5/4} (h/a)^{6/5}}{(a \Upsilon)^{3/4} r_{\rm pl}^{3/10} \rho^{11/20} },
\end{equation}
where $k=1.16$ is a numerical constant, $\Delta\approx 10$ the separation in Hill spheres, $C_{\rm D}\approx 1$ the drag coefficient, $f_{\rm dg}$ the dust-to-gas ratio in the disk (about \cm{0.015} for solar metallicity), $r_{\rm pl}$ the planetesimal radius, and $\rho$ the material density of the planetesimals and protoplanets. For the accretion timescale, we have assumed that planetesimal eccentricities are in an equilibrium between viscous stirring (by the protoplanet) and damping (by gas drag). This expression also assumes that the planetesimal mass reservoir is large enough to allow growth to  $M_{\rm eq}$, which, for in situ growth, only holds if  $M_{\rm eq}\leq M_{\rm iso}$. The opposite dependencies of $M_{\rm eq}$ and $M_{\rm iso}$ on orbital distance (the first decreasing, the latter increasing with distance) leads to a critical orbital distance ($\sim3$ AU in the current example) at which the nature of the growth tracks changes. Inside the critical distance, near in situ planetesimal growth is followed by giant impacts; outside, planetesimal growth is followed by orbital migration. Since the masses of the planets forming outside of the ice line are superior to those inside, they migrate inward faster and capture the many interior lower-mass protoplanets into large resonant convoys. This trend can be seen by the many parallel tracks in the right panel between $\sim 1-3$ Myr. In the context of pebble accretion models, the \cm{so-called turn-off mass \citep{Johansen2019} has the equivalent role as $M_{\rm eq}$ and is found by equating the pebble accretion timescale to the migration timescale. }

 
\textbf{Saturation Mass.} Another mass scale that can set the masses of migrating planets is the saturation mass $M_{\rm sat}$. It is shown by green solid and dashed lines, which are calculated with the disk properties at $t=0$ and 1 Myr, respectively. Type I orbital migration (\S\ref{sec:migrate}) comes in different sub-regimes, dictated by the disk thermodynamics \citep{paardekooper2008,kley2009}. Some sub-regimes lead to the existence of zero-torque locations in the disk that act as traps for migrating planets \citep{Lyra2010,Hasegawa2013} if they are in the appropriate mass regime (about $1-10 \mearth$). The upper mass limit is given by the saturation of the positive corotation torque and can be found by equating the viscous timescale across the corotation region \citep{hellary2012} with the libration timescale (e.g. \citealt{dittkrist2014}), 
\begin{equation}
    M_{\rm sat}=\left(\frac{8 \pi \alpha_{\rm \nu}}{3}\right)^{2/3} \left(\frac{h}{a}\right)^{7/3} \frac{\gamma^{1/2} M_\star}{k_{\rm xs}^2},
\end{equation}
where $\alpha_{\rm \nu}$ is the disk turbulence viscosity parameter \citep{Shakura1973} and $k_{\rm xs}$=1.16 is a numerical constant. Once the mass of a captured planet exceeds this limit, it will start to migrate inward, similarly as for $M_{\rm eq}$. For evaluating $M_{\rm eq}$ and $M_{\rm sat}$, the relevant orbital distance $a \sim3-7$ AU, where the tracks bend inward. At these locations, the grey and the green lines provide reasonable analytical estimates of the results seen numerically. $M_{\rm sat}$ is shown at two moments in time ($t=0$ and 1 Myr). It decreases over time because $h$ decreases as less viscous heating occurs with a smaller disk surface density. As can be seen in the right panel, the start of migration lies within this time interval.   
 
\textbf{Goldreich Mass.} The horizontal brown dashed-dotted line is the Goldreich mass $M_{\rm gold}$. In the post-oligarchic phase (mainly after gas disk dissipation),  planets can grow further by giant impacts. The resulting mass can be estimated \citep{Goldreich2004} assuming that the protoplanets increase their random velocities to a point where they are comparable to their escape velocities $v_{\rm esc}$. This balance holds if eccentricity damping (dynamical friction) by residual smaller bodies is weak\footnote{In the simulations, dynamical friction is neglected. The opposite effect, the excitation of the planetesimal random velocities by the protoplanets, is in contrast included \citep{Fortier2013}.}. Using the associated eccentricities $e\sim v_{\rm esc}/v_{\rm Kep}$ and width of the feeding zone of $2 a e$, one can derive a ``Goldreich mass'' 
\begin{equation}\label{eq:Mgold}
    M_{\rm gold}=16\left(\frac{2 \pi^7 \rho}{3}\right)^{1/4} \frac{(\Sigma_{\rm pl})^{3/2} a^{15/4}}{M_\star^{3/4}}.
\end{equation}
For the MMSN, $M_{\rm gold}$ yields inner solar system masses that are comparable to those of Earth and Venus, which is about one order of magnitude larger than $M_{\rm iso}$. 

\textbf{Growth Mass.} Finally, the red dotted line shows $M_{\rm grow}$. At even larger initial orbital distances $a\gtrsim7$ AU, the planetesimal accretion timescale is so slow that planet masses are limited by the growth time itself, rather than the availability of building blocks or orbital migration. Here, we use the the aforementioned oligarchic growth rate to estimate the mass to which protoplanets can grow by the moment of disk dispersal. After disk dissipation, the planetesimal random velocities increase, and the accretion rate becomes even smaller, so that this estimate provides a fair match to the numerically obtained masses at 100 Myr. Over Gyr timescales, however, higher masses might still be reached, limited eventually by the ejection of planetesimals \citep{Ida2004}. This latter effect is neglected in the model.

Figure \ref{fig:SyntSim090} shows that the innermost 6 planets, which formed inside of the water ice line, obtain their final masses through a series of giant impacts. We thus expect that their final masses should be comparable to $M_{\rm gold}$. However, equation (\ref{eq:Mgold}), when evaluated at the final planet positions ($\lesssim0.6$ AU) and the initial $\Sigma_{\rm pl}$ at these final positions, predicts masses that are smaller than the ones in the numerical simulation. As shown by the simulations, this discrepancy arises because inward migration leads to net inward transport of the building blocks. For the planets inside the ice line, the orbital distances shrink by about a factor 2-4 when comparing the embryo starting positions to the final planet location. Furthermore, growth via giant impacts is inherently a multi-body process. To take these two points into account, when calculating $M_{\rm gold}$ for the figure, we evaluate it at all starting positions of the embryos inside of the ice line and take the initial $\Sigma_{\rm pl}$ there, and then take the mean of these masses. Calculated in this way, $M_{\rm gold}$ seems to provide a useful estimate for the masses seen in the simulations. 

In the simulation, the masses have a tendency to increase slightly with orbital distance (albeit with scatter). This trend is \cm{as mentioned} an imprint of the initial planetesimal surface density $\Sigma_{\rm pl} \propto a^{-3/2}$ where more mass ($\propto \Sigma_{\rm pl} a^2$) is available per annulus with increasing distance. This pattern is typically seen in the synthetic systems where the giant impact phase sets the masses \citep{Mishra2021}.

The right panel of Figure \ref{fig:SyntSim090} illustrates the consequences of orbital migration. The disk torques on the planets are given in $\S\ref{sec:planet-formation-theory}$ for individual planets. However, the actual migration rates might differ substantially from what is estimated for single planets. The reason is that the planets are not migrating alone. In particular, they can get captured into MMRs \citep{Cresswell2008, alibert2013}, and the resulting torque is distributed among all planets in the resonant convoy. Some planets in the convoy might experience a negative torque, acting against the general inward migration. Examples include planets at the inner edge of the gas disk \citep{Masset2006} or in parts of the disk with positive torques \citep{Baruteau2016}. The resulting migration rates can then be much smaller than the single planet estimates \citep{Emsenhuber2020b}. Instead of fast inward migration of individual planets, the resulting pattern of the collective evolution of the orbits is rather similar to an accordion (right panel of Figure \ref{fig:SyntSim090}, before disk dispersal): inside of about 10 AU, the orbits are increasingly pushed together. When two orbits get sufficiently close, giant impacts occur, which re-establishes larger orbital spacing, and the process restarts. 

The final synthetic system contains 5 low-mass ($1.5-3 \mearth$) planets with $P<100$ days and radii in the range $1.1 - 1.5 \rearth$. Their bulk composition is similar to Earth (silicate-iron planets) without ices. The outermost planet (of the five) still contains a small amount (about $10^{-3} \mearth$) of primordial H/He. This gas increases its radius by about 30\%. The other planets either lost their primordial envelope because of XUV-driven escape, or because of impact stripping. The five planets have inclinations between 1$^\circ$ and 3$^\circ$, eccentricities of about 0.05, and are not in resonances.\footnote{Specifically, the period rates are 3.26, 1.78, 1.78 and 1.67, so the last pair is close to the 5/3 MMR.} These properties are in good agreement with those  typically observed for compact multis (\S\ref{sec:observations}).

\subsection{Nuances of Planetesimal Accretion}
\cm{Some models coupling N-body interactions with orbital migration have found fast inward migration, such that most protoplanets migrate all the way to the inner disk edge \citep[e.g.,][]{Matsumura2017,Izidoro2017}. These models are directly or indirectly based on pebble accretion, in contrast to the planetesimal case considered here. This introduces important differences in how orbital migration couples with solid accretion. Generally speaking, pebble accretion allows planets to grow massive on a short timescale over a wide range of orbital distances (see Fig. \ref{fig:pebble}). Early on, the gas surface density in the disk is still high. The Type I migration timescale is shorter for more massive planets and for higher gas surface density. This means that orbital migration will be fast. In planetesimals-based models like the one here, the situation is different, at least for typical initial conditions (see Fig. \ref{fig:SyntSim090}). Two factors work together to reduce migration rates: in the inner disk, oligarchic growth is fast, but the masses that can be reached are low because of the limited local planetesimal reservoir, too low for fast migration. Damping of eccentricity by nebular gas makes growth via giant impacts inefficient as a channel extending the local reservoir (although not impossible; see Fig. \ref{fig:SyntSim090}). In the outer disk, beyond the ice line, the local planetesimal reservoir is larger, but the oligarchic growth rate is slower. It can take $\sim$Myr until masses are reached where migration becomes significant. At this time, the gas surface density has decreased, and less time is left to migrate before complete disk dissipation. The combination of these effects thus reduces the efficacy of migration (and also gas accretion).}

\cm{Because of significant gaps in knowledge about how planetesimals assemble (see \ref{sec:dust to plantesimals}), the characteristic size of the planetesimals is poorly constrained. In the model used here, the size was assumed to be $\sim1$ km. In the literature, arguments can be found both in favor of both small \citep[e.g.,][]{Krivov2021}, and much larger ($\sim100$ km) sizes \citep[e.g.,][]{morby2009}. It is interesting to note that of the five mass scales discussed here, only the equality mass and the growth mass explicitly depend on planetesimal size, but the latter is not relevant in the inner system. The planetesimal isolation mass and the Goldreich mass depend on the availability of building blocks (assuming that growth is fast enough that the mass scales can be reached).  Planetesimal accretion speeds up with decreasing planetesimal size, because smaller planetesimals are more efficiently captured by the gas in the envelope of the protoplanets \citep{Podolak1988}, and because the disk gas more efficiently damps their random velocities, which in turn increases the Safronov factor (Eq. \ref{eqn:Mdot}). In the inner system, the growth timescale is sufficiently short that planetesimals as large as 100 km can be accreted \citep{Voelkel2020}. This indicates that the formation of close-in low-mass planets is not strongly affected by the assumed planetesimal size. Instead, the building block availability is what matters.  In contrast, the formation of protoplanets in the outer system depends sensitively on planetesimal size.  Accretion needs to outperform orbital migration \citep{Johansen2019}, which is relevant at large orbital distances because the planetesimal accretion timescale scales with the orbital timescale (Eq. \ref{eqn:Mdot}). For protoplanets to form by planetesimal accretion beyond the ice line, it is necessary to assume smaller km-sized planetesimals and/or to reduce orbital migration (e.g., with migration traps, \citealt{Coleman2016b,Cridland2019}). Pebble accretion, on the other hand, represents an important alternative explanation for the emergence of protoplanets in the outer disk because of the rapidity of mass growth in the 2D regime. In the inner disk, in the 3D regime, it is much less efficient (Sect. \ref{sec:planetesimals to planets}).} 

\subsection{Origins of Intra-System Uniformity and Diversity.}

\begin{figure}[h]
\centering
\includegraphics[width=0.5\textwidth]{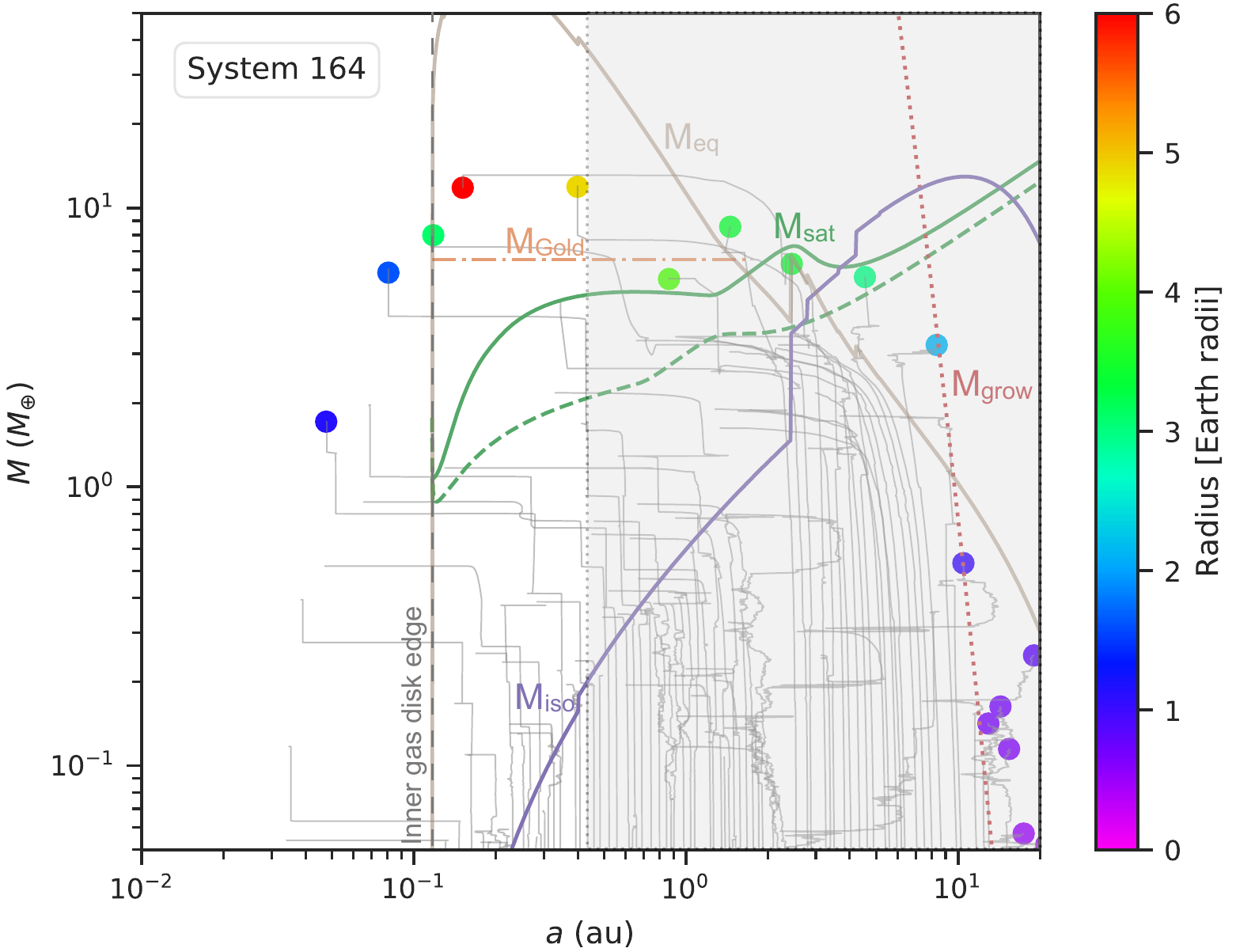}
\caption{Formation tracks of a synthetic system with planets of dissimilar sizes (System 164 with a $\sigma_{\rm R}=0.32$ dex in Figure \ref{fig:SyntARpeasord}). The plot is analogous to Figure \ref{fig:SyntSim090}, left panel. The tracks share many similarities with those in Figure \ref{fig:SyntSim090}. However, because of the slightly higher initial solid content, the embryos starting beyond the ice line in this case grow to somewhat higher masses during the lifetime of the disk. Three volatile-rich planets therefore migrate into the inner system ($P<100$ d). The inner system  additionally contains two lower mass, silicate-iron planets that formed through a series of giant impacts inside of the water ice line. This system thus contains close-in planets with distinct formation modes and associated mass scales and compositions, leading to radius diversity within one system.  }
\label{fig:SyntSim164}
\end{figure}

\paragraph{\textbf{Uniformity.}} The Goldreich mass $M_{\rm gold}$, the equality mass $M_{\rm eq}$, and the saturation mass $M_{\rm sat}$ are all candidates for setting the final mass scale of close-in low-mass planets in a planetary system. They all have typical values between 3 and 15 $\mearth$, compatible with the observed mass range. To understand the origins of uniformity, it is useful to consider the separate cases of systems with smaller versus larger initial solid contents. 

In systems with a lower initial content of planetesimals, $M_{\rm gold}$ is most relevant. The protoplanets outside of the ice line grow quite slowly, so that the extent of their inward orbital migration during the lifetime of the gas disk remains limited. Recall that lower planet masses correspond to longer migration times. As a result, the planets do not migrate all the way to the inner edge of the gas disk. For example, this applies to the system shown in Figure \ref{fig:SyntSim090}, where the ice-rich planets stay outside of 0.7 AU, in a zone not usually probed with transits. The planets inside of the ice line then grow by giant impacts, similarly to the case for the terrestrial planets in the solar system. One important difference, however, is that orbital migration leads to an inward shift of the building blocks, {\color{red}resolving the discrepancy between the initial disk profile and the mass profile of the peas-in-a-pod architecture noted in \S3}. This common growth mode (giant impacts) and the similar composition of all planets lead to uniformity in this first type of giant impact-dominated formation mode with radii typically below $\sim 1.7 \rearth$.  

A second distinct formation mode occurs in systems with a longer disk lifetime and/or a higher solid content, caused either by a higher dust-to-gas ratio (metallicity) or a higher initial disk gas mass. Here, the protoplanets outside of the ice line become more massive at  earlier times and readily migrate to the inner edge of the gas disk. As they migrate, these planets destroy the inner rocky oligarchs by accreting them or by pushing them into the star via resonant migration. In this case, the mass scale is set by $M_{\rm eq}$ or $M_{\rm sat}$, and is again quite uniform among all the close-in planets in a given system. These planets are typically a bit more massive and contain volatiles (ices and H/He) which they can retain over Gyr timescales due to their higher masses \citep{Owen2019}. This process also leads to systems with similar radii, but now greater than $\sim1.7 \ \rearth$. Examples of this formation channel are the Systems 188 or 22 in Figure \ref{fig:SyntARpeasord}. 

To summarize, when all observed planets have their masses set by the same mass scale, planets with similar masses and radii arise, leading to the observed intra-system similarity.  In other words, although a number of physical mechanisms can dominate the planet formation process in different settings, uniformity can be achieved when all of the inner planets are formed through {\it the same} mechanism. Moreover, mass scales of order $\sim3-10M_\oplus$ arise from several different mechanisms, so that many systems can produce planets in this mass range. 

\paragraph{\textbf{Diversity.}}

In contrast with intra-system similarity, diversity of radii within one system may result from several evolutionary processes, as described below. 

\textbf{(1) Diversity of Formation Channels.} One possibility is that the masses of the planets were not all governed by the same physical process, i.e. are not all set by the same mass scales. This effect is shown in Figure \ref{fig:SyntSim164}, corresponding to System 164 from Figure \ref{fig:SyntARpeasord}. \cm{The initial conditions are here a gas disk mass of 0.02 $M_\odot$ and a total planetesimal mass of 128 $M_\oplus$. The disk lifetime is 2.82 Myr.} It is the system with the highest $\sigma_{\rm R}$ of 0.32 dex. The $a-M$ tracks in Figure \ref{fig:SyntSim164} share many similarities with the one in the uniform System 90 (Figure \ref{fig:SyntSim090}). However, there is one crucial difference: in this system, the protoplanets starting outside of the ice line have grown to masses approaching 10 $\mearth$ instead of only about 5 $\mearth$ as in Figure \ref{fig:SyntSim090}. Given the $M_{\rm p}^{-1}$ dependence of the Type I migration timescale (equation \ref{eqn:mig}), this has the consequence that three volatile-rich sub-Neptunes have migrated to final orbital periods $P<100$ days. These planets have larger radii, $3-6 \ \rearth$ and belong to the migration-dominated formation pathway. At the same time, the system also contains two inner planets with lower masses and radii between 1 and 2 $\rearth$. They formed via giant impacts (as all planets inside 100 days in System 90) and have a rocky composition. The existence of different formation pathways and the different resulting compositions within one system thus leads to the larger spread in radii. The difference between System 90 and 164 is caused by the different initial conditions: System 164 contains initially 27 $\mearth$ of planetesimals more than System 90. This has the consequence that $M_{\rm eq}$ is higher in System 164 than 90, and the tracks indeed bend inward at higher masses. The mass difference is initially small (just a few Earth masses), but it is amplified because gas accretion becomes relevant once planet core masses approach 10 $\mearth$. As a result, small mass differences can be  sufficient to instigate different formation pathways for some close-in planets. 

\textbf{(2) Late-stage Giant Impacts.} Another source of diversity arises when giant impacts affect only one or two planets within a system. When protoplanets attain their final masses via many giant impacts starting from numerous low-mass oligarchs, similar masses result (usually with a slight positive trend with orbital distance), as seen in System 90 in Figure \ref{fig:SyntSim090}. In contrast, when only a few ($\lesssim 5$) close-in planets exist in a system at the moment of disk dissipation,  a merger of two of them results in a more massive planet that stands out from the others, leading to diversity. Examples of this process are System 93 in Figure \ref{fig:SyntARpeasord}, as well as the most massive planet at about 1 AU in System 90 (Figure \ref{fig:SyntSim090}).

\textbf{(3) Atmospheric Loss.} Finally, diversity is increased when evolutionary effects such as atmospheric evaporation (\S\ref{sec: atmospheric mass loss}) strongly modify the radius of some --- but not all --- planets in a system. This is the case in System 154 in Figure \ref{fig:SyntARpeasord}, where the largest planet has kept its H/He envelope, in contrast to the other planets in the system. Observed examples of this process \citep{Lopez2013,owen2016} might be Kepler-36 \citep{carter2012} or TOI-402/HD 15337 \citep{Dumusque2019}.

It is clear that the specific outcomes shown here are model dependent and would change with variations of important model parameters, such as the planetesimal size, \cm{the spatial distribution of planetesimals,} or the adopted $\alpha_\nu$ viscosity parameter. On the other hand, more quantitatively speaking, many of the processes governing the outcomes are related to fundamental principles of planet formation, like the mass dependencies of planetesimal accretion or orbital migration. The two formation pathways (giant impact-dominated versus migration-dominated) is also seen in a comparable way in pebble-based scenarios \citep{Lambrechts2019}, where the pebble flux plays the same role as the planetesimal surface density in the model considered here. The existence of two formation channels has also been discussed elsewhere (e.g., \citealt{swain2019,venturini2020}). In the work presented here, the two formation channels arise automatically from the global model, which also predicts outcomes that can be directly compared to observations.

\paragraph{\textbf{Temporal Evolution of the Peas-in-a-Pod Patterns.}} A valuable feature of the synthetic population is that it preserves a complete record of the temporal evolution of the peas-in-a-pod patterns. It is found that the mass/size similarity already exists early during the formation of the systems when planets mainly grow via oligarchic planetesimal accretion. This growth mode has the tendency to lead to comparable masses for neighbouring protoplanets, except at special places like the ice line. Later on, dynamical effects like collisions cause the planets within one system to grow at different rates and decrease the degree of the correlations. The spacing and packing follow an opposite temporal evolution: these trends are absent at early times. They only arise later over Myr timescales from the dynamical N-body interactions encoding the requirements of dynamical stability. Observations of the temporal evolution of the properties of the close-in population \citep{Berger2020,Sandoval2021}, and also the dependency on stellar mass \citep{Mulders2018,Wu2019,burn2021}, will be helpful to further improve the theoretical models.

\section{Conclusion} 
\label{sec:conclusion}

The main feature of the observed {\color{red}compact multi-planet} systems is their high degree of regularity (\S \ref{sec:observations}). The constituent planets are observed to have unexpected uniformity in planet size, mass, and orbital spacing  {\color{red}(the ``peas-in-a-pod'' pattern)}. The mutual inclinations are small, nearly zero, and the orbital eccentricities are much smaller than would be necessary for orbital stability. With some exceptions, planetary pairs are not found in mean motion resonance. Taken as a whole, these characteristics indicate that the planet formation process can often produce highly ordered systems. 

This class of structured planetary systems stands in stark contrast to the well-known diversity of planetary systems that characterizes the sample as a whole. The distributions of planet sizes and orbital spacing are much tighter for planets within {\color{red}individual} multi-planet systems than for the entire collection of planets found in all of the multi-planet systems.  In addition, these distributions are much wider for the entire sample of all exoplanets -- not just those found in compact multi-planet systems of interest -- consistent with the overall sample being more diverse than the {\color{red}orderly } peas-in-a-pod {\color{red}patterns that emerge in the compact multis}. 

The remarkable properties of these planetary systems poses an interesting question -- what basic physical principles acting during planet formation lead to such configurations? These systems form within the circumstellar disks produced during star formation (\S \ref{sec:stardisk}). The resulting disk properties are broadly consistent with those needed to produce the peas-in-a-pod pattern: The disks have enough total mass, a sufficient inventory of rocky material, and surface density distributions that are close to those required.\footnote{We note that the disks are also capable of producing other types of planetary systems, where the planet properties are less ordered.}  In order to produce the observed planetary systems from the disks, however, some of the available rocky material must move inward from its initial location and the surface density must become steeper. Another important clue is that such systems --- with nearly equal masses, low eccentricities, and co-planar orbits --- represent the lowest energy state accessible to a forming planetary system (\S \ref{sec:pp interactions}). This finding provides a general understanding of why these planetary systems have their observed properties, but does not describe the detailed path by which they achieve such structured configurations. The current theory of planet formation provides several specific mechanisms that can produce the observed planets with characteristic masses $\mp\sim10\mearth$ (\S \ref{sec:planet-formation-theory}) and population synthesis simulations can successfully reproduce the observed architectures of multi-planet systems (\S \ref{sec:popsynth}). These simulations, in conjunction with the underlying theory, suggest that uniformity can arise when all of the planets in a system have their masses determined by the same mechanism. Nonetheless, a fully predictive theory of planet formation remains elusive. 

The ubiquity of well-ordered {\color{red}compact multis} poses a number of interesting problems for future work. On the observational side, one key issue is to determine masses of the constituent planets in these systems (as only measurements of planetary radii are available in many cases). A related issue is to determine the mass-radius relationship for this class of planets (keeping in mind that it might not be a single-valued function). It is also important to more fully characterize the planetary mass function to see more clearly how the {\color{red}compact multis} fit into the bigger picture. In particular, it will be important to extend the {\color{red}observed mass distribution} down to lower masses, well below 1 $M_\oplus$. When such measurements become available, we can determine whether or not the apparent preference for making $\sim5$ -- 10 $M_\oplus$ planets is real, and if a significant number of additional planets lie between those currently detected. Another interesting study will be to extend observations of multi-planet systems to include longer period orbits to see if the peas-in-a-pod pattern observed in compact systems (typically with $P\le100$ days) persists. The presence or absence of the peas-in-a-pod phenomenon in outer regions of planetary systems will place important constraints on planet formation theory {\color{red} (e.g., see Section \ref{sec:eoptimize}, \citealt{Adams2020}, and references therein).}  {\color{red} Many of the observational goals described above will become tractable in the next decade, via radial velocity follow-up of apparently peas-in-a-pod like planetary systems (to measure the planet masses and search for non-transiting planets), and also the discovery of high-multiplicity systems containing sub-Earth sized planets with long-duration, wide, and deep space-based transit surveys such as the PLATO Mission.}


Our theoretical understanding of planet formation remains incomplete. On one hand, we are starting to understand the large number of sub-processes involved, such as streaming instabilities to form planetesimals and pebble accretion to produce super-Earth-class rocky bodies (\S \ref{sec:planet-formation-theory}). In addition, as proof of principle, population synthesis models that include algorithms for the myriad of required processes can produce planetary systems with the observed properties (\S \ref{sec:popsynth}). On the other hand, a detailed, first-principles understanding remains elusive. While the peas-in-a-pod {\color{red}architecture} is well-described as a minimum energy state (\S \ref{sec:eoptimize}), and any type of energy dissipation will move systems toward lower energy, the mechanisms that lead to such dissipation require further study. 

Another important unresolved issue is that of planet migration. After becoming part of the circumstellar disk, the rocky material that forms planets must move inward. However, both the timing and distances traveled in this process are not fully known. In addition to moving, the solids must evolve from the size to dust grains (0.1 -- 1 $\mu$m), to pebbles (0.1 -- 1 cm), to planetesimals (10 -- 100 km), and finally to planet-sized bodies. The rocky material could move inward during any of the latter three phases. Moreover, the distances migrated, as measured by $(\Delta a)/a$, remain unknown (although population synthesis models are starting to address this issue -- see Fig. \ref{fig:SyntSim090}). As an added complication, migration mechanisms can move planets (and/or smaller bodies) outward as well as inward, depending on disk structure. 

Taking a step back, we can think of the striking uniformity found in compact multi-planet systems as an example of {\it self organization}. In general, any self-organizing system has a primary driving force that acts to create structure and some additional `counter-force' that acts as a stabilizing influence \citep{aschwanden2018}. The compromise between the two processes results in a quasi-equilibrium state with complex structure. In the present context, the peas-in-a-pod {\color{red}architecture typical of compact multis contains} nearly equal planet masses and regular orbital spacing. Gravity acts as an organizing agent that drives the accumulation of mass, but some additional process\footnote{In the case of pebble accretion, for example, the Bernoulli principle increases pebble speeds and ultimately prevents further accretion.} must act in opposition to keep one planet from acquiring all of the available material. Similarly, planetary migration acts an organizing agent to push planetary orbits together, but some additional process\footnote{For example, turbulence in a gaseous disk and/or quasi-random N-body interactions from resonance overlap could play this role.} must prevent the orbits from becoming too close and hence unstable. With the peas-in-a-pod pattern becoming increasingly well-established, the challenge for the future is to identify (and study) the driving forces and counter-forces that lead to the production of such interesting planetary systems.


\bigskip
\textbf{Acknowledgments.} We would like to thank the Kavli Institute for Theoretical Physics at the University of California Santa Barbara for hosting a collaboration workshop that facilitated this work. We also thank Yann Alibert for many useful conversations at the beginning of this collaboration. We thank Remo Burn, Lokesh Mishra, and Alexandre Emsenhuber for help with the preparation and interpretation of the figures for the synthetic planet populations.  {\color{red} AMB was supported by NSF grant DMS-2103026.  This research was supported in part by the National Science Foundation under Grant No. NSF PHY-1748958.  Finally, we thank two anonymous referees for detailed comments that helped improve the manuscript.} 

\bigskip

\bibliographystyle{pp7}
\bibliography{pp7}

\end{document}